\documentclass[11pt]{article}
\usepackage[a4paper,left=3.0cm,right=3.0cm,top=3.0cm,bottom=3.0cm]{geometry} 
\usepackage{amsmath,amsthm,amsfonts,amssymb,natbib,graphicx,booktabs,float,caption,subcaption,multirow,dsfont,setspace,chngcntr}
\usepackage{enumerate}
\usepackage{natbib}
\usepackage{url} 

\usepackage{hyperref} 
\usepackage{xr} 

\graphicspath{{./graphics/}} 
\usepackage{MnSymbol} 
\usepackage{econometrics} 
\usepackage{bm}
\usepackage{dcolumn} 
\usepackage{bookmark}

		\usepackage[colorinlistoftodos,textsize=tiny]{todonotes}
		\newcommand{\Comments}{1}
		\newcommand{\mynote}[2]{\ifnum\Comments=1\textcolor{#1}{#2}\fi}
		\newcommand{\mytodo}[2]{\ifnum\Comments=1%
			\todo[linecolor=#1!80!black,backgroundcolor=#1,bordercolor=#1!80!black]{#2}\fi}

		\ifnum\Comments=1               
		\fi

		\ifnum\Comments=1               
		\fi

		\ifnum\Comments=1               
		\fi

\newcommand{\Var}{\mathrm{Var}}

\newcommand{\Cov}{\mathrm{Cov}}

\newcommand{\bs}[1]{\bm{#1}}

		\theoremstyle{definition}

\theoremstyle{plain}
\newtheorem{algorithm}{Algorithm}
\newtheorem{assumption}{Assumption}
\newtheorem{proposition}{Proposition}
\newtheorem{lemma}{Lemma}
\newtheorem{corollary}{Corollary}

\usepackage{appendix}
\usepackage{chngcntr}
\usepackage{apptools}
\AtAppendix{\counterwithin{equation}{section}}
\AtAppendix{\counterwithin{figure}{section}}
\AtAppendix{\counterwithin{table}{section}}
\AtAppendix{\counterwithin{lemma}{section}}
\AtAppendix{\counterwithin{assumption}{section}}
\AtAppendix{\counterwithin{proposition}{section}}
\AtAppendix{\counterwithin{corollary}{section}}

\usepackage[edges]{forest}
\usetikzlibrary{arrows.meta}
\tikzset{line/.style={draw,-Stealth}}
\forestset{declare boolean={my special}{0}}

\usepackage{authblk}

\title{Simultaneous Inference Bands for Autocorrelations}

\author[1]{Uwe Hassler}
\author[2,3]{Marc-Oliver Pohle}
\author[1]{Tanja Zahn}

\affil[1]{Goethe University Frankfurt, Faculty of Economics and Business}
\affil[2]{University of Wuppertal, Schumpeter School of Business and Economics}
\affil[3]{Heidelberg Institute for Theoretical Studies, Computational Statistics Group}

\affil[ ]{\small 

\textit{E-mail: }
\texttt{\href{mailto:hassler@wiwi.uni-frankfurt.de}{hassler@wiwi.uni-frankfurt.de}},
\texttt{\href{mailto:pohle@uni-wuppertal.de}{pohle@uni-wuppertal.de}},
\texttt{\href{mailto:tzahn@wiwi.uni-frankfurt.de}{tzahn@wiwi.uni-frankfurt.de}}
}

\begin{document}

\maketitle	

\begin{abstract}

Sample autocorrelograms typically come with significance (non-rejection) bands for the null hypothesis of no temporal correlation. These bands have three shortcomings.  First, they build on pointwise intervals and suffer from joint undercoverage (overrejection) under the null hypothesis. Second, in cases where this null is clearly violated one would rather prefer to see confidence bands to quantify estimation uncertainty. Third, they are invalid under conditional heteroskedasticity. We propose both simultaneous significance and confidence bands for time series and series of regression residuals. Our simultaneous inference bands are as easy to construct as their pointwise counterparts and at the same time provide an intuitive and visual quantification of sampling uncertainty as well as valid statistical inference. For residuals from dynamic regressions, we show how our inference bands from the case of observed time series and static regressions need to be adjusted due to a correction term in the asymptotic variances. For all our bands, we also provide robust versions, allowing for conditional heteroskedasticity. We analyse the finite-sample performance of our inference bands in a simulation study and illustrate their use in applications to monthly US inflation, Fama-French excess returns and residuals from Phillips curve regressions.

\end{abstract}

\noindent%
{\it Keywords:}  Autocorrelogram, confidence bands, joint significance, regression residuals.


\section{Introduction}	\label{sec:introduction}

\begin{figure}
                            \noindent \centering{}\includegraphics[width=\textwidth]{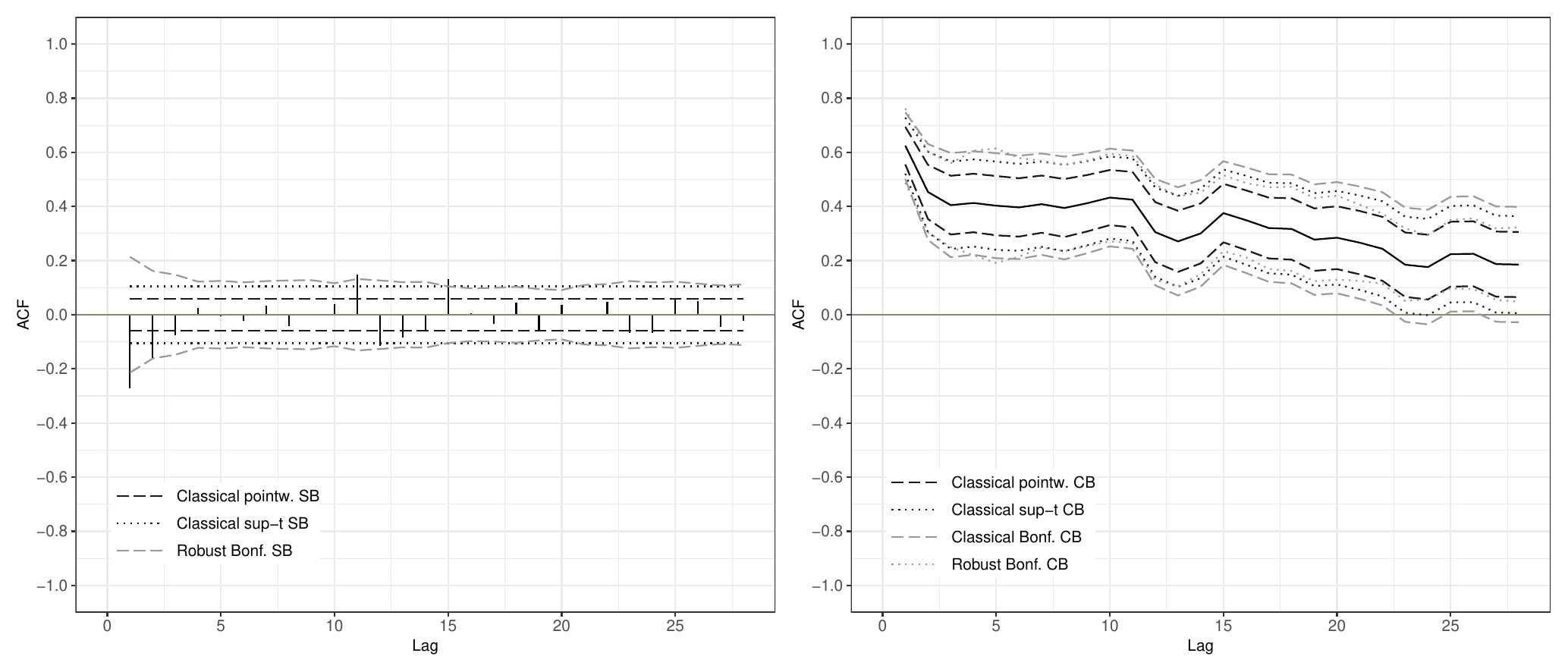}
							\caption{Left: ACF with  different types of $90\%$ significance bands for regression residuals (from a static Phillips curve).  Right: ACF with different types of $90\%$ confidence bands for monthly US inflation.}	
							\label{fig:phillips_and_inflation}
					\end{figure}

			The sample autocorrelation function (ACF) is  the common starting point to the statistical analysis of a time series or a series of regression residuals, giving a first overview of temporal dependence. The sample autocorrelogram is usually accompanied by a pointwise non-rejection band for the null hypothesis of temporal independence. 
            Under this hypothesis, the  empirical autocorrelation at a certain lag falls inside this band with a probability of $1 - \alpha$, for example 0.9; see the narrowest band (labelled classical pointw.\ SB) in the left graph of Figure~\ref{fig:phillips_and_inflation} for an example for regression residuals from a static Phillips curve regression, for details see Section \ref{sec:case_studies}. We call such bands, which represent regions of insignificance of a certain null, significance bands, following e.g.\ \citet{FrancqZakoian2009}.
            Those pointwise significance bands for ACFs are discussed in basically every textbook on time series analysis (e.\,g.,  \citet{BrockwellDavis91}, \citet{Hamilton94}, \citet{Fuller96}, \citet{shumway2017}) and are added to plots of empirical ACFs by default in most statistical software.

			There are three shortcomings of those significance bands. First, they are pointwise bands, arising as the Cartesian product of significance intervals of a certain level $1 - \alpha$ for individual autocorrelations. As the null hypothesis refers to the whole ACF, we are rather interested in joint inference, for which the pointwise bands are not valid. Simultaneous (or joint or uniform) bands are appropriately scaled-up versions of pointwise bands and provide valid joint inference, that is, we can reject the temporal independence hypothesis on the significance level $\alpha$ if the empirical ACF (up to a certain maximum lag) leaves the simultaneous significance band of level $1 - \alpha$ at any point. Amended with those bands, the correlogram provides an overview of the dependence structure and valid inference on the hypothesis of temporal independence at a glance. Figure~\ref{fig:phillips_and_inflation} (left graph) contains a simultaneous significance band for this null hypothesis, too (labelled classical sup-t SB). 
			
			The second issue concerns significance bands in general. They are tied to a certain null hypothesis, which might not always be of particular interest. For example, for an inflation series no one would seriously doubt that there is serial dependence. What is always of interest is a quantification of sampling uncertainty as provided by confidence bands. While significance bands are much more common, sometimes pointwise confidence bands are used. Like the pointwise significance bands, they are only valid for one autocorrelation at a single lag, while one is usually interested in multiple autocorrelations and joint inference for them. Again, simultaneous confidence bands are a natural choice here, see the right graph in Figure~\ref{fig:phillips_and_inflation}, which shows the empirical ACF for monthly US inflation alongside a pointwise and two types of simultaneous confidence bands under classical assumptions (labelled classical sup-t CB and classical Bonf.\ CB), for details see again Section \ref{sec:case_studies}. 
            The simultaneous confidence bands cover the whole ACF up to a chosen maximum lag with a specified (asymptotic) coverage probability of at least $1-\alpha$. 
			While simultaneous confidence bands are widespread, for example, in impulse response analysis or in nonparametric regression (see \citet{montielolea2019} and references therein), they surprisingly have not been used in the context of autocorrelations to the best of our knowledge. The same seems to be the case for simultaneous significance bands, which have recently been introduced for impulse responses by \citet{inoue2025} and for probability integral transform histograms by \citet{Demetrescu2025}. We subsume both significance and confidence bands under the term inference bands.

			The third issue is that the classical assumptions, under which (pointwise) significance and confidence bands for autocorrelations are usually constructed, do not allow for conditional heteroskedasticity, which is problematic for financial data. For significance bands, the null hypothesis of interest is usually white noise (WN), i.e., uncorrelatedness, or the slightly stronger martingale difference sequence (MDS) hypothesis, but under the classical assumptions it amounts essentially to the process being independent and identically distributed (iid). 
            Similarly, for confidence bands, the classical assumptions exclude conditional heteroskedasticity. Figure \ref{fig:Fama_french} in Section \ref{sec:case_studies} shows the empirical ACF of daily Fama-French excess returns alongside heteroskedasticity-robust significance bands (labelled robust Bonf. SB). They are substantially wider than the usual pointwise significance bands, but also than the simultaneous bands constructed under classical assumptions.
						 
			In this paper, we introduce simultaneous inference bands for ACFs of stationary time series as well as regression errors. The first key ingredient of the bands is asymptotic normality of empirical autocorrelations and, in particular, consistent covariance estimation. 
            For time series, we distinguish between two formulas for the asymptotic covariance matrix: Bartlett's formula \citep{Bartlett1946}, which holds under classical assumptions, and the general formula put forward by \citet{RomanoThombs1996}, which holds under mild assumptions. For their estimation, we rely on nonparametric estimators, which resemble long-run variance or heteroskedasticity and autocorrelation consistent (HAC) estimation, see \citet{melard1987} for the first case and  \citet{lobato2002} for the second. We construct our classical confidence bands relying on the first estimator, our heteroskedasticity-robust confidence bands relying on the second. The second ingredient for our inference bands is their construction principle, see \citet{montielolea2019} for a comprehensive treatment of different types of confidence bands under asymptotic normality. For all of our bands, we discuss two variants, sup-t and Bonferroni, differing in their scaling factor of the standard error. 
            While the sup-t bands have exact asymptotic coverage, the Bonferroni bands are always wider and thus asymptotically conservative. 
            For our classical bands, we prefer the sup-t variant. For our robust bands, we recommend the Bonferroni variant since, with its asymptotic conservativeness, it is able to mitigate the notorious problems associated with HAC-type estimators, which the estimator of \citet{lobato2002} shares. 
            Furthermore, it only requires estimation of the diagonal elements of the covariance matrix, giving it a certain robustness to estimation error in the many off-diagonal elements.
            
            Simultaneous significance bands under the iid null 
            are particularly easy to construct and do not even require variance estimation since the covariance matrix reduces to the identity matrix. 
            We also introduce robust significance bands for the hypotheses of MDS and WN. Under MDS, the asymptotic covariance matrix simplifies considerably as well \citep{guo2001} 
            and circumvents HAC-type estimation. 
            
            For regression errors, we distinguish between static and dynamic regressions and several assumptions regarding (non)stationarity, and derive 
            the formulas for the covariance matrices, partly building on results by \citet{CumbaHuizinga1992}. For static regressions under stationarity and exogeneity or under nonstationarity and cointegration, we show that the covariance matrices and thus the construction of inference bands are unaffected by having to replace true regression errors with residuals. For dynamic regressions (with lagged endogenous regressors), significance bands are particularly important since the errors being WN is often imposed to ensure consistent estimation of the regression coefficients. For them, the asymptotic covariance matrices need to be modified. We propose a shrinkage algorithm that ensures positive definiteness of their plug-in estimators.  
            Interestingly, under the classical assumptions, the bands ignoring the residual effect are also valid, but wider than the corrected bands and thus conservative.             

			Our inference bands show a good finite-sample performance in our extensive simulation study. In particular, our classical simultaneous significance bands 
            show a comparable or even better performance than the classical tests of temporal independence of \citet{BoxPierce70} and \citet{LjungBox78}. The same holds true for our robust significance bands compared to robust variants of the Box-Pierce test (\citet{guo2001}, \citet{Shao2010}). Regarding the coverage properties of our confidence bands, remarkably the variance estimator of \citet{melard1987} does not seem to suffer from the same problems as other HAC-type estimators (including the one by \citet{lobato2002}), which are well-known to cause oversizing of tests and undercoverage of confidence bands (see e.\,g., \citet{lazarus2018}). 
			
			We illustrate the use of our inference bands in applications to monthly US inflation, Fama-French excess returns, and regression residuals from static and dynamic Phillips curve regressions.

\begin{figure}[h]
\centering
{\scriptsize
\scalebox{0.93}{ 
            \begin{forest}
  warn/.append style = {
    font = \sffamily\footnotesize\bfseries,
  },
  box/.append style = {
    rectangle,
    font = \sffamily\scriptsize\bfseries,
    draw          = black,
    align         = center,
  },
  special/.append style = {
    rectangle,
    font = \sffamily\scriptsize\bfseries,
    draw          = black,
    align         = center,
    child anchor  = 140,
    tikz+={
                            \draw[-{Stealth[scale=1,angle'=45]}] (A)
           .. controls +(right:2cm) and +(west:2cm) ..
           node[midway, fill = white]{no}
           (2);
    }
  },
  for tree = {
    font = \sffamily\scriptsize,
    l sep'           = 10mm,
    s sep'           = 2.5mm,
    minimum width   = 4.0em,
    line width    = 1pt,
    child anchor  = north,
    parent anchor = south,
    tier/.option=level,
    l sep*=2,
  },
      forked edges,
                     edge={-Straight Barb}, 
                                        EL/.style = {edge label={node[below,  fill=white, align = center, pos = 0.6]{#1}},},                                                                               
  box,
  before typesetting nodes={%
    for descendants={%
      if={> O_= O_= | {content}{}{!u.content}{}}
      {
        edge=dotted,
        if n children=0 {
          edge=line 
        } {
          coordinate 
        }
      }{
        edge=line,
        if={> {O} {my special} }%
        {
          special 
        }{
          box 
        }
      }
    },
  }
    [data
      [lagged endogenous  regressors?, EL=regression residuals, name = A
          [testing which hypothesis \\ on the errors?, EL = yes
              [corrected robust \\ Bonf. SB \\ from \eqref{eq:significance_band_dynamic_regression_robust}, 
              EL =  MDS
            ]
                          [corrected classical \\ sup-t SB  \\ from \eqref{eq:significance_band_dynamic_regression}, EL = homosk. MDS or iid
            ]
          ]
        ]
        [main goal, EL = observed time series, name=2, my special
            [which hypothesis?, EL = {testing a white noise hypothesis}
            [robust Bonf. SB \\ from \eqref{eq:simultaneous_significance_band_rho_Bartlett_star}, 
            EL = MDS]
            [classical sup-t SB \\ from \eqref{eq:simultaneous_significance_band_rho}, EL = iid]
            ]
            [heteroskedastic?, EL = quantifying sampling uncertainty
                [robust Bonf. CB \\ from \eqref{eq:Bonferroni_rho} with \\ bandwidth \\ $(T-H)^{1/3}$, EL = yes]
                [classical sup-t CB \\ from \eqref{eq:supt_rho} with \\ bandwidth \\ $T^{1/2}$, EL = no]
            ]
        ]
    ]
\end{forest}
}
}
\caption{Decision tree for recommended bands and bandwidth choice, depending on the setting. Time series length is denoted by $T$ and the length of the band is denoted by $H$.}
\label{fig:decision_tree}
\end{figure}

            Figure \ref{fig:decision_tree} offers guidance in the form of a decision tree as to which bands we recommend for which setting. If the ACF at hand stems from an observed time series or from residuals of a static regression, one has to decide if one is primarily interested in testing the WN hypothesis. If yes, then one should choose significance bands, if not, confidence bands. If the ACF stems from regression residuals from a dynamic regression, significance bands are the right choice. The last question one has to answer in all these cases is if substantial conditional heteroskedasticity is likely to be present in the series or not, which amounts to deciding between the MDS and iid null hypotheses for the tests of white noise. If yes, as usually in financial time series, one should choose the robust version of our inference bands, if not, as in many macroeconomic applications, one should go with the classical version of our bands. We also include a recommendation for the bandwidth choice for the HAC-type variance estimators used for our confidence bands, for details see Section \ref{subsec:simulation_CB}.
						
			The rest of the paper is structured as follows. In Section \ref{sec:inference_bands_general} we discuss inference bands under asymptotic normality. In Section \ref{sec:inference_bands} we introduce our inference bands for autocorrelations for the time series case. Section \ref{sec:residuals} treats the asymptotic distribution of empirical autocorrelations of regression residuals and introduces the inference bands for regression errors. Section \ref{sec:simulations} presents the simulation study, Section \ref{sec:case_studies} the empirical applications, and Section \ref{sec:conclusion} concludes. The Appendix contains proofs of all propositions and lemmas in Section \ref{Sect:Proof} and additional material referenced throughout the paper.
           We provide the accompanying R package \texttt{ACFbands} at \url{https://github.com/TanjaZahn/ACFbands} and replication material at \url{https://github.com/TanjaZahn/ACFbands_replication}. 

\section{Inference Bands under Asymptotic Normality}
\label{sec:inference_bands_general}

 To fix notation, let $\lfloor x \rfloor = \max\left\lbrace m \in \mathbb{Z} \, | \, m \leq x \right\rbrace $, $x \in
			\mathbb{R}$. Further, bold capital letters stand for matrices with the $H$-dimensional identity matrix $\mI_H$, and bold lower case letters denote vectors. 
            Let $T$ stand for the sample size, 
            and $z_\tau$ for the $\tau$-quantile of the standard normal distribution. As $T\rightarrow \infty$, let $\Rightarrow$ stand for weak convergence and $\stackrel{d}{\to}$ specifically for convergence in distribution; $\stackrel{p}{\to}$ is short for convergence in probability.

Assume an $H$-dimensional parameter vector of interest $\vtheta = (\theta_1,\ldots,\theta_H)^\prime$ with a limiting normal estimator $\widehat \vtheta = (\widehat{\theta}_1,\ldots, \widehat{\theta}_H)^\prime$:
\begin{equation} \label{eq:Normal_theta}
	{\sqrt{T}} \left( \widehat{\bs{\theta}}  - \bs{\theta}	\right) \ \stackrel{d}{\to} \ \mV \sim \mathcal{N}_H (\bs{0}, \, \bs{\Sigma}) \, .
\end{equation}
Here, $\bm{\Sigma}  = \left(\sigma_{gh}\right)_{g,h = 1, \ldots, H}$ is the covariance matrix of the Gaussian random vector $\mV$. Let $M$ denote  the maximum of the absolute values of correlated standard normal variates:
\begin{equation} \label{eq:M_max_normal}
	M :=  \max_{h = 1, ..., H} \sigma_{hh}^{-1/2} |  V_h| \ \mbox{ with }  P(M \leq q_\tau (\mSigma)) = \tau \, ,
\end{equation}
and $q_\tau (\mSigma)$ is short for the $\tau$-quantile of $M$. Note that $q_\tau (\mSigma)$ is sometimes called an equicoordinate quantile of the multivariate normal distribution with covariance matrix $\bs \Sigma$ as it fulfils $ P \left(\sigma_{11}^{-1/2} |  V_1| \leq q_\tau (\mSigma), \ \dots \ ,  \sigma_{HH}^{-1/2} |   V_H| \leq q_\tau (\mSigma) \right)  = \tau \, $.
Equicoordinate quantiles can be easily calculated, 
e.\,g., by the R package \texttt{mvtnorm} \citep{genz2023}, which we use. 

%

A univariate confidence interval for $\theta_h$ at confidence level $1-\alpha$ is given by $\left[ \widehat{\theta}_{h} \pm  z_{1-\alpha/2} \cdot \sqrt{\frac{\sigma_{hh}}{T}}\right]$. 
A simultaneous rectangular confidence region of confidence level $1-\alpha$, which amounts to a confidence band when plotting $\widehat \theta_h$ against $h$, is given by the Cartesian product of scaled-up univariate confidence intervals: $\bigtimes_{h=1}^H \left[\widehat \theta_h \pm c \cdot  \sqrt{\frac{\sigma_{hh}}{T}}\right]$.
Just combining the confidence intervals for every $\theta_h$, that is, choosing the scaling factor $c$ as $z_{1-\alpha/2}$, would lead to serious undercoverage. Thus, these pointwise confidence bands need to be scaled up appropriately to achieve a joint or simultaneous coverage of at least $1 - \alpha$. We consider two choices for $c$. 
First, we 
use so-called sup-t bands \citep{montielolea2019}. They build on equicoordinate quantiles and are (asymptotically) exact by construction:
\begin{equation} \label{eq:sup-t_general}
	{CB}_{\vtheta}^{supt}(\mSigma)  = \bigtimes_{h=1}^H \left[\widehat \theta_h \pm q_{1-\alpha} (\mSigma) \sqrt{\frac{\sigma_{hh}}{T}}\right]  
\end{equation}
with $\lim_{T \to \infty} P \left(\vtheta \in {CB}_{\vtheta}^{supt} (\mSigma) \right) =  1 - \alpha$.
Second, we consider Bonferroni bands closely related to the Bonferroni correction in multiple testing:
\begin{equation} \label{eq:Bonferroni_general}
{CB}_{\vtheta}^{bf} (\mSigma) = \bigtimes_{h=1}^H \left[\widehat \theta_h \pm z_{1-\alpha/(2H)}  \sqrt{\frac{\sigma_{hh}}{T}}\right]
\end{equation}
such that $\lim_{T \to \infty} P \left(\vtheta \in {CB}_{\vtheta}^{bf} (\mSigma) \right) \geq 1 - \alpha$, see Lemma \ref{lemma:asymptotic_coverage_general}. 
Bonferroni bands are always at least as wide as sup-t bands due to $q_{1-\alpha} (\mSigma) \leq  z_{1-\alpha/(2H)}$, see Lemma \ref{lemma:width} in Appendix \ref{appsubsec:additional_theoretical_results} and the  discussion in \cite{montielolea2019}. Under independence (i.\,e., under $\mSigma=\mI_H$), the Bonferroni band becomes virtually identical to the sup-t band \citep{proschan2011}, but under dependence it can be considerably wider and thus conservative in the sense of having asymptotic overcoverage, see Appendix \ref{subsec:theoretical_example_time_series} for further discussion and illustration. A possible advantage of the Bonferroni band is that it only requires knowledge (estimation) of the diagonal elements of $\mSigma$.  

Consider now a null hypothesis $H_0$ that implies a certain value $\vtheta_0 = (\theta_{1,0},\ldots,\theta_{H,0})^\prime$ for the parameter vector and possibly a certain restriction on the covariance matrix, which we then call $\bm{\Sigma}_0  = \left(\sigma_{gh,0} \right)_{g,h = 1, \ldots, H}$. A simultaneous rectangular region of non-rejection, which we call simultaneous significance band, is given by the Cartesian product of scaled-up univariate non-rejection regions: 
$\bigtimes_{h=1}^H \left[\widehat \theta_{h,0} \pm c \cdot  \sqrt{\frac{\sigma_{hh,0}}{T}} \right]$. Similarly to the case of confidence bands, choosing $c=z_{1-\alpha/2}$ leads to a pointwise significance band. 
Choosing $c$ as the respective equicoordinate quantile leads to the sup-t simultaneous significance band:
\begin{equation} \label{eq:S_general}
{SB}^{supt}_{\vtheta} (\mSigma_0) = \bigtimes_{h=1}^H \left[ \theta_{h,0} \pm q_{1-\alpha} (\mSigma_0) \sqrt{\frac{\sigma_{hh,0}}{T}}\right] \, .
\end{equation}
Picking $c=z_{1-\alpha/(2H)}$ yields the Bonferroni significance band:
\begin{equation} \label{eq:S_general_bf}
{SB}^{bf}_{\vtheta} (\mSigma_0) = \bigtimes_{h=1}^H \left[ \theta_{h,0} \pm z_{1-\alpha/(2H)} \sqrt{\frac{\sigma_{hh,0}}{T}}\right] \, .
\end{equation}
The sup-t significance band covers $\widehat{\bs{\theta}}$ with probability $1-\alpha$ as $T \rightarrow \infty$ under $H_0$, i.e., the asymptotic size is $\alpha$. The Bonferroni band is again always wider and thus asymptotically conservative. 
We state the coverage properties of the inference bands \eqref{eq:sup-t_general} to \eqref{eq:S_general_bf} in a lemma.

\begin{lemma} \label{lemma:asymptotic_coverage_general}
    Assume that \eqref{eq:Normal_theta} holds 
    with positive definite and finite $\mSigma$ and $\mSigma_0$. It holds that
    \begin{enumerate}
    \renewcommand{\labelenumi}{(\roman{enumi})}
    \item $\lim_{T \to \infty} P \left(\vtheta \in {CB}_{\vtheta}^{bf} (\mSigma) \right) \geq \lim_{T \to \infty} P \left(\vtheta \in {CB}_{\vtheta}^{supt} (\mSigma) \right) =  1 - \alpha,$
    \item $\lim_{T \to \infty} P \left(\widehat \vtheta \notin {SB}^{bf}_{\vtheta} (\mSigma_0) \right) \leq \lim_{T \to \infty} P \left(\widehat \vtheta \notin {SB}^{supt}_{\vtheta} (\mSigma_0) \right) = \alpha$ under $H_0$.
    \end{enumerate}
\end{lemma}




\section{Inference Bands for Autocorrelations} \label{sec:inference_bands}

	

\subsection{Asymptotic Distribution and Variance Estimation}

Throughout, we work under the assumption that the underlying stochastic process $\{y_t\}_{t \in \mathbb{Z}}$ is covariance stationary. Define the autocovariances and autocorrelations  as $\gamma_y (h) := \Cov \left( y_t, \, y_{t+h} \right)$  and  $\rho_y (h) := \frac{\gamma_y (h)}{\gamma_y (0)}$  for  $h \in \mathbb{Z}$. 
 Denote the  vector of  autocorrelations up to a maximum lag $H \in \mathbb{N}$ by 
$\bm{\rho}_y := (\rho_y(1),...,\rho_y(H))^\prime$, while $\bm{\gamma}_y := (\gamma_y(0), \gamma_y(1),...,\gamma_y(H))^\prime$ has $H+1$ entries. 
			Given a time series $\{y_t\}_{t=1}^T$ of length $T$, we estimate the autocovariances and autocorrelations via $\widehat{\gamma}_y (h) := \frac{1}{T} \sum_{t=1}^{T-h} (y_t - \overline y) (y_{t+h} - \overline y)$  for $h \geq 0$ with $\overline y := \frac 1 T \sum_{t=1}^T y_t$  and $\widehat{\rho}_y (h) := \frac{\widehat{\gamma}_y (h)}{\widehat{\gamma}_y (0)}$.
			Denote the empirical autocorrelation vector up to order $H$ by 
            $\widehat{\bm{\rho}}_y := \left( \widehat{\rho}_y(1),...,\widehat{\rho}_y(H) \right)^\prime$. 
We work under high-level assumptions of limiting normality with a certain form of the covariance matrix and briefly discuss sufficient conditions implying our assumptions. 
Define the auxiliary $(H+1) \times 1$ vector process $\{\bm{w}_t\}$ by $\bm{w}_t := (w_{0,t}, w_{1,t}, \ldots, w_{H,t})^\prime$ with $w_{h,t} := (y_t - \mu) (y_{t-h} - \mu) ,\ h=0,1, \ldots, H$, and its long-run covariance matrix $\bm{C}$:
\begin{equation} \label{CovC}
    \bm{C} =(c_{gh})_{g,h=0, \ldots, H} := \sum_{d= - \infty}^\infty \left( E[\bm{w}_t \bm{w}_{t+d}^\prime] - E[\bm{w}_{t}] E[\bm{w}_{t+d}^\prime]\right) \, .
\end{equation}

	\begin{assumption} \label{ass:process}
Let the covariance stationary process $\{y_t - \mu\}_{t \in \mathbb{Z}}$ satisfy the following central limit theorem (CLT)
\begin{equation} \label{eq:NormalV_generalBartlett}
				{\sqrt{T}} \left( \widehat{\bs{\rho}}_y  - \bs{\rho}_y	\right) \ \stackrel{d}{\to} \ \mathcal{N}_H (\bs{0}, \, \bm{B}^* ) \quad \mbox{as } T \to \infty \, ,
			\end{equation}
where the asymptotic covariance matrix $\bm{B}^* =(b_{gh}^*)_{g,h=1, \ldots, H} $ is positive definite with
\begin{equation} \label{CovBartlett_star}
b_{gh}^* := \frac{1}{(\gamma_y(0))^2} (c_{gh} - \rho_y(g) c_{0h} - \rho_y(h) c_{0g}   + \rho_y(g)\rho_y(h) c_{00} ).
\end{equation}
			\end{assumption}

\citet{RomanoThombs1996} were the first to establish this general formula for the asymptotic variance and showed that Assumption \ref{ass:process} is implied by assuming an $\alpha$-mixing process balancing the degree of allowed temporal dependence and the order of required  moments beyond finite fourth moments; 
see also \citet{lobato2002} 
under the assumption of near epoch dependence. 

We also work under the stronger assumption that the asymptotic covariance matrix follows the classical \citet{Bartlett1946} formula.


\begin{assumption} \label{ass:Bartlett} Let Assumption \ref{ass:process} hold with the asymptotic covariance matrix $\bm{B}^*$ being equal to $\bm{B} =(b_{gh})_{g,h=1, \ldots, H} $ with
\begin{equation} \label{eq:Bartlett}
				b_{gh} := \sum_{k=1}^\infty \left[ \rho_y (k+g)+\rho_y (k-g)-2 \rho_y (k)\rho_y (g)\right] \left[\rho_y (k+h)+\rho_y (k-h)-2 \rho_y (k)\rho_y (h)\right] \, .
			\end{equation}
			\end{assumption}

 The original formula for $\mB$ by \citet{Bartlett1946} has been expressed conveniently as given in (\ref{eq:Bartlett}) by \citet[eq.\,(7.2.5)]{BrockwellDavis91}.  Limiting normality with $\bm{B}$ is a classical textbook result. It was established for linear processes, i.e., $y_t - \mu = \sum_{j=0}^\infty \varphi_j \varepsilon_{t-j}$ with iid innovations $\{\varepsilon_{t}\}$ with finite variance, by  \citet{AndersonWalker1964} relying on $\sum_{j=0}^\infty j \varphi_j^2 < \infty$; see also \citet[Cor.\,6.3.6.1]{Fuller96} or \citet[Thm.\,7.2.2]{BrockwellDavis91}. The restriction to iid innovations was relaxed by \citet[Thm.~3]{HannanHeyde1972} who assumed a conditionally homoskedastic MDS and $\sum_{j=0}^\infty \sqrt{j} \varphi_j^2 < \infty$. Correlation of the squared noise sequence, however, is not covered by Assumption~\ref{ass:Bartlett}, excluding, e.g., generalized autoregressive conditional heteroskedasticity (GARCH) noise, see \citet{FrancqZakoian2009}.

 

To estimate $\bm{B}^*$, we first replace all $\rho_y (h)$ in $b_{gh}^*$ in \eqref{CovBartlett_star} by their empirical analogues $\widehat{\rho}_y (h)$. Second, we take $\widehat c_{g h}$ 
from $\widehat{\bm{C}}$ estimated as long-run variance matrix of $\bm{w}_t$ motivated by (\ref{CovC}). To this end, define $\widehat{\bm{w}}_t := (\widehat{w}_{0,t}, \widehat{w}_{1,t} , \ldots, \widehat{w}_{H,t})^\prime$ for $t=H+1, \ldots, T$ with $\widehat{w}_{h,t} := (y_t - \overline y) (y_{t-h} - \overline y)$ and $\overline{\widehat{\vw}} := \frac{1}{T-H} \sum_{t=H+1}^T  \widehat{\vw}_t$. Further, define $ \widehat{\bm{\Gamma}}_w (d) := \frac{1}{T-H} \sum_{t=1+H}^{T-d} (\widehat{\bm{w}}_t - \overline{\widehat{\vw}}) (\widehat{\bm{w}}_{t+d} - \overline{\widehat{\vw}})^\prime$ for $d \geq 0$ and $ \widehat{\bm{\Gamma}}_w (d) := \frac{1}{T-H} \sum_{t=1+H-d}^{T} (\widehat{\bm{w}}_t - \overline{\widehat{\vw}}) (\widehat{\bm{w}}_{t+d} - \overline{\widehat{\vw}})^\prime$ for  $d < 0$.
Then, we estimate
\begin{equation} \label{C_hat}
    \widehat{\bm{C}} := \sum_{d=-T+1+H}^{T-1-H} K \left( \frac{d}L \right) \widehat{\bm{\Gamma}}_w (d) \, 
\end{equation}
where $K(\cdot)$ denotes a symmetric kernel and $L$ a bandwidth. We call the resulting estimator $\widehat{\bm{B}}^*$ and its elements $\widehat{b}_{gh}^*$. This estimator is a slight modification of the estimator by \citet{lobato2002}, 
see also Subsection \ref{subsec:significance_bands}, which has been applied by \citet{Shao2010}, too. Note that we divide by $T-H$ when calculating the covariance matrices above since this is the sample size of the auxiliary time series $\widehat{\bm{w}}_t$.

\citet{melard1987} proposed to estimate $\bs B$ by replacing each $\rho_y(h)$ in \eqref{eq:Bartlett}  by $K(h/L) \widehat{\rho}_y(h)$, 
where $K(\cdot)$ and $L$ again denote a kernel and a bandwidth. 
The M\'elard-Roy estimator amounts to:
\begin{align*} \label{eq:Melard_Roy_estimator}
	\widehat{b}_{gh} =& \sum_{k=1}^{T-1} \left[ K \left( \frac{k+g}{L} \right) \widehat{\rho}_y (k+g)+ K \left( \frac{k-g}{L} \right) \widehat{\rho}_y (k-g)-2 K \left( \frac{k}{L} \right) K \left( \frac{g}{L} \right) \widehat{\rho}_y (k)  \widehat{\rho}_y (g)\right] \nonumber \\ 
	&\cdot \left[  K \left( \frac{k+h}{L} \right) \widehat{\rho}_y (k+h)+ K \left( \frac{k-h}{L} \right) \widehat{\rho}_y (k-h)-2 K \left( \frac{k}{L} \right) \widehat{\rho}_y (k) K \left( \frac{h}{L} \right) \widehat{\rho}_y (h)\right] \, .
\end{align*}
 
Under standard assumptions on the kernel and the bandwidth \citet{melard1987} show that the resulting estimator $\bs{\widehat{B}}$ is positive semidefinite and consistent, and consistency of $\widehat{\bm{B}}^*$ is established in \citet{lobato2002}. 
Following this paper, we use the Bartlett kernel $K (\frac{h}{L}) = \left( 1 - \frac{|h|}{L+1} \right) \mathds{1}\{ |h| \leq L \}$ with indicator $\mathds{1}\{ \cdot \}$ for $\widehat{\bm{C}}$; note, however, that we adopt  the slight modification $ \left( 1 - \frac{|h|}{L} \right) \mathds{1}\{ |h| \leq L \}$ following \citet{melard1987} for $\bs{\widehat{B}}$.
 Even though both estimators are HAC-type estimators, $\bs{\widehat{B}}$ is rather semiparametric, making use of the structure of the Bartlett formula from \eqref{eq:Bartlett}, while $\widehat{\bm{B}}^*$
is essentially a classical HAC estimator for the long-run covariance matrix 
of the auxiliary process $\{\bm{w}_t\}$, 
so fully nonparametric. Thus, not necessarily the same bandwidth choice will work well for the two estimators. Based on our simulations, where we try different bandwidth choices suggested in the literature, we recommend and use the rule-of-thumbs $L = T^{1/2}$ for $\bs{\widehat{B}}$ and $L = (T-H)^{1/3}$ for $\widehat{\bm{B}}^*$. 
Furthermore, $\widehat{\bm{B}}^*$ exhibits the well-known problems of HAC-type estimators \citep{lazarus2018} in our simulations, that is, oversizing of tests and undercoverage of confidence bands, getting worse with smaller samples and stronger temporal dependence, while $\widehat{\bm{B}}$ remarkably does not.

\subsection{Confidence Bands} 
\label{subsec:confidence_bands}

Combining the construction principles of simultaneous inference bands from Section \ref{sec:inference_bands_general} with these variance estimators yields the desired inference bands for autocorrelations. If we work under the classical Bartlett formula from Assumption \ref{ass:Bartlett} and use the variance estimator $\bs{\widehat{B}}$, we call the resulting confidence bands classical bands. When we work under the more general Assumption \ref{ass:process}, which in particular allows for conditional heteroskedasticity, and use the variance estimator $\widehat{\bm{B}}^*$, we call the resulting bands robust bands. As the effective sample size when estimating $\widehat{\mC}$, the key ingredient of $\bs{\widehat{B}}$, is $T-H$, we also use $T-H$ in the construction of the robust bands. 
Regarding the two variants of simultaneous confidence bands, we recommend the sup-t variant of the classical bands because they have exact asymptotic coverage and the estimator $\widehat{\bm{B}}$ works very well. For the robust bands, we recommend the Bonferroni variant since its conservativeness mitigates the undercoverage caused by the HAC-estimator $\widehat{\bm{B}}^*$ and it additionally does not require estimation of the off-diagonal elements, thus being robust to estimation errors for them. \citet{pohle2026} give a similar reasoning for the use of Bonferroni bands.

Combining the sup-t principle from (\ref{eq:sup-t_general}) with $\bs{\widehat{B}}$ leads to the classical sup-t band 
\begin{equation} \label{eq:supt_rho}
	{CB}_{\bm{\rho}_y}^{supt} (\widehat \mB) = \bigtimes_{h=1}^H \left[\widehat  \rho_y (h) \pm q_{1-\alpha} (\widehat \mB)  \sqrt{\frac{\widehat b_{hh}}{T}}\right] \, ,
\end{equation}
while the classical versions of the Bonferroni and pointwise band just replace $q_{1-\alpha} (\widehat \mB)$ by $z_{1-\alpha/(2H)}$ and $z_{1-\alpha/2}$, respectively. 
Replacing $\bs{\widehat{B}}$ by $\widehat \mB^*$ and $T$ by $T-H$ leads to the robust versions of these bands, in particular the robust Bonferroni band:
\begin{equation} \label{eq:Bonferroni_rho}
	{CB}_{\bm{\rho}_y}^{bf} (\widehat \mB^*) = \bigtimes_{h=1}^H \left[\widehat \rho_y (h) \pm z_{1-\alpha/(2H)}  \sqrt{\frac{\widehat b_{hh}^*}{T-H}}\right] \, .
\end{equation}
We collect the coverage properties of the confidence bands in a corollary to Lemma \ref{lemma:asymptotic_coverage_general}, for which we slightly abuse notation and let $\widehat \mB$ and $\widehat \mB^*$ in \eqref{eq:supt_rho} and \eqref{eq:Bonferroni_rho} now denote any consistent estimator of $\mB$ and $\mB^*$, respectively.
\begin{corollary} \label{coro:asymptotic_coverage_ac_confidence_bands}
    \ 
    \begin{enumerate}
    \renewcommand{\labelenumi}{(\roman{enumi})}
    \item Under Assumption \ref{ass:Bartlett} and $ \bs{\widehat{B}} \stackrel{p}{\to} \bs{B}$ it holds that $\lim_{T \to \infty} P \left( \bm{\rho}_y \in {CB}_{\bm{\rho}_y}^{supt} (\widehat \mB) \right) =  1 - \alpha$.
    \item Under Assumption \ref{ass:process} and $ \widehat \mB^* \stackrel{p}{\to} \mB^*$ it holds that $\lim_{T \to \infty} P \left( \bm{\rho}_y \in {CB}_{\bm{\rho}_y}^{bf} (\widehat \mB^*) \right) \geq  1 - \alpha$.
    \end{enumerate}
\end{corollary}


We assumed the number of lags $H$ to be finite, following common practice in the literature on asymptotic inference for autocorrelations discussed above. 
In practice, often $H$ is chosen ad hoc or following a rule-of-thumb that lets $H$ grow slowly with $T$, like the default choice in the R function \texttt{acf}, $\lfloor 10 \, \log_{10} (T) \rfloor$, which we use in our applications. 
Indeed, it is sometimes natural to let $H$ grow with $T$ as larger samples make a more precise estimation of $\widehat{\rho}_y(h)$ for larger $h$ possible, and the interest is often in the whole ACF, in particular for significance bands, where the usual null hypotheses concern all lags. A theoretical treatment for $H \to \infty$ is available from \citet{XiaoWu2014}. With certain deterministic sequences $a_T$ and $b_T$ we know from \citet[Cor.~1]{XiaoWu2014}  under general serial dependence that
\begin{equation} \label{eq:H_to_infty}
   a_T^{-1} \left[\sigma_0^{-\frac 1 2} \max_{1 \leq h \leq H_T} \sqrt{T} |\widehat \rho_y (h) - (1- h/T) \rho_y (h)| - b_T \right] \, \stackrel{d}{\to} \ \mbox{standard Gumbel} ,
\end{equation} 
where the maximum lag $H_T$ may go off to infinity marginally more slowly than $T$. 
This lends itself to construct simultaneous inference bands. Since the extreme value (Gumbel) distribution may provide a poor approximation in finite samples, \citet{XiaoWu2014} discussed a block-of blocks bootstrap method, see also \citet{BraumannKreissMeyer2021} for improved inference via an autoregressive sieve bootstrap. 
While the theoretically appropriate assumption on $H$ ultimately depends on the goal one has in a certain application, an advantage of assuming a fixed $H$ 
is that it allows for inference bands of varying width via the standard deviations $\sqrt{b_{hh}}$ at single lags (as seems to be relevant in applications, see Section \ref{sec:case_studies}), whereas there is a common $\sqrt{\sigma_0}$ in the extreme value asymptotics in \eqref{eq:H_to_infty}. 

In the Appendix (Section~\ref{subsec:theoretical_example_time_series}) we discuss, compare and illustrate the properties  of the different types of confidence bands assuming a stationary autoregressive process of order 1 (AR(1)).

\subsection{Significance Bands} \label{subsec:significance_bands}

The strongest null hypothesis for which we construct significance bands is $H_0^{iid}: y_t \sim iid(0,\sigma^2)$. Under $H_0^{iid}$, Assumption \ref{ass:Bartlett} is fulfilled and Bartlett's formula \eqref{eq:Bartlett} reduces to the identity matrix, $\mB = \mI_H$. Under independence, the equicoordinate quantile from \eqref{eq:M_max_normal} equals $q_\tau (\mI_H) = z_{(1+\tau^{1/H})/2}$, which also shows up in the \v{S}id{\'a}k-correction for multiple testing \citep{sidak1967}. 
Together with \eqref{eq:S_general}, this leads to what we call the classical sup-t significance band, 
which does not require any estimation:
			\begin{equation} \label{eq:simultaneous_significance_band_rho}
				{SB}^{supt}_{\bm{\rho}_y} (\mI_H)  =  \bigtimes_{h=1}^H \left[\pm z_{(1+(1-\alpha)^{1/H})/2} \sqrt{\frac{1}{T}}\right] \, .
			\end{equation}
The classical pointwise band usually accompanying autocorrelograms is
\begin{equation} \label{eq:pointwise_significance_band_rho}
	{SB}^{pw}_{\bm{\rho}_y} (\mI_H)   =  \bigtimes_{h=1}^H \left[\pm z_{1-\alpha/2} \sqrt{\frac{1}{T}}\right] \, .
\end{equation}
Both significance bands are equally simple to construct, but under $H_0^{iid}$, ${SB}^{supt}_{\bm{\rho}_y} (\mI_H)$ provides a valid test with asymptotic size $\alpha$, while the pointwise band is all the more oversized the larger $H$ is: $\lim_{T \to \infty} P \left( \widehat \vrho_y \notin {SB}^{pw}_{\bm{\rho}_y} (\mI_H)   \right) = 1 - (1-\alpha)^H > \alpha \mbox{ for } H \geq 2$.
The pointwise band is arguably a nice and simple graphical diagnostic tool to inspect serial independence by the classical textbook recommendation of counting the number of empirical autocorrelations that fall out of the band and checking if it is close to  $H \alpha$, the expected number under independence. However, the simultaneous band is certainly an at least as nice and simple diagnostic tool, where one simply has to check if the ACF leaves the band, and at the same time a valid inferential procedure. Compared to the classical tests of $H_0^{iid}$ by \cite{BoxPierce70} and \citet{LjungBox78}, the simultaneous significance band adds interpretability due to its graphical representation.

Often, we want to test weaker hypotheses than $H_0^{iid}$ 
and, in particular, allow for conditional heteroskedasticity. The literature on this topic is reviewed in \citet{escanciano2009}. A central issue in this literature is how the asymptotic covariance matrix $\bm{B}^*$ simplifies under different assumptions. Working under the weak hypothesis of only WN, 
$H_0^{WN}: y_t \sim WN(0,\sigma^2)$, the general formula from \eqref{CovBartlett_star} simplifies to ${b}_{gh}^{WN} :=   \frac{c_{gh}}{(\gamma_y(0))^2}$. However, it does not lead to a substantial simplification of the estimation since one still has to estimate the long-run covariance matrix $\bm{C}$ from \eqref{CovC} via $\widehat{\bm{C}}$ from \eqref{C_hat} -- only being able to set $\E[\bm{w}_{t}]=\bm{\gamma}_y$ to its value under the null: $(\gamma_y(0),0,\ldots,0)'$. This is the approach proposed by \citet{lobato2002}. Using this estimator, one can build significance bands for $H_0^{WN}$ via \eqref{eq:S_general_bf}. 

We instead recommend robust significance bands constructed under the slightly stronger null hypothesis of a weakly stationary MDS, $H_0^{MDS}: y_t \sim MDS(0,\sigma^2)$, since the practical relevance of distinguishing between WN and MDS seems limited and for an MDS the estimation of a long-run covariance matrix is no longer necessary. Thus, we can avoid HAC estimation and $\bm{B}^*$ turns into \citep{guo2001}:
\begin{equation} \label{eq:B_MDS}
    \bm{B}^{MDS} =(b^{MDS}_{gh})_{g,h=1, \ldots, H} \quad \text{ with } \quad b_{gh}^{MDS}:= \frac{E[y_t^2 y_{t-g} y_{t-h}]}{(\gamma_y(0))^2}.
\end{equation}
A natural estimator $\widehat{\bm{B}}^{MDS}$ is given by deletion of the first row and the first column of $\widehat{\bm{\Gamma}}_w (0)$ and division of the resulting matrix by $(\widehat{\gamma}_y(0))^2$. More compactly, consider the $H \times (H+1)$ selection matrix $\mS$ selecting the second up to the last row of an $(H+1) \times (H+1)$ matrix, $\mS = (\vzeros_H, \mI_H)$, where $\vzeros_H=(0,\ldots,0)^\prime$. We than have
\begin{equation} \label{eq:Bhat_MDS}
\widehat{\mB}^{MDS} = \frac{\mS \widehat{\mGamma}_w (0) \mS^\prime}{(\widehat{\vgamma}_y(0))^2} .
\end{equation}
Often, more restrictive assumptions that lead to $\bm{B}^{*}$ becoming a diagonal matrix are employed (see \citet{lobato2001} and references therein). This would be unnecessarily restrictive for us since the Bonferroni band does not require estimation of the off-diagonal elements anyway. Combining the estimator $\widehat{\bm{B}}^{MDS}$ with \eqref{eq:S_general_bf} leads to our Bonferroni significance band for $H_0^{MDS}$ using the effective sample size $T-H$:
\begin{equation} \label{eq:simultaneous_significance_band_rho_Bartlett_star}
				{SB}^{bf}_{\bm{\rho}_y} (\widehat{\bm{B}}^{MDS})  =  \bigtimes_{h=1}^H \left[ \pm z_{1-\alpha/(2H)} \sqrt{\frac{\widehat{{b}}_{hh}^{MDS}}{T-H}}\right] \, . 
\end{equation}
Of course, sup-t and pointwise variants of this significance band are also immediate. Heteroskedasticity-robust pointwise significance are discussed in \citet{FrancqZakoian2009}, employing a more restrictive variance formula than ${\bm{B}}^{MDS}$, which they call generalized Bartlett formula. We collect the coverage properties of our significance bands in a corollary to Lemma \ref{lemma:asymptotic_coverage_general}, again slightly abusing notation for the variance estimator.

\begin{corollary} \label{coro:asymptotic_coverage_ac_significance_bands}
    \ 
    \begin{enumerate}
    \renewcommand{\labelenumi}{(\roman{enumi})}
    \item Under $H_0^{iid}$ it holds that $\lim_{T \to \infty} P  \left( \widehat \vrho_y \notin {SB}^{supt}_{\bm{\rho}_y} (\mI_H)  \right) =  \alpha$.
    \item Under $H_0^{MDS}$ and $ \widehat \mB^{MDS} \stackrel{p}{\to} \mB^{MDS}$ it holds that $\lim_{T \to \infty} P  \left( \widehat \vrho_y \notin {SB}^{bf}_{\bm{\rho}_y} (\widehat \mB^{MDS})  \right) \leq  \alpha$.
    \end{enumerate}
\end{corollary}




			\section{Regression Residuals} \label{sec:residuals}
			
Consider the regression model
			\begin{equation} \label{linearModel}
				y_t= a + {\vx}_t^\prime \boldsymbol{\beta}+e_t \, , \quad t=1, \ldots, T \, ,
			\end{equation}
			with ${\vx}_t^\prime = (x_{1,t}, \ldots, x_{K,t})$ being a vector process of dimension $K$. In (\ref{linearModel}), the effective sample size is $T$ such that we assume additional starting values in case that the vector ${\vx}_t$ contains lagged values. 
			We are interested in simultaneous inference bands for the true autocorrelation structure $\vrho_{ e} $, where now the vector $\vrho_{ e} $ contains the theoretical error autocorrelations. 
            One could proceed as in Section \ref{sec:inference_bands} if the regression errors $\{e_t\}$ were observable. Since they are not, we investigate  the least squares (LS) residual autocorrelations $\widehat \vrho_{\widehat e}  := (\widehat \rho_{\widehat e}  (1), \ldots, \widehat \rho_{\widehat e}  (H))^\prime$ with $\widehat \rho_{\widehat e} (h) := \frac{\sum_{t=h+1}^T \widehat e_t \widehat e_{t-h}}{\sum_{t=1}^T \widehat e_t^2}$ and $\widehat e_t := y_t - \widehat a - \widehat \vbeta^\prime \vx_t$.
			In this section, we establish the limiting distributions under stationarity and construct simultaneous inference bands building on them. Two cases have to be distinguished. First, we treat static regressions. Second, we turn to stationary dynamic regressions (lagged endogenous regressors). 


			\subsection{Static Regressions}  \label{subsec:static_regressions}

            
			We first assume that $\vx_t$ from (\ref{linearModel}) does not contain lagged endogenous regressors while 	lagged exogenous regressors are not ruled out. To simplify the exposition and save space, we make the following high-level assumption and do not spell out technical details with respect to the underlying processes that are sufficient.
			
			\begin{assumption}[Stationarity] \label{Ass:stationary}
				Let $\{e_t\}$ be a zero mean process meeting Assumption~\ref{ass:process}, such that (\ref{eq:NormalV_generalBartlett}) holds for $\widehat \vrho_{e} - \vrho_{e}$. Further, $\{(e_t, \vx_t^\prime)^\prime\}$ is strictly stationary and ergodic  with $\vmu_x :=\E[\vx_t]$,  $\mathbf{\Sigma}_{x} :=\E[(\vx_t - \vmu_x ) (\vx_t - \vmu_x )^\prime ]$ is positive definite, $\mOmega_{xe} := \sum_{j=- \infty}^{\infty} \E[(\vx_t - \vmu_x ) (\vx_{t-j} - \vmu_x )^\prime  e_t e_{t-j}]$ is positive definite and finite. 
	Moreover,  the regressors are  exogenous in that $\E [\vx_t e_{t-j}] = \vzeros$, $ j \in \mathbb{Z}$, and a CLT holds: $\frac{1}{\sqrt{T}} \sum_t (\vx_t - \vmu_x) e_t \ \stackrel{d}{\to} \ \mathcal{N}_K (\vzeros, \mOmega_{xe})$.
			\end{assumption}

			The finding of this subsection is: The limit result (\ref{eq:NormalV_generalBartlett}) is recovered notwithstanding the replacement of  $\{ e_ t  \}$ by the residual process $\{ \widehat e_ t \}$.
			
			\begin{proposition} \label{Prop:Static}
				Consider the regression model (\ref{linearModel}) under Assumption\,\ref{Ass:stationary}.  With LS  residual autocorrelations $\widehat \vrho_{\widehat e}$ the limiting distribution from (\ref{eq:NormalV_generalBartlett}) continues to hold as $T \to \infty$, i.\,e., ${\sqrt{T}} \left( \widehat{\bs{\rho}}_{\widehat e}  - \bs{\rho}_e	\right) \ \stackrel{d}{\to} \ \mathcal{N}_H (\bs{0}, \, \bs{B}^*)$, where $\bs{B}^*$ is  given in terms of $\{e_t\}$.
			\end{proposition}
			
        The case of static cointegration regressions treated in Appendix \ref{appsubsec:cointegration} results in identical findings as in Proposition~\ref{Prop:Static}, and this continues to hold in the presence of linear time trends, see Appendix \ref{appsubsec:integration_with_drift}. In other words: the results for static regressions do not hinge on the stationarity assumption. 

			\subsection{Dynamic Regressions: Lagged Endogenous Regressors} 	\label{subsec:dynamic_regressions}

			Now we allow for lagged endogenous regressors in ${\vx}_t$,  which covers autoregressions and the autoregressive distributed lag (ARDL) model. Consequently, we restrict the sequence of regression errors $\{e_t\}$ from (\ref{linearModel}) to be WN to ensure LS consistency in this setting. Further, the limiting theory relies on  predetermined regressors relative to the errors defined in terms of the information set
            \begin{equation} \label{eq:info_set}
                \mathcal{I}_t := \sigma(e_{t}, e_{t-1}, \ldots , \vx_{t+1}, \vx_t, \vx_{t-1}, \ldots) .
            \end{equation}
			We assume stationary processes and maintain Assumption~\ref{ass:process}.


			\begin{assumption}[Predetermined] \label{Ass:MDS}
				Let $\{e_t\}$ from (\ref{linearModel}) be an MDS meeting Assumption~\ref{ass:process}, such that $\sqrt{T} \widehat{\vrho}_e$ is asymptotically normal with covariance matrix $\mB^{MDS}$ from \eqref{eq:B_MDS}. Further, $\{(e_t, \vx_t^\prime)^\prime\}$ is strictly stationary and ergodic with finite fourth moments. Assume that $\mathbf{\Sigma}_{x}$ and ${\mSigma}_{x e} :=\E[e_t^2(\vx_t - \vmu_x ) (\vx_t - \vmu_x )^\prime ]$ are positive definite. 
				 Moreover, assume that $\E[e_{t+1} | \mathcal{I}_t] =0$ 
				with $\sigma^2_e := \E [e_{t}^2] >0$.
			\end{assumption}

By Assumption \ref{Ass:MDS} it follows that $\{\vx_t e_t \}$ forms a vector MDS, and the regressors are predetermined in that $\E[\vx_t e_{t+h}] = \vzeros$, $ h \geq 0$, although they may correlate with past errors:
			\begin{equation} \label{Gamma}
				\vc_h := \E[\vx_t e_{t-h}]  \, , \ h > 0 \, , \quad \mGamma := (\vc_1, \ldots, \vc_H)^\prime \, .
			\end{equation}
			By an MDS CLT and the continuous mapping theorem it follows for the LS estimator under Assumption \ref{Ass:MDS} that $\sqrt{T} (\widehat \vbeta - \vbeta)  \stackrel{d}{\to}  \mathcal{N}_K (\vzeros, \mSigma_x^{-1} \mSigma_{x e} \mSigma_x^{-1} )$ and we have the following result, which follows from \citet[Prop.\ 1]{CumbaHuizinga1992}.
			
			\begin{proposition} \label{Prop:ARDL}
				Consider the regression model (\ref{linearModel}) under Assumption\,\ref{Ass:MDS}. As $T \to \infty$ it holds that
				\begin{equation}  \label{eq:Sigma_rho}
				\sqrt{T} \widehat \vrho_{\widehat e} \ \stackrel{d}{\to} \ \mathcal{N}_H \left(\vzeros, \mSigma_\rho^{het}\right) \, , \quad \mSigma_\rho^{het} :=	  \mB^{MDS} - 2 \, \frac{\mGamma \mSigma_x^{-1} \mGamma^\prime}{\sigma^2_e} + \frac{\mGamma {\mSigma}_x^{-1} {\mSigma}_{x e} {\mSigma}_x^{-1} \mGamma^\prime}{\sigma^4_e} .
				\end{equation}
			\end{proposition}

			If there are no lagged endogenous regressors, $\mGamma = \mZeros_{HK}$, then the residual effect vanishes: $\mSigma_\rho^{het}= \mB^{MDS}$. With endogenous regressors, the asymptotic covariance matrix $\mSigma_\rho^{het}$ differs from $\mB^{MDS}$ by a correction term, which is in contrast to the cases of observed time series and regression residuals from static regressions. $\mSigma_\rho^{het}$ allows for conditional heteroskedasticity with respect to $\mathcal{I}_t$. We also consider the restriction to the classical assumption of conditional homoskedasticity. 
			
			\begin{assumption}[Conditional Homoskedasticity] \label{Ass:homoskedastic_errors}
			
			It holds that
			$\E[e_{t+1}^2 | \mathcal{I}_t] = \sigma^2_e > 0$.
			\end{assumption}

            Under Assumption \ref{Ass:homoskedastic_errors}, $\{e_t \}$ is in particular a conditionally homoskedastic MDS with respect to its own past and by \citet[Thm.~3]{HannanHeyde1972} $\mB^{MDS}$ turns into $\mI_H$. Furthermore, it follows from Assumption \ref{Ass:homoskedastic_errors} that $\mSigma_{x e} = {\mSigma}_x \sigma^2_e$ and $\mSigma_\rho^{het}$ turns into 
			\begin{equation} \label{eq:asymp_variance_homoskedastic_errors}
				\mSigma_\rho^{hom} := \mI_H - \frac{\mGamma \mSigma_x^{-1} \mGamma^\prime}{\sigma^2_e}.
			\end{equation}

			The matrix $\mSigma_\rho^{hom}$ is (weakly) smaller than $\mI_H$ in every entry since  $\mSigma_x$ is positive definite. 

Consistent estimation of $\mSigma_\rho^{het}$ 
            and $\mSigma_\rho^{hom}$ is straightforward, using plug-in estimation based on \eqref{eq:Sigma_rho} and \eqref{eq:asymp_variance_homoskedastic_errors} and the following estimators: $\widehat{\bm{B}}^{MDS} \stackrel
			{p}{\to} \bm{B}^{MDS}$ from \eqref{eq:Bhat_MDS}, $\widehat \mSigma_{x} := \frac{1}{T}\sum_{t=1}^T  ( \vx_t - \overline \vx ) (\vx_t - \overline \vx )^\prime \stackrel
			{p}{\to} \mSigma_{x}$, $\widehat \mSigma_{x e} := \frac{1}{T}\sum_{t=1}^T \widehat e_t^2 ( \vx_t - \overline \vx ) (\vx_t - \overline \vx )^\prime \stackrel
			{p}{\to} \mSigma_{x e}$, $\widehat{\sigma}^2_e := \frac{1}{T}\sum_{t=1}^T \widehat e_t^2  \stackrel{p}{\to} \sigma^2_e$, $\widehat \vc_h := \frac{1}{T}\sum_{t=h+1}^T  \vx_t \widehat e_{t-h} \stackrel
			{p}{\to} \vc_h$, 
where $\vc_h$ is from (\ref{Gamma}). 
 Denote the resulting estimators by $\widetilde \mSigma_\rho^{het}$ and $\widetilde \mSigma_\rho^{hom}$. These consistent plug-in estimators are not guaranteed to be positive definite. In fact, they tend to be negative definite quite often, see Section \ref{sec:simulations} for simulation evidence. For an explanation of this behaviour and to understand our proposed solution, note that usually current regressors only correlate strongly with recent errors. Thus, $\vc_h$ from \eqref{Gamma} tends to be very small except for small $h$, leading to the correction matrix in \eqref{eq:Sigma_rho} and \eqref{eq:asymp_variance_homoskedastic_errors} being close to the zero matrix except for its upper left part. Due to estimation uncertainty, the many elements that are approximately zero can become quite large and lead to $\widetilde \mSigma_\rho^{hom}$ or $\widetilde \mSigma_\rho^{het}$ becoming negative definite. This is especially easy to see for $\widetilde \mSigma_\rho^{hom}$, where the matrix that is subtracted is nonnegative in all of its entries, that is, can only fluctuate in the positive direction and is subtracted from the identity matrix, which leads to all off-diagonal elements being nonpositive. 
 We propose a simple shrinkage algorithm, which guarantees positive definiteness of the resulting estimator and at the same time reduces estimation uncertainty (as it avoids estimation of many entries of a matrix that are essentially zero) by shrinking the correction matrix towards the zero matrix, while keeping as many elements in the upper left as possible. The algorithm works very well in our simulations.
		 	
		 	\begin{algorithm} \label{alg:shrinkage}
		 		Inputs: Plug-in estimator $\widetilde \mSigma_\rho$ (either $\widetilde \mSigma_\rho^{het}$ or $\widetilde \mSigma_\rho^{hom}$) and $\widehat{\bm{B}}^{MDS}$ for $\widetilde \mSigma_\rho^{het}$.
		 		\begin{enumerate}
		 			\item Set $k=1$.
		 			\item Consider the upper left $k \times k$ submatrix of $\widetilde \mSigma_\rho$ and check if it is positive definite (if all its eigenvalues are positive). If this is the case and if $k<H$, set $k=k+1$ and repeat step 2. If not, proceed to step 3.
		 			\item Take $\widehat{\bm{B}}^{MDS}$ (or $\mI_H$ for $\widetilde \mSigma_\rho^{hom}$) and replace its upper left $(k-1) \times (k-1)$ submatrix with the corresponding elements of $\widetilde \mSigma_\rho$ and return the resulting matrix, call it $\widehat \mSigma_\rho$.		 			
		 		\end{enumerate}
		 		Output: $\widehat \mSigma_\rho$, more precisely $\widehat \mSigma_\rho^{hom}$ if the input was $\widetilde \mSigma_\rho^{hom}$ and $\widehat \mSigma_\rho^{het}$ if the input was $\widetilde \mSigma_\rho^{het}$.
		 	\end{algorithm}

We close this subsection by mentioning that Proposition \ref{Prop:ARDL} continues to hold for the cointegrated error-correction model; see Appendix \ref{appsubsec:ECM}.
					

			\subsection{Inference Bands for Autocorrelations of Regression Errors} 
			\label{sec:inference_bands_residuals}

%
			
			Proposition \ref{Prop:Static} implies that for error autocorrelations $\bm{\rho}_{e}$ from a static regression model, 
            inference bands can be constructed as laid out in Section \ref{sec:inference_bands}, 
            just replacing observed time series by regression residuals $\widehat{e}_t$ in the calculation of empirical autocorrelations and variance estimators. 
			
%
			
For error autocorrelations from dynamic regressions, 
confidence bands do not really make sense as one typically imposes the assumption of WN for consistency in a dynamic regression model. In contrast, significance bands are all the more important since testing $H_0^{WN}$ for the regression errors $\{e_t\}$ amounts to a test of model misspecification. Proposition \ref{Prop:ARDL} shows how the formulas for our significance bands from Section \ref{subsec:significance_bands} need to be adjusted to account for the effect of having to replace true errors by residuals. We call the resulting bands corrected bands in contrast to the naive approach of not accounting for the residual effect and just using the bands for observed time series. We formulate the null hypothesis for our classical corrected bands as being a conditionally homoskedastic MDS with respect to the information set $\mathcal{I}_{t-1}$ from \eqref{eq:info_set}, i.e., with respect to current and past regressors and past errors, i.e., $\E[e_{t+1} | \mathcal{I}_t] =0$ and $\E[e_{t+1}^2 | \mathcal{I}_t] = \sigma^2_e$, and call it $H_0^{hom}$. We could alternatively formulate the null as $e_t \sim iid(0,\sigma_e^2)$ under Assumption 5 since both lead to the covariance matrix $\mSigma_\rho^{hom}$ from \eqref{eq:asymp_variance_homoskedastic_errors}. The resulting corrected sup-t band is
\begin{equation} \label{eq:significance_band_dynamic_regression}
		{SB}^{supt}_{\bm{\rho}_{e}} (\widehat \mSigma_\rho^{hom}) =   \bigtimes_{h=1}^H \left[\pm q_{1-\alpha} (\widehat \mSigma_\rho^{hom}) \sqrt{\frac{\widehat \sigma^{hom}_{\rho,hh}}{T}}\right]. 
\end{equation}
Interestingly, the corresponding naive band ${SB}^{supt}_{\bm{\rho}_e} (\mI_H)$ from \eqref{eq:simultaneous_significance_band_rho} is always at least as wide as ${SB}^{supt}_{\bm{\rho}_{e}} (\widehat \mSigma_\rho^{hom})$ and thus also valid, but conservative, as we show in Proposition \ref{Prop:significance_bands_dynamic_regression} in the Appendix. It tends to be considerably wider only for the first few lags 
due to the correction term mainly influencing the variances $\sigma^{hom}_{\rho,hh}$ for small $h$. We shed further light on this and more generally on the structure of the asymptotic covariance matrix and the behaviour of the corrected and naive bands in a theoretical example in Appendix \ref{subsec:theoretical_example_residuals}.

Consider as further null hypothesis $H_0^{het}$ of $\{e_t\}$ being an MDS with respect to $\mathcal{I}_{t-1}$. The corrected Bonferroni significance band for $H_0^{het}$ reads
\begin{equation} \label{eq:significance_band_dynamic_regression_robust}
		{SB}^{bf}_{\bm{\rho}_e} (\widehat \mSigma_\rho^{het}) =   \bigtimes_{h=1}^H \left[\pm z_{(1+(1-\alpha)^{1/H})/2} \sqrt{\frac{\widehat \sigma^{het}_{\rho,hh}}{T-H}}\right]. 
\end{equation}

We collect the coverage properties of the significance bands for regression errors in a corollary to Lemma \ref{lemma:asymptotic_coverage_general} and Propositions \ref{Prop:Static}, \ref{Prop:ARDL} and \ref{Prop:significance_bands_dynamic_regression}, again abusing notation for the variance estimators.

\begin{corollary} \label{coro:asymptotic_coverage_ac_significance_bands_residuals}
    Consider the regression model (\ref{linearModel}). 
    \begin{enumerate}
    \renewcommand{\labelenumi}{(\roman{enumi})}
    \item Under $H_0^{iid}$ on $\{e_t\}$ and Assumption \ref{Ass:stationary} it holds that $\lim_{T \to \infty} P  \left( \widehat \vrho_{\widehat e} \notin {SB}^{supt}_{\bm{\rho}_e} (\mI_H)  \right) =  \alpha$.
    \item Under $H_0^{MDS}$ on $\{e_t\}$, Assumption \ref{Ass:stationary}, and $ \widehat \mB^{MDS} \stackrel{p}{\to} \mB^{MDS}$ it holds that\\
    $\lim_{T \to \infty} P  \left( \widehat \vrho_{\widehat e} \notin {SB}^{bf}_{\bm{\rho_e}} (\widehat \mB^{MDS})  \right) \leq  \alpha$.
    \item Under $H_0^{hom}$, Assumption \ref{Ass:MDS}, and $\widehat \mSigma_\rho^{hom} \stackrel{p}{\to} \mSigma_\rho^{hom}$ it holds that\\
    $\lim_{T \to \infty} P  \left( \widehat \vrho_{\widehat e} \notin {SB}^{supt}_{\bm{\rho}_e} (\mI_H)  \right) \leq  P  \left( \widehat \vrho_{\widehat e} \notin {SB}^{supt}_{\bm{\rho}_{e}} (\widehat \mSigma_\rho^{hom})  \right) =  \alpha$.
    \item Under $H_0^{het}$, Assumption \ref{Ass:MDS}, and $\widehat \mSigma_\rho^{het} \stackrel{p}{\to} \mSigma_\rho^{het}$ it holds that\\
    $\lim_{T \to \infty} P  \left( \widehat \vrho_{\widehat e} \notin {SB}^{bf}_{\bm{\rho}_e} (\widehat \mSigma_\rho^{het})  \right) \leq  \alpha$.
    \end{enumerate}
\end{corollary}

			\section{Simulations} \label{sec:simulations}
			
			We analyse the finite-sample performance of our inference bands in an extensive simulation study. Tabulated simulation results are contained in appendices \ref{Sect:sims_ts} for time series and \ref{Sect:sims_residuals} for regression errors. For the time series case we simulate from AR(1) processes, $y_t = \phi y_{t-1} + \varepsilon_t$, with $\phi=0,0.25,0.5,0.75,0.95$ and different innovation processes $\{\varepsilon_t\}$: On the one hand, we use iid normal innovations, $\varepsilon_t \sim ii\mathcal{N} (0,1)$, such that the classical Assumption \ref{ass:Bartlett} is fulfilled. On the other hand, we use GARCH innovations, $\varepsilon_t = h_t u_t$ with $h_t^2 = \alpha_0 + \alpha_1 \varepsilon_{t-1}^2 + \alpha_2 h_{t-1}^2$, 
            with the same parameter choices and error distributions as in the simulation study of \citet{lobato2002}, i.e., parameters $\alpha_0=0.001$, $\alpha_1=0.05$, $\alpha_2=0.9$, and standard normal errors, $u_t \sim ii\mathcal{N} (0,1)$, or standardized chi-square errors with 3 degrees of freedom, $u_t = \frac{v_t - 3}{\sqrt{6}}$ with $v_t \sim ii\chi^2(3)$.  We use the sample sizes $T=50,200,800$, the maximum lags $H=1,10,25$, a nominal coverage of $1-\alpha=0.9$, a significance level of $\alpha=0.1$, and 1000 simulation runs. 

\subsection{Testing White Noise}

We begin with testing for WN and report the rejection frequencies of $H_0^{WN}$ in our tables, where $\phi=0$ corresponds to size and the other values of $\phi$ to power. Our significance bands, even though they are constructed under $H_0^{iid}$ and $H_0^{MDS}$, can of course also be viewed as tests of $H_0^{WN}$ under additional assumptions. For the iid innovations (Table \ref{tab:significance_bands_coverage}), both the classical sup-t band ${SB}^{supt}_{\bm{\rho}_y} (\mI_H)$ from \eqref{eq:simultaneous_significance_band_rho} and the robust Bonferroni band ${SB}^{bf}_{\bm{\rho}_y} (\widehat{\bm{B}}^{MDS})$ 
from \eqref{eq:simultaneous_significance_band_rho_Bartlett_star} perform well in terms of size and tend to be conservative for smaller $T$ in combination with large $H$. The classical band expectedly is more powerful, in particular for $T=50$. The default classical pointwise band ${SB}^{pw}_{\bm{\rho}_y} (\mI_H)$ from \eqref{eq:pointwise_significance_band_rho} is all the more oversized the larger $H$ as expected. For the GARCH innovations with normal errors (Table \ref{tab:significance_bands_coverage_GARCH_normal}), the robust Bonferroni band shows a good performance with mild oversizing for $H=25$, which might be due to the normal approximation requiring larger samples for larger $H$ with this more involved DGP as has been conjectured by \citet{lobato2002} after observing similar behaviour for robust Box-Pierce tests. The classical band expectedly runs into problems under heteroskedasticity, i.e., shows oversizing for all $H$ except for $T=50$. The oversizing is not too strong, but it grows with $T$. This undesirable feature has also been observed for the classical Box-Pierce test, see again \citet{lobato2002}. For the chi-squared errors (Table \ref{tab:significance_bands_coverage_GARCH_chisqu}), the results are similar. 
Unsurprisingly, the width of all simultaneous bands increases with $H$ and decreases with $T$, and the robust bands are wider on average (averaged over the lags $h=1,...,H$ and simulation runs) than their classical counterparts, 
see the tables in \ref{appsubsec:width_inference_bands}.

We also consider Box-Pierce-type tests as competitors for our significance bands (see tables \ref{tab:tests_iid}, \ref{tab:tests_GARCH_normal}, and \ref{tab:tests_GARCH_chisqu}), where the classical tests of WN of \citet{BoxPierce70} and its modification by \citet{LjungBox78} are the relevant competitors for our classical band and the test by \citet{guo2001}, for which we use the estimator $\widehat{\bm{B}}^{MDS}$ from \eqref{eq:Bhat_MDS}, and the self-normalizing test by \citet{Shao2010} are the competitors for our robust band. The first three tests show similar and often even worse performance than our bands. The test by \citet{Shao2010} seems to have severe problems with larger $H$ and is often heavily oversized.

Confidence bands can be used for testing $H_0^{WN}$ as well (by checking whether $\bf{0}$ is covered by the band) and we analyse their performance for this task as well (see again tables \ref{tab:significance_bands_coverage}, \ref{tab:significance_bands_coverage_GARCH_normal}, and \ref{tab:significance_bands_coverage_GARCH_chisqu}). 
The classical sup-t confidence band ${CB}_{\bm{\rho}_y}^{supt} (\widehat \mB)$ from \eqref{eq:supt_rho} shows very similar size and power to its significance band counterpart ${SB}^{supt}_{\bm{\rho}_y} (\mI_H)$. This is interesting since the two types of bands have quite different properties. The confidence bands are on average (averaged over simulation runs and lags) wider than the significance bands (Table \ref{tab:width_significance_bands_iid}). They tend to get wider with increasing $h$, see the theoretical discussion in Appendix \ref{subsec:theoretical_example_time_series}, in particular Figure \ref{fig:conf_bands_width_over_h}, and the inflation case study in Section \ref{sec:case_studies}. They also depend on an estimated covariance matrix, while the classical significance bands have constant width and do not depend on estimated quantities. Thus, the observed similarity in empirical size and power properties already hints at the remarkable performance of the estimator of \citet{melard1987}, $\bs{\widehat{B}}$, which is discussed in Section \ref{subsec:simulation_CB}. For $h=1$, 
the confidence bands are narrower than the significance bands, which is probably the reason why they perform as good in terms of power. For the robust bands, however, the confidence band ${CB}_{\bm{\rho}_y}^{bf} (\widehat \mB^*)$ shows stronger oversizing than the significance band ${SB}^{bf}_{\bm{\rho}_y} (\widehat{\bm{B}}^{MDS})$, probably caused by the HAC-type estimator $\widehat{\bm{B}}^*$. For this reason, we prefer 
the estimator $\widehat{\bm{B}}^{MDS}$ for constructing significance bands, which avoids HAC-estimation by imposing the slightly stronger MDS assumption (see the discussion in Section \ref{subsec:significance_bands}).


			\subsection{Confidence Bands} \label{subsec:simulation_CB}
            
			We maintain the previous setup and now analyse the empirical coverage of our confidence bands (by checking the relative frequency with which they cover the true ACF 
            $\bm{\rho}_y := (\phi,\phi^2,...,\phi^H)^\prime$).         Concerning the bandwidth choice for the variance estimators, \citet{melard1987} proposed to use $m=1,3,5$ with $L = m T^{\frac 1 2 }$ for their estimator $\widehat{\bm{B}}$. We consider these choices (excluding $m=5$ from the tables since it usually works worst), the textbook rule for standard HAC-estimators from \citet{stock2020}, $L = 0.75 T^{\frac 1 3}$, and additionally $L = T^{\frac 1 3}$. For the estimator $\widehat{\bm{B}}^*$, we replace $T$ by the effective sample size $T-H$ (tables \ref{tab:confidence_bands_coverage_bandwidth_classical_iid}, \ref{tab:confidence_bands_coverage_bandwidth_robust_iid}, \ref{tab:confidence_bands_coverage_bandwidth_robust_GARCH_normal}, and \ref{tab:confidence_bands_coverage_bandwidth_robust_GARCH_chisqu}). 
            Indeed, $L = T^{\frac 1 2 }$ tends to work best with $\widehat{\bm{B}}$ and $L = (T-H)^{\frac 1 3 }$ with $\widehat{\bm{B}}^*$, so we recommend these choices and continue to work with them.
            
             
           The coverage of the classical sup-t band ${CB}_{\bm{\rho}_y}^{supt} (\widehat \mB)$ from \eqref{eq:supt_rho} for the iid innovations is very satisfactory (Table \ref{tab:confidence_bands_coverage}), demonstrating that $\widehat{\bm{B}}$ does hardly suffer from the problems of standard HAC estimators (except for the extreme cases $T=50$ and $\phi=0.95$). 
           For the robust sup-t band in turn, which uses $\widehat{\bm{B}}^*$, these problems are visible, i.e., undercoverage increasing with length of the band $H$, strength of temporal dependence $\phi$, and improving with sample size $T$. The robust Bonferroni band ${CB}_{\bm{\rho}_y}^{bf} (\widehat \mB^*)$ from \eqref{eq:Bonferroni_rho} mitigates this problem and improves performance due to its asymptotic  conservativeness, which also becomes more pronounced with stronger temporal dependence (for details see the next paragraph). For the robust bands, a similar picture arise with the GARCH innovations (tables \ref{tab:confidence_bands_coverage_GARCH_normal} and \ref{tab:confidence_bands_coverage_GARCH_chisqu}), with worse undercoverage for the smaller samples due to the more involved DGP. Here, the classical bands even sometimes show a better performance for the smaller samples. However, they exhibit the undesirable behaviour of stronger undercoverage with rising sample size, similar to the classical significance bands.


        The average width of the confidence bands increases with their length $H$, and decreases with the degree of persistence $\phi$ and the sample size $T$. In line with the theoretical discussion in Appendix \ref{sec:theoretical_example_time_series} (see in particular Figure \ref{fig:conf_bands_relative_width_ar1}), the Bonferroni bands are generally very close in coverage and width to the sup-t bands under independence and for weak temporal dependence and markedly wider and have substantially higher coverage for $\phi=0.5$ and especially for $\phi=0.75$ and $\phi=0.95$.


			\subsection{Testing White Noise for Dynamic Regression Errors} \label{subsec:simulation_dynamic}
        
            
            Finally, we analyse the finite-sample performance of our significance bands for the null hypothesis that the errors in a dynamic regression model are WN. We simulate time series of length $T$ from an AR(2) process $y_t = \phi_1 y_{t-1} + \phi_2 y_{t-2} +  \varepsilon_t$, with ${\varepsilon_t}$ being either $ii\mathcal{N} (0,1)$ or stemming from the normal GARCH model described above. We estimate an AR(1) model  
            by LS: $y_t =  \widehat a + \widehat \phi y_{t-1} + \widehat e_t$. Thus, with the true error being $e_t = \phi_2 y_{t-2} + \varepsilon_{t}$, size arises for $\phi_2=0$ and we assess power with $\phi_2=0.125$ and $\phi_2=0.25$.

            For the iid errors (Table \ref{tab:coverage_dynamic_regression_iid}), the corrected classical sup-t band ${SB}^{supt}_{\bm{\rho}_{e}} (\widehat \mSigma_\rho^{hom})$ from \eqref{eq:significance_band_dynamic_regression} and the corrected robust Bonferroni band ${SB}^{bf}_{\bm{\rho}_e} (\widehat \mSigma_\rho^{het})$ from \eqref{eq:significance_band_dynamic_regression_robust} are well-sized, being conservative for the smaller samples, and very similar in power. The naive classical band ${SB}^{supt}_{\bm{\rho}_e} (\mI_H)$ is expectedly more conservative and has lower power. For $H=1$ it is very conservative, which is explained by the way too large variance of the naive simultaneous band for $h=1$, see the analytical example in Appendix \ref{subsec:theoretical_example_residuals}. The naive classical pointwise band ${SB}^{pw}_{\bm{\rho}_e} (\mI_H)$ is again heavily oversized. For GARCH errors (Table \ref{tab:coverage_dynamic_regression_GARCH}), the robust band works well, being mildly oversized for $H=25$, similarly to the closely related robust band ${SB}^{bf}_{\bm{\rho}_y} (\widehat{\bm{B}}^{MDS})$ 
            in the time series case. The corrected and naive classical bands again show the undesirable behaviour of being more oversized in larger samples. 
            
            These results show that the Algorithm \ref{alg:shrinkage} works well. We also report the fraction of times that it actually needs to shrink the plug-in estimators $\widetilde \mSigma_\rho^{het}$ and $\widetilde \mSigma_\rho^{hom}$ to make them positive definite and the median lag, at which the adjustment takes place (tables \ref{tab:dyn_iid_adjust_frac}, \ref{tab:dyn_iid_adjust_lag}, \ref{tab:dyn_garch_adjust_frac}, and  \ref{tab:dyn_garch_adjust_lag}). The algorithm is needed quite frequently, and often a large part of the plug-in covariance matrix is shrunk (as expected from the discussion in Section \ref{sec:residuals} and the theoretical example in Appendix \ref{subsec:theoretical_example_residuals}). 

            We again consider Box-Pierce-type tests as competitors (tables \ref{tab:coverage_dynamic_regression_iid} and \ref{tab:coverage_dynamic_regression_GARCH}). As classical competitors, we use the Ljung-Box test and the Breusch-Godfrey test \citep{Breusch78, Godfrey78}. To account for the effect of AR(1) residuals, the Ljung-Box test is applied with a $\chi^2(H-1)$ distribution, such that no results are available for $H=1$.  As a robust competitor, we employ a generalized Box-Pierce test in the sense of \citet{lobato2002} using our corrected covariance matrix $\widetilde \mSigma_\rho^{het}$. As for the time series case, our bands show a similar or even better performance than the respective tests. In particular, the robust bands are less oversized than the robust Box-Pierce test  under GARCH errors.

			\section{Case Studies} \label{sec:case_studies} 
            
			
			
		We construct inference bands for inflation, excess returns, and regression residuals from various specifications of a Phillips curve regression. For our  analysis, we use monthly inflation and unemployment data from the FRED-MD database \citep{mccracken2016} from 1961:01 to 2024:06 and daily Fama-French (three factors) excess returns \citep{fama1993} from the Fama-French data library \citep{fama2023} from 2006, January 3, to 2026, April 30.
         Plots of these series can be found in figures \ref{fig:monthyl_inflation_unemployment} and \ref{fig:fama_return_series} in the Appendix. For all our autocorrelogram plots, we use the afore-mentioned rule-of-thumb $H = \lfloor 10 \, \log_{10} (T) \rfloor$, for the confidence bands a level of $90\%$, and for the significance bands a size of $10\%$.
		
			
		For the two macroeconomic case studies, we use classical inference bands as we do not expect considerable heteroskedasticity, following our recommendation from Figure \ref{fig:decision_tree}, but we also plot robust bands and discuss them at the end as robustness checks.	Figure~\ref{fig:phillips_and_inflation} displays the empirical ACF with confidence bands for monthly US inflation (month-on-month, seasonally adjusted). 
        The width of all the bands increases from the first lag to the 
        third and from then on almost stays constant over the lags. This behaviour is due to the smaller asymptotic variance for the first lags as we also observe in the AR(1) example in Appendix \ref{subsec:theoretical_example_time_series}. As expected, the pointwise band is the narrowest, but essentially useless due to its heavy undercoverage, and the sup-t band is in turn narrower than the Bonferroni band.
		
\begin{figure}
				\noindent \centering{}\includegraphics[scale=0.5]{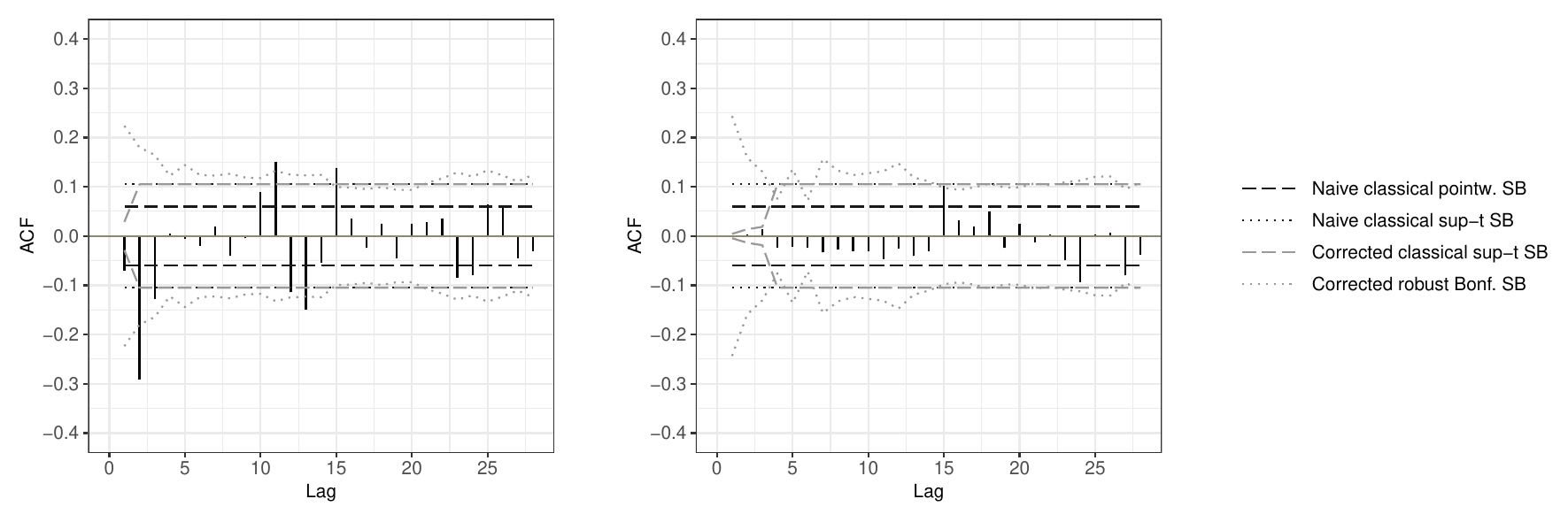}
				\caption{Empirical autocorrelations with $90\%$ significance bands for regression residuals from a Phillips curve regression in differences as in Equation \eqref{eq:phillips_dyn_diff}. Left: $p=1$ and $r=0$. Right: $p=r=12$.}	
				\label{fig:phillips_sig_dyn_diff}
			\end{figure}

        The Phillips curve is a widespread econometric model used to study the relation between unemployment and inflation (e.\,g., \citet{Gordon2013}, \citet{Blanchard2016}, \citet{Ball2019}). Another strand of the literature employs the model for forecasting (e.\,g., \citet{Stock1999, Stock2008}). 
            Throughout, we assume that the 
            non-accelerating inflation rate of unemployment (NAIRU) is constant, and approximate expected inflation by past inflation. Coefficient estimates as well as plots of the residuals for the following regression models are reported in tables \ref{tab:phillips_estimates_diff} and \ref{tab:phillips_estimates_level} as well as figures \ref{fig:phillips_res} to \ref{fig:phillips_res_against_x_diff} in the Appendix. We begin by estimating a static Phillips curve by LS, i.\,e., $p=0$ and $r=0$ in
			\begin{equation} \label{eq:phillips_dyn_diff}
				\Delta \pi_t = a + \sum_{i=1}^{p} \phi_i \Delta \pi_{t-i} + \sum_{j=0}^{r} \gamma_j u_{t-j}  + \varepsilon_t \, , \quad t=1, \ldots, 762  \, .
			\end{equation}
Here, $\pi_t$ is inflation, $\Delta \pi_t = \pi_t - \pi_{t-1}$, and $u_t$ is the unemployment rate. The residual autocorrelogram is shown in Figure~\ref{fig:phillips_and_inflation} together with the pointwise significance band ${SB}^{pw}_{\bm{\rho}_e} (\mI_H)$ from \eqref{eq:pointwise_significance_band_rho} and the sup-t band ${SB}^{supt}_{\bm{\rho}_e} (\mI_H)$ according to (\ref{eq:simultaneous_significance_band_rho}). From the latter we can conclude that the errors from the static Phillips curve regression still contain significant (on the $10\%$ level) dynamics, which are not explained by the model. Thus, we move to dynamic Phillips curve specifications. 
 Figure~\ref{fig:phillips_sig_dyn_diff} displays the empirical ACF with significance bands for two specifications alongside three classical significance bands: the naive pointwise band ${SB}^{pw}_{\bm{\rho}_e} (\mI_H)$, the naive sup-t band ${SB}^{supt}_{\bm{\rho}_e} (\mI_H)$, 
 and the corrected sup-t band ${SB}^{supt}_{\bm{\rho}_{e}} (\widehat \mSigma_\rho^{hom})$ 
 from \eqref{eq:significance_band_dynamic_regression}. The left side of the figure shows that using only one lag of the dependent variable ($p=1$) and the contemporaneous unemployment rate ($r = 0$) as regressors still yields significant 
 autocorrelations of the error. Apart from the first couple of lags, both the sup-t bands (naive and exact) are very similar.  Adding a full year of lags of both variables on the right-hand side of the figure ($p=r=12$) visibly reduces the empirical autocorrelations. The WN hypothesis for the regression errors is no longer rejected according to both sup-t bands. In contrast, the invalid pointwise band would lead to the opposite conclusion.
			
Now we check if our choice of the classical bands in the macroeconomic applications is justified, i.e., if conditional homoskedasticity is a reasonable assumption: Indeed, there seems to be no substantial heteroskedasticity in the residuals from our three regression specificiations with respect to their own past (see again figures \ref{fig:phillips_res} and \ref{fig:phillips_res_dyn_diff}) and also with respect to the most recent regressors in our final specification (Figure \ref{fig:phillips_res_against_x_diff}). The respective robust bands, which can also be found in the main plots, speak the same language in that they are not markedly wider than the classical bands (except for the first three lags for the regression errors).  

While the forecasting literature tends to formulate the Phillips curve in terms of differences, the traditional Phillips curve is usually studied with inflation in levels. We provide a corresponding analysis in Appendix \ref{appsubsec:additional_material_case_studies}.	
	
  			\begin{figure}
				\noindent \centering{}\includegraphics[scale=0.5]{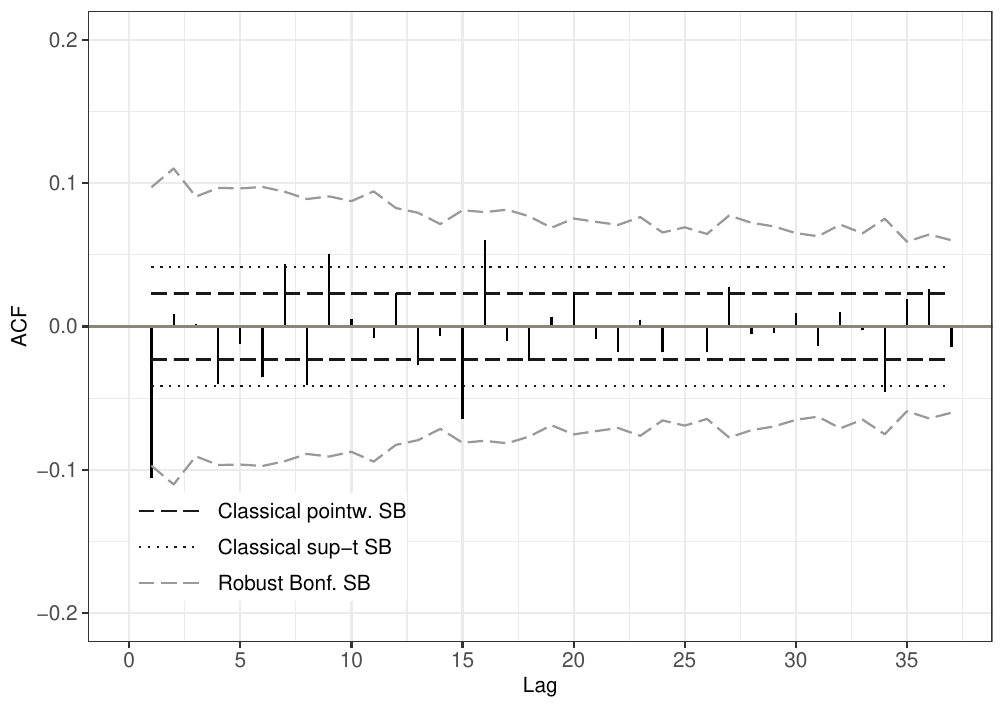}
				\caption{Empirical autocorrelations with $90\%$ significance bands for daily excess returns.}	
				\label{fig:Fama_french}
			\end{figure}		

Figure \ref{fig:Fama_french} displays the autocorrelogram of the Fama-French excess return series accompanied by three significance bands. As white noise hypotheses, in particular the MDS hypothesis, are of key interest for financial returns \citep{escanciano2009}, significance bands are the natural choice here, providing a first test of this hypothesis directly with the autocorrelogram. Due to the considerable heteroskedasticity present in financial returns, which this series also exhibits (see Figure \ref{fig:fama_return_series} again), the robust simultaneous significance band ${SB}^{bf}_{\bm{\rho}_y} (\widehat{\bm{B}}^{MDS})$ from \eqref{eq:simultaneous_significance_band_rho} is the right choice, being even especially tailored to $H_0^{MDS}$. The usual classical pointwise significance band ${SB}^{pw}_{\bm{\rho}_y} (\mI_H)$ from \eqref{eq:pointwise_significance_band_rho} does not only lack simultaneity, but also does not account for heteroskedasticity. Its sup-t version ${SB}^{supt}_{\bm{\rho}_y} (\mI_H)$ from \eqref{eq:simultaneous_significance_band_rho} cures the first issue, but not the second. Indeed, it is considerably less wide than the robust band and the empirical autocorrelations at many lags surpass it considerably. For the robust band ${SB}^{bf}_{\bm{\rho}_y} (\widehat{\bm{B}}^{MDS})$, a different picture arises. Only the first empirical autocorrelation falls out of it by a tiny margin. Thus, the MDS null hypothesis is rejected on the $10\%$ level, which certainly calls for a more detailed investigation, but there is no overwhelming evidence for a breach of the efficient market hypothesis as the other two bands would suggest.

			\section{Conclusion} \label{sec:conclusion}
			
			We introduce asymptotic simultaneous significance and confidence bands for autocorrelations to supersede the classical pointwise non-rejection bands that are usually added to autocorrelograms. We construct classical and heteroskedasticity-robust variants of our bands for observed time series as well as for regression errors, discuss their theoretical properties, analyse their finite sample properties in an extensive simulation study, and illustrate their usefulness in case studies. We give guidance on which variant of our inference bands to use in which applied setting, see in particular Figure \ref{fig:decision_tree}, and offer an R package that plots the  desired variant alongside the autocorrelogram. 

            Extensions of interest include further time series correlations such as cross-correlations, partial autocorrelations, and autocorrelations for residuals from vector autoregressions, see Appendix \ref{Sect:extensions_time_series_correlations} for a sketch of these extensions. Furthermore, it would be fruitful to extend our bands to generalizations of autocorrelations, which capture other forms of time series dependence, for example, to the quantilogram by \citet{linton2007}, which captures temporal dependence in the tails. As touched upon in Section \ref{subsec:confidence_bands}, another research endeavour could be a comprehensive comparison of our approach to constructing simultaneous inference bands assuming a fixed number of lags with the approach of \citet{XiaoWu2014}, where the number of lags grows with the sample size.

\section*{Data Availability Statement}

The data that support the findings of this study are openly available in the FRED-MD database \citep{mccracken2016} at \url{https://www.stlouisfed.org/research/economists/mccracken/fred-databases} and the Fama-French data library \citep{fama2023} at \url{https://mba.tuck.dartmouth.edu/pages/faculty/ken.french/data_library.html}.

\addcontentsline{toc}{section}{Data Availability Statement} 

\section*{Acknowledgements}

We thank Dietmar Bauer, Jörg Breitung, Matei Demetrescu, Timo Dimitriadis, Simon Freyaldenhoven, Daniel Gutknecht, Christoph Hanck, Carsten Jentsch, Helmut L\"utkepohl, Melanie Schienle, Tim Vogelsang, Michael Wolf, seminar participants at TU Dortmund University, Universitat de les Illes Balears, WHU Vallendar, Heidelberg Institute for Theoretical Studies, Goethe University Frankfurt and Free University Berlin and conference participants at the 2024 NBER-NSF Time Series Conference, the 2025 Annual Meeting of the Standing Field Committee of Econometrics of the German Economic Association, the 3rd Vienna Workshop on Economic Forecasting 2025, the Annual Conference of the International Association for Applied Econometrics 2025, the Inaugural Workshop of the Institute of Statistics at Karlsruhe Institute of Technology, and the 12th International Conference on Time Series and Forecasting 2026 for helpful comments. Marc-Oliver Pohle is grateful for support by the Klaus Tschira Foundation, Germany.

\bibliographystyle{chicago}

\bibliography{literature_autocorrelation_bands}
\addcontentsline{toc}{section}{\refname}

			\newpage
			
			\appendix
			\appendixpage

The Appendix contains eight sections. Section~\ref{Sect:Proof} provides mathematical proofs of the lemma and the propositions in the paper. Proposition~\ref{Prop:Static} is extended to cover cointegration and time trends in Section~\ref{Sect:Ext_Coint}. Additional  theoretical results are given in Section~\ref{Sect:Additiona_Theoretical_Results} and the properties of the new inference bands are discussed and illustrated for AR(1) models in Section~\ref{sec:theoretical_example_time_series}. Tabulated simulation results are contained in Section~\ref{Sect:sims_ts} for time series and in Section~\ref{Sect:sims_residuals} for regression residuals. Section~\ref{appsubsec:additional_material_case_studies} complements our case studies in that it offers a study of the Phillips curve in levels and some robustness checks. Section~\ref{Sect:extensions_time_series_correlations} sketches extensions of our inference bands to further time series correlations.
            
			\section{Proofs} \label{Sect:Proof}

\subsection{Proof of Lemma \ref{lemma:asymptotic_coverage_general}}

            
We first show (i): 
\begin{align*} 
\lim_{T \to \infty} P \left(\vtheta \in {CB}_{\vtheta}^{supt} (\mSigma) \right)  =& \lim_{T \to \infty} P \left( \sqrt{T} \frac{\left| \widehat{\theta}_1 - \theta_1 \right|}{\sqrt{\sigma_{11}}} \leq q_{1-\alpha} (\mSigma),\ldots, \sqrt{T} \frac{\left| \widehat{\theta}_H - \theta_H \right|}{\sqrt{\sigma_{HH}}} \leq q_{1-\alpha} (\mSigma) \right)\\
=& P \left(\sigma_{11}^{-1/2} |  V_1| \leq q_{1 - \alpha} (\mSigma), \ \dots \ ,  \sigma_{HH}^{-1/2} |   V_H| \leq q_{1 - \alpha}  (\mSigma) \right) = 1 - \alpha
\end{align*}

This establishes the equality, the inequaity follows by $q_{1-\alpha} (\mSigma) \leq  z_{1-\alpha/(2H)}$, see Lemma \ref{lemma:width}.

The proof of (ii) is analogous.
			
			\subsection{Proof of Proposition~\ref{Prop:Static}} \label{appsubsec:proof_prop1}
		
		{\sc Notation}
			
			To facilitate the exposition  we will work with demeaned variables, $\underline v_ t := v_t - \overline v$, or demeaned vectors:
			\[
			\underline \vv_ t := \vv_t - \overline \vv \, , \quad \overline \vv = \frac{1}{T} \sum_{t=1}^T \vv_t \, .
			\]
			Sums run from $t=1$ to $T$ if not indicated otherwise. 
			By the Frisch-Waugh-Lovell Theorem, the LS estimator of the slope parameters becomes
			\begin{eqnarray*}
				\widehat \vbeta - \vbeta &=&  \left(\sum_t \underline \vx_t \underline \vx_t^\prime \right)^{-1} \sum_t \underline \vx_t \underline e_t \\
				&=& \left(\underline \mX^\prime \underline \mX \right)^{-1} \sum_t \underline \vx_t e_t  \, ,
			\end{eqnarray*}
			where $\underline \mX$ is a $T \times K$ matrix of demeaned regressors. The LS residuals are:
			\[
			\widehat e_t = \underline e_t - (\widehat \vbeta - \vbeta )^\prime \underline \vx_t \, .
			\]
			
			We have to analyse ($h \geq 0$)
			\begin{eqnarray} 
				\sum_{t=h+1}^T \widehat e_t \widehat e_{t-h} &=&  \sum_{t} (\underline e_t - (\widehat{\vbeta}-{\vbeta})^\prime \underline \vx_t) (\underline e_{t-h} - (\widehat{\vbeta}-{\vbeta})^\prime \underline \vx_{t-h}) \nonumber \\
				&=&A_T^{(h)} - B_T^{(h)} - C_T^{(h)} + D_T^{(h)} \, , \label{ABCD}
			\end{eqnarray}
			where
			\begin{eqnarray*} 
				A_T^{(h)} &:=&  \sum_{t}   \underline e_t \underline e_{t-h} \, , \\
				B_T^{(h)} &:=&  {(\widehat{\vbeta}-{\vbeta})^\prime}  \sum_{t}  \underline \vx_t \underline e_{t-h} \, , \\
				C_T^{(h)} &:=&  (\widehat{\vbeta}-{\vbeta})^\prime  \sum_{t} \underline \vx_{t-h} \underline e_t  \, , \\
				D_T^{(h)} &:=& (\widehat{\vbeta}-{\vbeta})^\prime  \sum_{t} \underline \vx_t \underline \vx_{t-h}^\prime  (\widehat{\vbeta}-{\vbeta})  \, .
			\end{eqnarray*}
		
{\sc Under Assumption~\ref{Ass:stationary}}

By Assumption, 
			\[
			\frac{1}{\sqrt{T}} \sum_t (\vx_t - \vmu_x) e_t \ \stackrel{d}{\to} \ \mathcal{N}_K (\vzeros, \mOmega_{xe}) \, .
			\]
			Note that
			\begin{eqnarray*}
				\widehat \vbeta -  \vbeta &=&  \left(\underline \mX^\prime \underline \mX \right)^{-1} \sum_t  (\vx_t - \vmu_x) e_t -  \left(\underline \mX^\prime \underline \mX \right)^{-1} (\overline \vx - \vmu_x) \sum_t   e_t  \\
				&=&  \left(\underline \mX^\prime \underline \mX \right)^{-1} \sum_t  (\vx_t - \vmu_x) e_t -   O_p(T^{-1}) \, O_p(T^{-0.5}) \, O_p(T^{0.5})\\  &=& O_p(T^{-0.5}) \, .
			\end{eqnarray*}
			Further, by exogeneity and ergodicity,
			\[
			\frac{1}{T}  \sum_{t}  \underline \vx_t \underline e_{t-j} = o_p(1) \, , \quad j \in \mathbb{Z}.
			\]
			Consequently, 
			\begin{eqnarray*}
				\frac{B_T^{(h)}}{T} =  o_p(T^{-0.5}) \, , \quad \frac{C_T^{(h)}}{T} =  o_p(T^{-0.5}) \, , \quad \frac{D_T^{(h)}}{T} =  O_p(T^{-1}) \, .
			\end{eqnarray*}
			Therefore ($h \geq 0$),
            \begin{eqnarray*}
             \widehat \gamma_{\widehat e} (h) = \frac{A_T^{(h)}}{T} + o_p(T^{-0.5}) = \widehat \gamma_{e} (h) + o_p(T^{-0.5} )\stackrel{d}{\to} \gamma_e(h) \, ,  
            \end{eqnarray*}
and ($h \geq 1$),
			\begin{eqnarray*}
				\sqrt{T} (\widehat \gamma_{\widehat e} (h) - \gamma_e(h)) &=&  \sqrt{T} \left( \frac{A_T^{(h)}}{T} + o_p(T^{-0.5}) - \gamma_e(h) \right) \\ &=& \sqrt{T} (\widehat \gamma_{ e} (h) - \gamma_e(h)) + o_p(1) \, .
			\end{eqnarray*}
			The limit of Proposition\,\ref{Prop:Static} under Assumption\,\ref{Ass:stationary} is established by  \eqref{eq:NormalV_generalBartlett}. Hence, the proof is complete.

			\subsection{Proof of Proposition~\ref{Prop:ARDL}} \label{appsubsec:proof_prop2}
			
			\citet{CumbaHuizinga1992} treated regression (\ref{linearModel}) with moving average errors of finite order $q$ that hence may correlate with $\vx_t$ contemporaneously. To account for endogeneity they consider instrumental variables (IV) estimation of $\vbeta$ and study residual autocorrelations only for lags larger than $q$: $\widehat{\rho}_{\widehat e} (q+1), \ldots, \widehat{\rho}_{\widehat e} (q+H)$. Under $q=0$, IV reduces to LS, which is our framework. Further, our Assumptions~\ref{Ass:MDS} and  \ref{ass:process} ensure that, first, the (infeasible) vector $\sqrt{T} \widehat \vrho_e$ is asymptotically normal by \eqref{eq:NormalV_generalBartlett}, and that, second, by an MDS central limit theorem
\[
\sqrt{T} (\widehat \vbeta - \vbeta) \stackrel{d}{\to} \mathcal{N}_K (\vzeros, \mSigma_x^{-1} \mSigma_{x \varepsilon} \mSigma_x^{-1} ) \, .
\]
From  that it follows that
\[
\sqrt{T} \widehat \vrho_{\widehat e} \stackrel{d}{\to} \mathcal{N}_H (\vzeros,\mSigma_\rho) \, ,
\]
where the shape of the asymptotic covariance matrix is given in \citet[Prop.~1]{CumbaHuizinga1992}. In their notation, $\mSigma_\rho$ becomes
\[
\mSigma_\rho:= \mV_r + {\mB \mV_d \mB^\prime} + {\mC {\mD}^\prime  \mB^\prime} + {\mB {\mD}  \mC^\prime}
\]
			It is straightforward to verify that with our notation for $q=0$ it holds that
			\begin{eqnarray*}
				\mV_r &=& {\mB}^*_0 \, , \\
				\mB &=& - \frac{\mGamma}{\sigma_\varepsilon^2}\, , \\
				\mV_d &=& \mSigma_x^{-1} \mSigma_{x \varepsilon} \mSigma_x^{-1} \, , \\
				\mC &=& \mGamma \, , \\
				\mD &=& \mSigma_x^{-1} .
			\end{eqnarray*}
			This yields $\mSigma_\rho$ as required and completes the proof.

			\section{Extensions of Proposition~\ref{Prop:Static} under Cointegration and Time Trends} \label{Sect:Ext_Coint}
			
The static regression results of Proposition~\ref{Prop:Static} do not hinge on  stationarity. They can be extended to the case of static regressions of cointegrated variables and linear time trends.

\subsection{Results under Cointegration} \label{appsubsec:cointegration}
		
Now, let $\mB (r) = (B_e (r), \mB_x^\prime (r))^\prime$ denote a vector Brownian motion of dimension $K+1$ to cover the case of regressors that are integrated of order 1, $I(1)$, but not cointegrated. This turns (\ref{linearModel}) into a cointegrating regression. In order not to burden the exposition with too many technicalities, we simply assume a functional central limit theorem to hold without discussing assumptions behind it; for details see, e.\,g., \citet{PhillipsDurlauf86}.
			
			\begin{assumption}[(Co)Integration] \label{Ass:integrated}
				Let  $\{\vx_t\}$ and $\{e_t\}$ from (\ref{linearModel}) meet
				\begin{equation} \label{FCLT}
					\frac{1}{\sqrt{T}} \left( \begin{array}{c} \sum_{t=1}^{\lfloor r T \rfloor} e_t \\ \vx_{\lfloor r T \rfloor }  \end{array} \right) \ \Rightarrow \ \left( \begin{array}{c} B_e(r) \\ \mB_x (r)  \end{array} \right)\, , \quad  r \in [0,1] \, ,
				\end{equation}
				and $\{(e_t, \Delta \vx_t^\prime)^\prime\}$ is strictly stationary and ergodic with $\Delta \vx_t = \vx_t - \vx_{t-1}$. Further, $\{\vx_t\}$ alone is not cointegrated.	
			\end{assumption}
			
			On top of being integrated, the regressors are allowed to be driven by linear time trends (``integrated with drift''); not all components of $\vdelta$ have to be different from 0.
			
			\begin{assumption}[Integration with drift] \label{Ass:drift}
				Let the regressors be $\vx_t := \vdelta \, t + \vxi_t$, $ \vdelta \neq \vzeros$, 
				where	$\{\vxi_t\}$ meets Assumption\,\ref{Ass:integrated}, i.\,e., $\{\vxi_t\}$ replaces $\{\vx_t\}$ in Assumption\,\ref{Ass:integrated}.
			\end{assumption}

For both cases we can prove one common result coinciding with Proposition \ref{Prop:Static}: The limit result (\ref{eq:NormalV_generalBartlett}) is recovered notwithstanding the replacement of  $\{ e_ t  \}$ by the residual process $\{ \widehat e_ t \}$.

			\begin{proposition} \label{Prop:Cointegration}
				Consider the regression model (\ref{linearModel}) where the error process $\{e_t\}$ is a zero mean process meeting Assumption~\ref{ass:process}, such that (\ref{eq:NormalV_generalBartlett}) holds for $\widehat \vrho_{e} - \vrho_{e}$.  In addition assume  Assumption\,\ref{Ass:integrated} or Assumption\,\ref{Ass:drift}.  With LS  residual autocorrelations $\widehat \vrho_{\widehat e}$ the limiting distribution from (\ref{eq:NormalV_generalBartlett}) continues to hold as $T \to \infty$,
					\begin{equation} \label{eq:NormalV_e_Bartlett_hat}
					{\sqrt{T}} \left( \widehat{\bs{\rho}}_{\widehat e}  - \bs{\rho}_e	\right) \ \stackrel{d}{\to} \ \mathcal{N}_H (\bs{0}, \, \bs{B}^*) \, ,
				\end{equation}
			where $\bs B^*$ is given in terms of $\{e_t\}$.
			\end{proposition}

Proposition~\ref{Prop:Cointegration} has been formulated for LS estimation. Similarly, under Assumption \ref{Ass:integrated} or \ref{Ass:drift}, one may allow for so-called efficient cointegrating regression as defined by \citet{saikkonen1991}, see also \citet{phillips1990}, \citet{park1992} or \citet{VogelsangWagner2014}.

\subsection{Proof of Proposition~\ref{Prop:Cointegration}} \label{appsubsec:proof_cointegration}

We maintain the notation of demeaned variables from the previous section. Further, all integrals are  from
		$0$ to $1$, unless stated otherwise and the integration variable may be suppressed. The proof consists of two steps according to Assumptions~\ref{Ass:integrated} and \ref{Ass:drift}, respectively.

{\sc Under Assumption\,\ref{Ass:integrated}}

Let $\vu_t^\prime :=(e_t, \Delta \vx_t^\prime)$ and assume that
\[
\mOmega := \lim_{T \to \infty} \E \left[ \sum_{t=1}^T \vu_t \sum_{t=1}^T \vu_t^\prime  \right] = \mSigma_0 + \mLambda + \mLambda^\prime
\]
is finite, where
\[
\mSigma_0 := \E [\vu_t \vu_t^\prime ] \, , \quad \mLambda := \sum_{j=1}^\infty \vu_t \vu_{t+j}^\prime \, .
\]
Further, partition
\[
\mOmega = \left( \begin{array}{cc} \omega_e & \vomega_{xe}^\prime  \\ \vomega_{xe} & \mOmega_x  \end{array} \right),
\]
where $\mOmega_x$ is positive definite ruling out cointegration among the regressors. Under further technical assumptions spelled out in \citet[Coro.\,2.2]{PhillipsDurlauf86}, it follows  that
\[
\frac{1}{\sqrt{T}} \sum_{t=1}^{\lfloor r T \rfloor } \vu_t \Rightarrow  \mB (r) = \left( \begin{array}{c} B_e(r) \\ \mB_x (r)  \end{array} \right) = \mOmega^{0.5} \mW (r),
\]
where $\mW$ is a standard vector Wiener process while $B_e$ and $\mB_x$ are in general correlated.
			
It follows by \citet[Lemma\,2.1]{ParkPhillips88} and the continuous mapping theorem that
			\[
			\frac{1}{T^{1.5}}  \sum_t \vx_t   \ \stackrel{d}{\to} \  \int \mB_x (r) \, \mbox{d} r  \, , \quad 
			\frac{1}{T^2}  \sum_t \underline \vx_t \underline \vx_t^\prime   \ \stackrel{d}{\to} \ \int \underline \mB_x \underline \mB_x^\prime \, ,
			\]
			\[
			\frac{1}{{T}}  \sum_t \underline \vx_t  \underline e_{t} = \frac{1}{{T}}  \sum_t \underline \vx_t   e_{t}  \ \stackrel{d}{\to} \ \mI^{(0)} := \int \underline \mB_x  \mbox{d} B_e + \sum_{j=0}^\infty \E[\Delta \vx_t e_{t+j}] \, ,
			\]
			where $\underline \mB (r) := \mB (r) - \int  \mB (s)  \, \mbox{d} s$ stands for a so-called demeaned Brownian motion. \citet[Thm.\,3.2]{ParkPhillips88} thus showed that
			\[
			{T} \, (\widehat{\vbeta}-{\vbeta}) \ \stackrel{d}{\to} \ \left(\int \underline \mB_x \underline \mB_x^\prime\right)^{-1} \mI^{(0)}.
			\]
			Because of $\vx_t = \vx_{t-h} + \Delta \vx_t + \ldots + \Delta \vx_{t-h+1}$ it is straightforward to obtain ($h \geq 0$)
			\[
			\frac{1}{{T}}  \sum_t \underline \vx_{t-h}   e_{t}  \ \stackrel{d}{\to} \ \mI^{(h)} := \mI^{(0)} - \sum_{j=0}^{h-1} \E[\Delta \vx_{t-j} e_{t}] \, ,
			\]
			\[
			\frac{1}{{T}}  \sum_t \underline \vx_{t}   e_{t-h}  \ \stackrel{d}{\to} \ \mI^{(-h)} := \mI^{(0)} + \sum_{j=0}^{h-1} \E[\Delta \vx_{t-j} e_{t-h}] \, .
			\]
			For $B_T^{(h)}$ through $D_T^{(h)}$ from (\ref{ABCD}) one hence observes
			\[
			B_T^{(h)}  = O_p(1) \, , \ C_T^{(h)}  = O_p(1) \, , \ D_T^{(h)}  = O_p(1) \, .
			\]
			Consequently ($h \geq 0$),
            \begin{eqnarray*}
             \widehat \gamma_{\widehat e} (h) = \frac{A_T^{(h)}}{T} + O_p(T^{-1}) = \widehat \gamma_{e} (h) + O_p(T^{-1} )\stackrel{d}{\to} \gamma_e(h) \, ,  
            \end{eqnarray*}
and ($h \geq 1$),
\begin{eqnarray*}
				\sqrt{T} (\widehat \gamma_{\widehat e} (h) - \gamma_e(h)) &=&  \sqrt{T} \left( \frac{A_T^{(h)}}{T} + O_p(T^{-1}) - \gamma_e(h) \right) \\ &=& \sqrt{T} (\widehat \gamma_{ e} (h) - \gamma_e(h)) + O_p(T^{-0.5}) \, .
                \end{eqnarray*}
The limit of (\ref{eq:NormalV_e_Bartlett_hat}) under Assumption\,\ref{Ass:integrated} is established by (\ref{eq:NormalV_generalBartlett}).
			
\bigskip		
			
			{\sc Under Assumption\,\ref{Ass:drift}}
			
			Two cases have to be distinguished: scalar regressor ($K=1$) and $K > 1$.\\
			Case $K=1$: Note that
			\[
			\frac{1}{T^3} \sum_{t=1}^T \left(t - \frac{T+1}2 \right)^2 \to \frac{1}{12} \quad \mbox{and } \frac{1}{T^3} \sum_{t=1}^T \underline{x}_t^2 \stackrel{p}{\to} \frac{\delta^2}{12} \, .
			\]
			At the same time, by \citet[Lemma\,2.1]{ParkPhillips88},
			\[
			\frac{1}{T^{1.5}} \sum_{t=1}^T \underline{x}_t e_t \stackrel{d}{\to} \int r \mbox{d} B_e(r) - \frac{1}{2} B_e(1) \, .
			\]
			Consequently, $\widehat \beta - \beta  = O_p(T^{-1.5})$. Since
            \[
            \frac{1}{T^{1.5}} \sum_{t=1}^T \underline{x}_{t-h} e_t = \frac{1}{T^{1.5}} \sum_{t=1}^T \underline{x}_t e_t + o_p(1) \, ,
            \]
it follows that     
			\[
			B_T^{(h)} = O_p(1) \, , \quad C_T^{(h)} = O_p(1) \, , \quad D_T^{(h)} = O_p(1) \, , 
			\]
			and the argument is completed as under Assumption\,\ref{Ass:integrated}.\\
			Case $K>1$: Define $\vh_1 :=\vdelta/\sqrt{\vdelta^\prime \vdelta}$ of unit length. As a corollary to Gram-Schmidt orthogonalization there exists $\mH_2$ such that the $(K \times K)$-matrix $\mH$  with $\vh_1$ as first column is orthogonal, see e.\,g., \citet[Prop.\,2.50]{Dhrymes2000}:
			\[
			\mH := (\vh_1, \, \mH_2) \, , \quad \mH \mH^\prime = \mH^\prime \mH = \mI_K. 
			\]
			Consequently, $\mH_2^\prime \vdelta = \vzeros$, and the linear trend can be concentrated in a single component $z_{1,t}$:
			\begin{eqnarray*}
				y_t &=& a + \vbeta^\prime {\vx}_t +e_t\\
				&=& a + \vbeta^\prime \mH \left( \begin{array}{c} \vh_1^\prime \vx_t \\ \mH_2^\prime \vx_t \end{array} \right) +e_t\\
				&=& a + d z_{1,t} + \vb^\prime  {\vz}_{2,t}+e_t \, , 
			\end{eqnarray*}
			where $z_{1,t} := \vh_1^\prime \vx_t$, ${\vz}_{2,t}:=\mH_2^\prime \vx_t$,  $d:=\vbeta^\prime \vh_1$ and  $\vb:=\mH_2^\prime \vbeta$. While $z_{1,t}$ is dominated by the linear trend component, the $I(1)$  vector ${\vz}_{2,t}$ is without drift. It follows by \citet[Thm.\,3.6]{ParkPhillips88} that
			\begin{eqnarray*}
				\widehat d - d &=& (\widehat \vbeta - \vbeta)^\prime \vh_1 = O_p(T^{-1.5}) \\
				\widehat \vb - \vb &=& \mH_2^\prime (\widehat \vbeta - \vbeta) = O_p(T^{-1})\\
				\widehat a - a &=&  O_p(T^{-0.5}).
			\end{eqnarray*}
			Thus, the residuals turn into
			\[
			\widehat e_t = e_t - O_p(T^{-0.5}) -  O_p(T^{-1.5}) z_{1,t} -  O_p(T^{-1}) z_{2,t},
			\]
			where $z_{2,t}$ as a linear combination of the vector $\vz_{2,t}$ is I(1) without drift. Careful evaluation of $\sum_{t=h+1}^T \widehat e_t \widehat e_{t-h}$ yields (using repeatedly \citet[Lemma\,2.1]{ParkPhillips88})
			\[
			\frac{1}{\sqrt{T}} \sum_{t=h+1}^T \widehat e_t \widehat e_{t-h} = \frac{1}{\sqrt{T}} \sum_{t=h+1}^T  e_t  e_{t-h} + O_p(T^{-0.5}) \, , \quad h \geq 0 \, .
			\]
			Consequently, as under Assumption\,\ref{Ass:integrated},
			\begin{eqnarray*}
				\sqrt{T} (\widehat \gamma_{\widehat e} (h) - \gamma_e(h)) = \sqrt{T} (\widehat \gamma_{ e} (h) - \gamma_e(h)) + O_p(T^{-0.5}) \, .
			\end{eqnarray*}
The limit of (\ref{eq:NormalV_e_Bartlett_hat}) under Assumption\,\ref{Ass:drift} is established by (\ref{eq:NormalV_generalBartlett}).
           
Hence, the proof is complete.

		\subsection{Further Results in the Presence of Linear Time Trends} \label{appsubsec:integration_with_drift}

Propositions~\ref{Prop:Static} and \ref{Prop:Cointegration} continue to hold when detrending the regressors or when the processes are trend-stationary.

			
Let us briefly turn again to the $I(1)$ case with drift. The LS analysis under Assumption~\ref{Ass:drift} is complicated by
		\[
		\frac{1}{T^3} \sum_{t=1}^T \vx_t \vx_t^\prime \stackrel{p}{\to} \vdelta \vdelta^\prime ,
		\]
		which has reduced rank for $K>1$. To circumvent this problem, \citet{ParkPhillips88}  concentrate all linear trends in one scalar component $z_{1,t}$, such that (\ref{linearModel}) becomes
		\begin{equation*} \label{linModeldrift}
			y_t= a + d z_{1,t} + {\vz}_{2,t}^\prime \vb +e_t \, , \quad t=1, \ldots, T \, ,
		\end{equation*}
		where ${\vz}_{2,t}$ is an $I(1)$  vector without drift of dimension $K-1$ and $z_{1,t}$ is integrated with drift; details are given in the proof of Proposition \ref{Prop:Static}. Along the same lines one may study a detrended regression, i.\,e., replace (\ref{linearModel}) by
		\begin{equation} \label{linModeldetrend}
			y_t= a + d t + {\vx}_{t}^\prime \vbeta +e_t \, , \quad t=1, \ldots, T \, .
		\end{equation}
		Similar to the proof of Proposition\,\ref{Prop:Cointegration} under Assumption\,\ref{Ass:drift} one can show:
		The limit result (\ref{eq:NormalV_e_Bartlett_hat}) 
        continues to hold when replacing  $\{ e_ t  \}$ by the residual process $\{ \widehat e_ t \}$ upon detrending (i.\,e., including $dt$). We thus have the following corollary under Assumption\,\ref{Ass:stationary}, \ref{Ass:integrated} or \ref{Ass:drift}.
		
		\begin{corollary} \label{Coro:Drift}
			Consider LS residuals $\widehat e_t = y_t - \widehat a - \widehat d t - \widehat \vbeta^\prime \vx_t$ from (\ref{linModeldetrend}). Under the assumptions of Propositions\,\ref{Prop:Static} or \ref{Prop:Cointegration} the limit distribution from (\ref{eq:NormalV_e_Bartlett_hat}) continues to hold  for the  residual autocorrelations $\widehat \vrho_{\widehat e}$ as $T \to \infty$.
		\end{corollary}
			
		As a special case of $I(1)$ regressors with drift, consider the scalar case ($K=1$) with $\delta \neq 0$:
		\[
		x_{\lfloor r T \rfloor} = \delta \, \lfloor r T \rfloor + \xi_{\lfloor r T \rfloor} = O (T) + O_p (T^{0.5}) .
		\]
		Clearly, the linear trend dominates the stochastic component. In fact, the Brownian motion behind $\xi_t$ according to (\ref{FCLT}) does not show up in the limit, see \citet{West88}. Consequently, it may be replaced by a stochastic component of smaller order, e.\,g., by $\Delta \xi_t$, which is $I(0)$ and hence $O_p(1)$. This amounts to a so-called trend-stationary process where 
		\[
		x_t = \delta \, t + \Delta \xi_t \, .
		\]
		The limit distribution equals that of the $I(1)$ case with drift, see \citet{Hassler2000} for details. Hence, we have the following corollary.
		
		\begin{corollary} \label{Coro:Static}
			Consider the regression model (\ref{linearModel}) with a scalar trend-stationary regressor, $x_t= \delta \, t + \Delta \xi_t$, where $\{\xi_t\}$ meets Assumption\,\ref{Ass:integrated} and $\delta \neq 0$. Under the error assumption of Assumption\,\ref{Ass:stationary}  the limit result from Proposition\,\ref{Prop:Static} continues to hold.
		\end{corollary}
		
		The case of trend-stationary regressors with $K >1$ has attracted little attention in the literature and seems to be of little empirical relevance. This is why we do not discuss this case.

		\subsection{Results on Error-Correction Regressions} \label{appsubsec:ECM}

		Here, we briefly treat the error-correction model (ECM) that combines Assumption\,\ref{Ass:integrated} with Assumption\,\ref{Ass:MDS}. Let $y_t$ and $\vx_t$ be cointegrated such that
		\[
		\eta_t := y_t - a - \vbeta^\prime \vx_t
		\]
		is integrated of order 0. Further, assume that $y_t$ is error-correcting ($\gamma \neq 0$),
		\begin{equation} \label{ECM}
			\Delta y_t = \alpha + \gamma \, \eta_{t-1} +  \vtheta^\prime \vz_{t} + e_{t} \, , 
		\end{equation}
		\[
		\vz_{t} = (\Delta y_{t-1}, \ldots, \Delta y_{t-p}, \Delta \vx_{t-1}^\prime , \ldots, \Delta \vx_{t-q}^\prime)^\prime .
		\]
		In practice, $\eta_{t}$ is not observable but may be replaced by the LS residual of the cointegrating regression (\ref{linearModel}), now called $\widehat \eta_t$ to avoid confusion:
		\[
		\widehat \eta_t := y_t - \widehat a - \widehat \beta^\prime \vx_t \, .
		\]
		The LS regression considered in a second step is
		\begin{equation} \label{ECMhat}
			\Delta y_t = \alpha + \gamma \, \widehat \eta_{t-1} +  \vtheta^\prime \vz_{t} + e_{t} \, , \quad t=1, \ldots, T \, .
		\end{equation}
The effect of the first step is negligible since
\[
\eta_t - \widehat \eta_t = (\widehat a - a) + (\widehat \vbeta - \vbeta)^\prime \vx_t = O_p (T^{-0.5}) + O_p (T^{-1}) \vx_t = O_p (T^{-0.5}) \, ,
\]
		where the rates of $\widehat a$ and $\widehat \vbeta $ are known e.\,g., from \citet{Stock87} or \citet{ParkPhillips88}. Hence, the following corollary follows immediately.
		
		\begin{corollary} \label{Coro:ECM}
			Consider the regression model (\ref{ECMhat}) under Assumption\,\ref{Ass:integrated}, and let the regressors in (\ref{ECM}) be predetermined in that Assumption\,\ref{Ass:MDS} holds for $e_t$ and $(\eta_{t-1}, \vz_t^\prime)^\prime$. With the above notation, the limit result from Proposition\,\ref{Prop:ARDL} continues to hold for the LS residual autocorrelations computed from (\ref{ECMhat}).
		\end{corollary}

		\section{Additional Theoretical Results} \label{Sect:Additiona_Theoretical_Results}
			
		\subsection{Lemma on an Upper Bound for Equicoordinate Quantiles} \label{appsubsec:additional_theoretical_results}
		
		The following lemma is important for the comparisons of width and coverage of different types of confidence and significance bands that we consider in the paper. A related discussion can be found in the Appendix of \cite{montielolea2019}.
		
		\begin{lemma} \label{lemma:width}
			Let $\mSigma$ denote the covariance matrix of a Gaussian random vector and let $q_{1-\alpha} (\mSigma)$ denote its $1-\alpha$ equicoordinate quantile as defined around \eqref{eq:M_max_normal}. Let $\chi^2_{1-\alpha}(1)$ denote the $1-\alpha$ quantile of the $\chi^2$ distribution with one degree of freedom. It holds that
			$$q_{1-\alpha} (\mSigma) \leq q_{1-\alpha} (\mI_H) = z_{(1 + (1-\alpha)^{\frac{1}{H}})/2} = \sqrt{\chi^2_{(1-\alpha)^{\frac{1}{H}}}(1)} \leq \sqrt{\chi^2_{1-\frac{\alpha}{H}}(1)} = z_{1-\alpha/2H}.$$
		\end{lemma}
		
		{\sc Proof}
		The first inequality follows from Corollary 1 in \cite{sidak1967}, which implies that the coverage of rectangular confidence regions for a multivariate normal random vector is smallest under independence. This in turn implies that the $1-\alpha$ equicoordinate quantile has to be largest under independence. The second inequality follows because $(1-\alpha)^{\frac{1}{H}} \leq 1-\frac{\alpha}{H}$ for $0 < \alpha < 1$ and $H \in \mathbb{N}$.

        \subsection{Proposition on the Relation of Corrected and Naive Classical Significance Bands for Regression Errors} \label{appsubsec:prop_naive_vs_corrected}

        			\begin{proposition} \label{Prop:significance_bands_dynamic_regression}
				Consider the regression model (\ref{linearModel}) under Assumption\,\ref{Ass:MDS}.
				Denote the width of the generic sup-t significance band ${SB}^{supt}_{\vtheta} (\mSigma_0)$ from \eqref{eq:S_general} at lag $h$ by $w(h,{SB}^{supt}_{\vtheta} (\mSigma_0))  := 2 q_{1-\alpha} (\mSigma_0) \sqrt{\frac{\sigma_{hh,0}}{T}}$, see \eqref{eq:width_inference_band}.

				\begin{enumerate}
					\item Then it holds  that
						$w(h,{SB}^{supt}_{\bm{\rho}_{e}} (\widehat \mSigma_\rho^{hom})) \leq w(h,{SB}^{supt}_{\bm{\rho}_{e}} (\mI_H))$.
					\item Let additionally Assumption \ref{Ass:homoskedastic_errors} hold. Then it holds that 
					$w(h,{SB}^{supt}_{\bm{\rho}_{e}} (\mSigma_\rho)) \leq w(h,{SB}^{supt}_{\bm{\rho}_{e}} (\mI_H))$ and that $\lim_{T \to \infty} P \left( \widehat{\bs{\rho}}_{\widehat{e}} \in {SB}^{supt}_{\bm{\rho}_{e}} (\mI_H) \right)  \geq 1 - \alpha$.
				\end{enumerate}
		
			\end{proposition}  
			
			{\sc Proof}         
			Note that $\mSigma_{\rho}^{hom}$ from \eqref{eq:asymp_variance_homoskedastic_errors} is weakly smaller than $\mI_H$ in every entry due to the positive definiteness of $\mSigma_x$. This also holds true for its estimator $\widehat{\mSigma}_\rho^{hom}$ due to $\widehat \mSigma_x$ being positive semi-definite. Thus, firstly in particular the diagonal elements of $\widehat{\mSigma}_\rho^{hom}$ are smaller than 1 and by Lemma \ref{lemma:width} $q_{1-\alpha} (\widehat{\mSigma}_\rho^{hom}) \leq q_{1-\alpha} (\mI_H)$. Thus, the first part follows by using the width formula in the Proposition.
			
			The first statement of the second part follows by the same arguments, noting that under Assumption \ref{Ass:homoskedastic_errors} $\mSigma_{\rho}^{het} = \mSigma_{\rho}^{hom}$, see \eqref{eq:asymp_variance_homoskedastic_errors}.
			
			For the second statement of the second part we establish a lower bound for the asymptotic coverage probability of the naive significance band:
			\begin{align*} 
				\lim_{T \to \infty} P \left( \widehat{\bs{\rho}}_{\widehat{e}}  {SB}^{supt}_{\bm{\rho}_{e}} (\mI_H))   \right)
				&= \lim_{T \to \infty} P \left(  \left| \sqrt{T} \widehat{\rho}_{\widehat{e}} (1) \right| \leq q_{1-\alpha} (\mI_H), \ldots, \left| \sqrt{T} \widehat{\rho}_{\widehat{e}} (H) \right| \leq q_{1-\alpha} (\mI_H) \right)\\
				&= P \left( |V_1| \leq q_{1-\alpha} (\mI_H), \ldots, |V_H| \leq q_{1-\alpha} (\mI_H) \right)\\
				&\geq P \left( |V_1| \leq q_{1-\alpha} (\mSigma_\rho^{hom}), \ldots, |V_H| \leq q_{1-\alpha} (\mSigma_\rho^{hom}) \right)\\
				&\geq P \left( \frac{|V_1|}{\sqrt{\sigma_{\rho,11}}} \leq q_{1-\alpha} (\mSigma_\rho^{hom}), \ldots, \frac{|V_H|}{{\sqrt{\sigma_{\rho,HH}}}} \leq q_{1-\alpha} (\mSigma_\rho^{hom}) \right)\\
				&= \lim_{T \to \infty} P \left(  \frac{\left| \sqrt{T} \widehat{\rho}_{\widehat{e}} (1) \right|}{{\sqrt{\sigma_{\rho,11}}}} \leq q_{1-\alpha} (\mSigma_\rho^{hom}), \ldots, \frac{\left| \sqrt{T} \widehat{\rho}_{\widehat{e}} (H) \right|}{{{\sqrt{\sigma_{\rho,HH}}}}} \leq q_{1-\alpha} (\mSigma_\rho^{hom}) \right)\\
				&=1-\alpha,
			\end{align*}
			where $\bs V = \left( V_1, \cdots , V_H \right)' \sim \mathcal{N}_H \left(\vzeros, \mSigma_\rho^{hom} \right)$, $\sigma_{\rho,gh}$ denotes an element of $\mSigma_\rho^{hom}$. The first inequality follows from the first inequality in Lemma \ref{lemma:width}, the second inequality follows because under homoskedasticity the diagonal elements of $\mSigma_\rho^{hom}$ are smaller than 1, and in the second to last row the formula for the asymptotic coverage of ${SB}^{supt}_{\bm{\rho}_{\varepsilon}} (\mSigma_\rho^{hom})$ appears, which equals $1-\alpha$. Thus, the proof is complete.

            The first result of Proposition \ref{Prop:significance_bands_dynamic_regression} remarkably is a finite-sample result that states that as long as the exact bands are estimated using homoskedastic standard errors, the naive bands are always weakly wider (in every sample). The second result states that when the true covariance matrix is used, under homoskedasticity the naive bands are at least as wide as the exact bands. This in particular means that as long as a consistent estimator is used, this relation holds asymptotically, which also implies that the naive bands are conservative, that is, have asymptotic overcoverage.
		
\section{Theoretical Examples}
\label{sec:theoretical_example_time_series}

\subsection{Properties and Relations of the Inference Bands} \label{subsec:theoretical_example_time_series}

We now analyse and illustrate the properties (i.\,e., width and coverage) of our inference bands and the relations between the different types of confidence bands on the one hand and the confidence and significance bands on the other.

The width of the generic confidence band $\bigtimes_{h=1}^H \left[\widehat \theta_h \pm c \cdot  \sqrt{\frac{\sigma_{hh}}{T}}\right]$ at lag $h$ equals 
\begin{equation} \label{eq:width_inference_band}
	w(h) = 2 c \cdot  \sqrt{\frac{\sigma_{hh}}{T}}.
\end{equation}
In this Section, we work under Assumption \ref{ass:Bartlett}. Thus, for the classical confidence bands for $\bm{\rho}$ this translates to $\sigma_{hh} = \widehat b_{hh}$ and $c$ equals $z_{1-\alpha/(2H)}$ for the Bonferroni band, $q_{1-\alpha} (\widehat \mB)$ for the sup-t band \eqref{eq:supt_rho} and $z_{1-\alpha/2}$ for the pointwise band. The formula also holds for the classical significance bands with $\sigma_{hh} = 1$ and $c$ equalling $z_{(1+(1-\alpha)^{1/H})/2}$ for the sup-t band \eqref{eq:simultaneous_significance_band_rho}, $z_{1-\alpha/(2H)}$ for the Bonferroni band, and $z_{1-\alpha/2}$ for the pointwise band \eqref{eq:pointwise_significance_band_rho}. Thus, the width of all the bands grows with the scaling factor $c$ and shrinks with the sample size $T$. The only influence of the lag $h$ is through the variance $\sigma_{hh}$. Consequently, the width of the significance bands for $\bm{\rho}$ is constant over the lags. The width of the confidence bands depends on the data only through the estimated covariance matrix $\widehat \mB$, where for the Bonferroni bands the influence is only through the variance $\widehat b_{hh}$ and for the sup-t bands additionally through the equicoordinate quantile $q_{1-\alpha} (\widehat \mB)$.

We now analyse the properties and relations of the inference bands for a specific data generating process (DGP), the autoregressive process of order 1 (AR(1) process), 
\begin{equation} \label{DGP_AR}
	y_t = \phi \, y_{t-1} + \varepsilon_t \, , \quad \varepsilon_t \sim iid (0,1)
\end{equation}
with five different parameter values, $\phi=0, 0.25, 0.5, 0.75,0.95$, representing varying degrees of temporal dependence from white noise to very strong persistence. 
This DGP is simple in that it allows to control the degree of temporal dependence in one parameter, but at the same time it is realistic and relevant in that it yields a good approximation to the behaviour of many economic time series.

We do not estimate the Bartlett covariance matrix $\mB$, but assume that the AR(1) parameter $\phi$ is known and thus the covariance matrix can be calculated by the formulas in \citet{Cavazos-Cadena1994}; see  Section~\ref{appsubsec:additional_material_example} for the $\mB$-matrices for $\phi=0,0.25, 0.5, 0.75,0.95$. Thus, the following width comparisons can be regarded as asymptotic, complementing the width and coverage comparisons in the simulation study. Also the coverages we plot below are asymptotic, making use of the asymptotic normality of the empirical autocorrelations.

\begin{figure}
	\noindent \centering{}\includegraphics[scale=0.5]{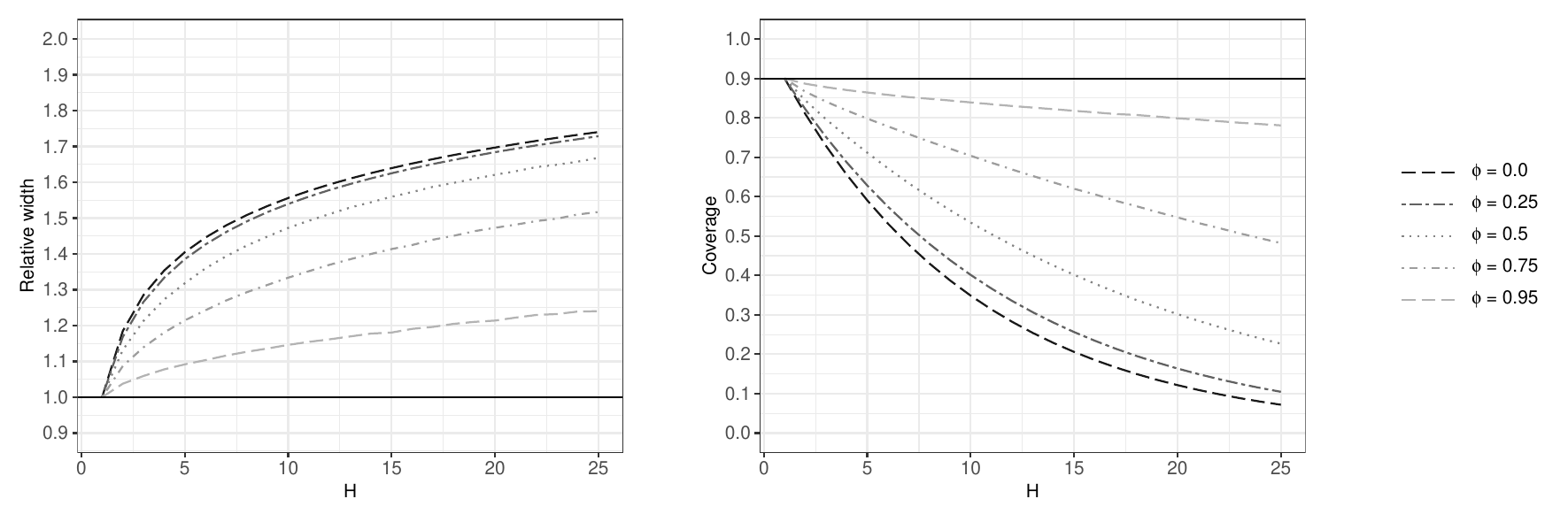}
	\vspace{-0.5cm}
	\caption{Relative width of $90\%$ sup-t and pointwise confidence bands and asymptotic coverage of pointwise bands for an AR(1) process with parameter $\phi$.}	
	\label{fig:conf_bands_width_and_coverage_pointwise}
\end{figure}

In the left panel of Figure \ref{fig:conf_bands_width_and_coverage_pointwise} we plot the relative width of the $90\%$ sup-t and pointwise confidence band, which equals $q_{0.9} (\mB)/z_{0.95}$ and does not change with $h$, against different values for the length of the band $H$ and for the different values of $\phi$. Note that the confidence bands for $\phi=0$ also represent the respective significance bands. Of course, the width of the sup-t band rises with its length $H$, but grows slowly and at a decreasing rate. The stronger the temporal dependence is, the smaller the relative width (as equicoordinate quantiles naturally get smaller under stronger dependence). Under independence and $H=25$, the sup-t band is roughly 1.75 times as wide as the pointwise band. The right panel of Figure \ref{fig:conf_bands_width_and_coverage_pointwise} depicts the asymptotic coverage of the pointwise bands. As expected, they show strong undercoverage increasing with $H$ and decreasing with the degree of temporal dependence, while the sup-t bands have exact asymptotic coverage of 0.9.

\begin{figure}[H]
	\noindent \centering{}\includegraphics[scale=0.5]{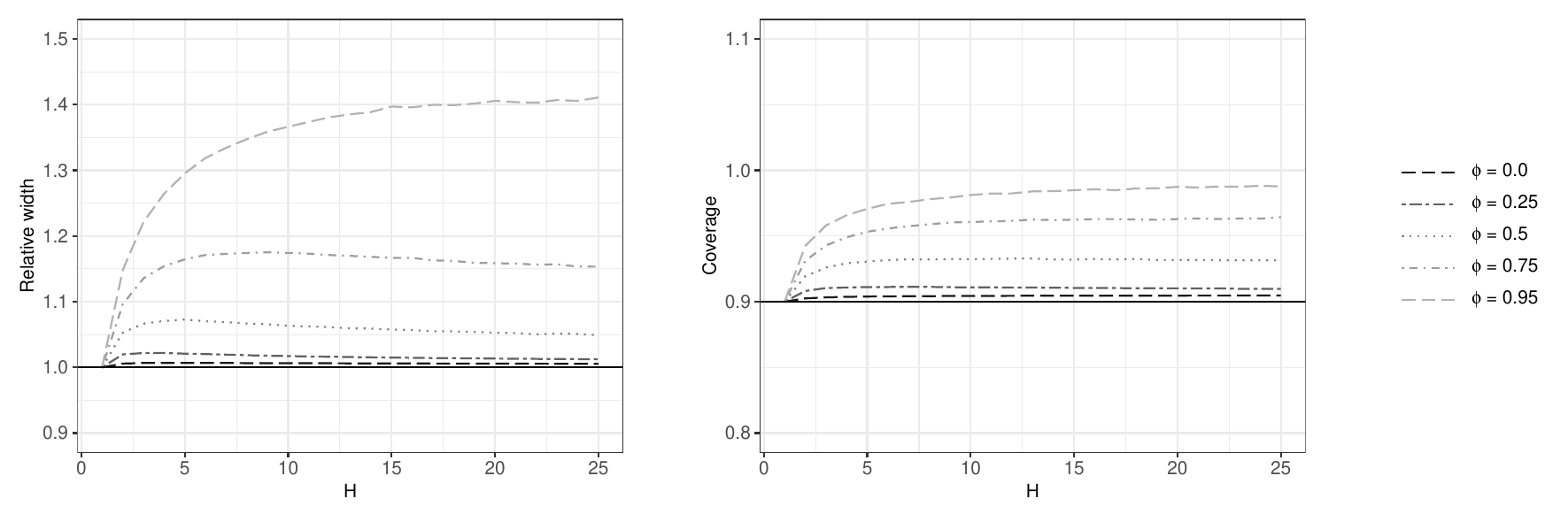}
	\vspace{-0.5cm}
	\caption{Relative width of $90\%$ Bonferroni and sup-t confidence bands and asymptotic coverage of Bonferroni bands for an AR(1) process with parameter $\phi$.}	
	\label{fig:conf_bands_relative_width_ar1}
\end{figure}

The left panel of Figure \ref{fig:conf_bands_relative_width_ar1} depicts the relative width of the $90\%$ Bonferroni and sup-t band, $z_{1-1/(20H)}/q_{0.9} (\mB)$, the right panel depicts the asymptotic coverage of the Bonferroni band. As mentioned in Section \ref{sec:inference_bands_general}, the Bonferroni band  is always at least as wide as the sup-t band. Its conservativeness in terms of width and coverage increases with the degree of persistence. For $\phi=0.75$ it is about $15\%$ wider than the sup-t band for most values of $H$ and has an asymptotic coverage of about $95\%$ instead of the desired $90\%$.

\begin{figure}
	\noindent \centering{}\includegraphics[scale=0.5]{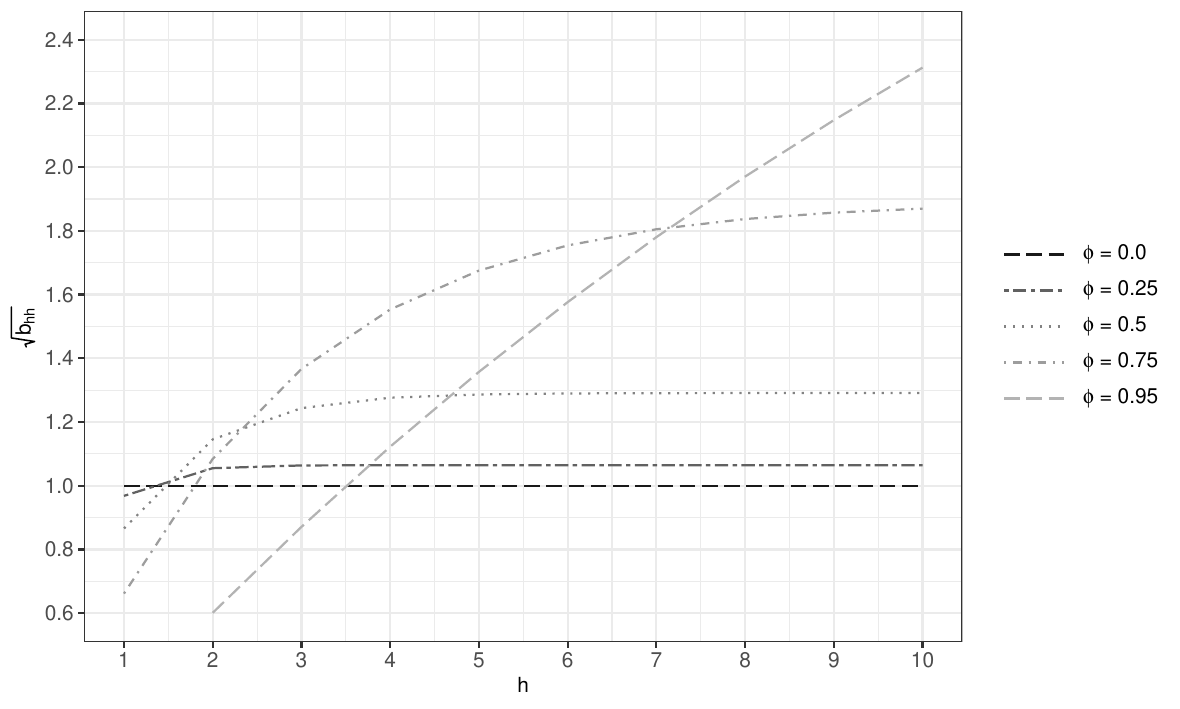}
	\vspace{-0.5cm}
	\caption{Plot of $\sqrt{b_{hh}}$ against $h$ determining the width of  sup-t confidence bands from (\ref{eq:supt_rho}) for $\phi=0,0.25,0.5,0.75, 0.95$.}	
	\label{fig:conf_bands_width_over_h}
\end{figure}

Now we analyze how the width of the sup-t  bands changes with the lag $h$. Figure~\ref{fig:conf_bands_width_over_h} plots the standard deviation $\sqrt{b_{hh}}$ against $h$, which determines the behaviour of the width for a certain value of $\phi$. An interesting  pattern emerges. While for the case $\phi=0$, the width of the band is constant over $h$, this is not the case for the other values of $\phi$. There, for $h=1$ the variance $b_{11}$ is smaller than 1 (the higher $\phi$, the smaller) 
and rising with $h$, quickly approaching a certain limit (the higher $\phi$, the larger). That is, the width of the confidence band first grows with $h$ and from a certain small lag on stays virtually constant. When computing the average width over $h$, the bands are the wider the stronger the degree of temporal dependence is, but the width for the first lag (which shows the opposite behaviour) can be especially important in terms of power (see our simulations in Section \ref{sec:simulations}).

Overall, the counteracting effects on the width, see \eqref{eq:width_inference_band}, of the equicoordinate quantile (which decreases with degree of temporal dependence) and standard deviation (which increases with degree of temporal dependence) is dominated by the standard deviation so that the average width of the confidence bands (averaged over h) increases with the degree of temporal dependence (see again also our simulation results in Section \ref{sec:simulations}) as one would expect. In particular, the significance bands (with unchanged width over $h$) are narrower than the corresponding confidence bands (again averaged over $h$).

\subsection{Bartlett Covariance Matrices for the AR(1) Case} \label{appsubsec:additional_material_example}
			
			The following contains examples for the matrix $\mB  = \left(b_{gh}\right)_{g,h = 1, \ldots, H}$ given by Bartlett's formula. We calculate  $\mB (\phi)$ for an AR(1) process following Cavazos-Cadena (1994) for $ H =10$. Results are rounded to three decimal places.
			
			\[
			\mB (0) = \begin{bmatrix} 1 & 0 & 0 & 0 & 0 & 0 & 0 & 0 & 0 & 0 \\ 0 & 1 & 0 & 0 & 0 & 0 & 0 & 0 & 0 & 0 \\ 0 & 0 & 1 & 0 & 0 & 0 & 0 & 0 & 0 & 0 \\ 0 & 0 & 0 & 1 & 0 & 0 & 0 & 0 & 0 & 0 \\ 0 & 0 & 0 & 0 & 1 & 0 & 0 & 0 & 0 & 0 \\ 0 & 0 & 0 & 0 & 0 & 1 & 0 & 0 & 0 & 0 \\ 0 & 0 & 0 & 0 & 0 & 0 & 1 & 0 & 0 & 0 \\ 0 & 0 & 0 & 0 & 0 & 0 & 0 & 1 & 0 & 0 \\ 0 & 0 & 0 & 0 & 0 & 0 & 0 & 0 & 1 & 0 \\ 0 & 0 & 0 & 0 & 0 & 0 & 0 & 0 & 0 & 1 \end{bmatrix}
			\] 
			
			\[
			\mB (0.25) = \begin{bmatrix} 0.938 & 0.469 & 0.176 & 0.059 & 0.018 & 0.005 & 0.002 & 0 & 0 & 0 \\ 0.469 & 1.113 & 0.527 & 0.194 & 0.064 & 0.02 & 0.006 & 0.002 & 0 & 0 \\ 0.176 & 0.527 & 1.132 & 0.533 & 0.196 & 0.065 & 0.02 & 0.006 & 0.002 & 0 \\ 0.059 & 0.194 & 0.533 & 1.133 & 0.533 & 0.196 & 0.065 & 0.02 & 0.006 & 0.002 \\ 0.018 & 0.064 & 0.196 & 0.533 & 1.133 & 0.533 & 0.196 & 0.065 & 0.02 & 0.006 \\ 0.005 & 0.02 & 0.065 & 0.196 & 0.533 & 1.133 & 0.533 & 0.196 & 0.065 & 0.02 \\ 0.002 & 0.006 & 0.02 & 0.065 & 0.196 & 0.533 & 1.133 & 0.533 & 0.196 & 0.065 \\ 0 & 0.002 & 0.006 & 0.02 & 0.065 & 0.196 & 0.533 & 1.133 & 0.533 & 0.196 \\ 0 & 0 & 0.002 & 0.006 & 0.02 & 0.065 & 0.196 & 0.533 & 1.133 & 0.533 \\ 0 & 0 & 0 & 0.002 & 0.006 & 0.02 & 0.065 & 0.196 & 0.533 & 1.133 \end{bmatrix}
			\]

			\[
			\mB (0.5) = \begin{bmatrix} 0.75 & 0.75 & 0.562 & 0.375 & 0.234 & 0.141 & 0.082 & 0.047 & 0.026 & 0.015 \\ 0.75 & 1.312 & 1.125 & 0.797 & 0.516 & 0.316 & 0.188 & 0.108 & 0.062 & 0.034 \\ 0.562 & 1.125 & 1.547 & 1.266 & 0.879 & 0.562 & 0.343 & 0.202 & 0.116 & 0.066 \\ 0.375 & 0.797 & 1.266 & 1.629 & 1.312 & 0.905 & 0.577 & 0.351 & 0.207 & 0.119 \\ 0.234 & 0.516 & 0.879 & 1.312 & 1.655 & 1.327 & 0.913 & 0.582 & 0.353 & 0.208 \\ 0.141 & 0.316 & 0.562 & 0.905 & 1.327 & 1.663 & 1.332 & 0.916 & 0.583 & 0.354 \\ 0.082 & 0.188 & 0.343 & 0.577 & 0.913 & 1.332 & 1.666 & 1.333 & 0.916 & 0.583 \\ 0.047 & 0.108 & 0.202 & 0.351 & 0.582 & 0.916 & 1.333 & 1.666 & 1.333 & 0.917 \\ 0.026 & 0.062 & 0.116 & 0.207 & 0.353 & 0.583 & 0.916 & 1.333 & 1.667 & 1.333 \\ 0.015 & 0.034 & 0.066 & 0.119 & 0.208 & 0.354 & 0.583 & 0.917 & 1.333 & 1.667 \end{bmatrix}
			\]

			\[
			\mB (0.75) = \begin{bmatrix} 0.438 & 0.656 & 0.738 & 0.738 & 0.692 & 0.623 & 0.545 & 0.467 & 0.394 & 0.328 \\ 0.656 & 1.176 & 1.395 & 1.43 & 1.361 & 1.237 & 1.09 & 0.939 & 0.796 & 0.665 \\ 0.738 & 1.395 & 1.868 & 2.017 & 1.975 & 1.828 & 1.631 & 1.419 & 1.21 & 1.017 \\ 0.738 & 1.43 & 2.017 & 2.413 & 2.485 & 2.37 & 2.157 & 1.902 & 1.64 & 1.39 \\ 0.692 & 1.361 & 1.975 & 2.485 & 2.807 & 2.813 & 2.641 & 2.379 & 2.083 & 1.786 \\ 0.623 & 1.237 & 1.828 & 2.37 & 2.813 & 3.078 & 3.035 & 2.821 & 2.524 & 2.199 \\ 0.545 & 1.09 & 1.631 & 2.157 & 2.641 & 3.035 & 3.258 & 3.18 & 2.938 & 2.618 \\ 0.467 & 0.939 & 1.419 & 1.902 & 2.379 & 2.821 & 3.18 & 3.375 & 3.274 & 3.012 \\ 0.394 & 0.796 & 1.21 & 1.64 & 2.083 & 2.524 & 2.938 & 3.274 & 3.45 & 3.333 \\ 0.328 & 0.665 & 1.017 & 1.39 & 1.786 & 2.199 & 2.618 & 3.012 & 3.333 & 3.497 \end{bmatrix}
			\]

            \[
            \mB (0.95) = \begin{bmatrix} 0.098 & 0.185 & 0.264 & 0.334 & 0.397 & 0.453 & 0.502 & 0.545 & 0.582 & 0.614 \\ 0.185 & 0.361 & 0.52 & 0.661 & 0.787 & 0.899 & 0.997 & 1.084 & 1.159 & 1.224 \\ 0.264 & 0.52 & 0.759 & 0.972 & 1.163 & 1.332 & 1.481 & 1.612 & 1.726 & 1.825 \\ 0.334 & 0.661 & 0.972 & 1.26 & 1.517 & 1.745 & 1.946 & 2.123 & 2.277 & 2.411 \\ 0.397 & 0.787 & 1.163 & 1.517 & 1.842 & 2.131 & 2.387 & 2.612 & 2.808 & 2.978 \\ 0.453 & 0.899 & 1.332 & 1.745 & 2.131 & 2.485 & 2.797 & 3.072 & 3.312 & 3.521 \\ 0.502 & 0.997 & 1.481 & 1.946 & 2.387 & 2.797 & 3.169 & 3.498 & 3.785 & 4.035 \\ 0.545 & 1.084 & 1.612 & 2.123 & 2.612 & 3.072 & 3.498 & 3.883 & 4.22 & 4.515 \\ 0.582 & 1.159 & 1.726 & 2.277 & 2.808 & 3.312 & 3.785 & 4.22 & 4.612 & 4.954 \\ 0.614 & 1.224 & 1.825 & 2.411 & 2.978 & 3.521 & 4.035 & 4.515 & 4.954 & 5.348 \end{bmatrix}
            \]

			\subsection{Analytical Example for Dynamic Regression} 
			\label{subsec:theoretical_example_residuals}
			
				We now analyse and illustrate the structure of the asymptotic variance $\mSigma_\rho^{hom}$ and the properties of the classical naive and corrected simultaneous significance band for dynamic regressions. We consider again an AR(1) process as in \eqref{DGP_AR}. We then estimate an AR(1) model with intercept by LS,
					\begin{align} \label{eq:AR1_regression}
						y_t = \widehat{a} + \widehat{\phi} y_{t-1} + \widehat{e}_t \, , \quad t=1, \ldots, T \, ,
					\end{align}
					compute the  residuals, $\widehat e_t = y_t - \widehat a  - \widehat \vbeta^\prime  \vx_t 
						= y_t - \widehat a  - \widehat \phi  y_{t-1}$, i.\,e., $\vx_t = y_{t-1}$, and obtain $\widehat{\vrho}_{\widehat e}$.
					We now calculate the covariance matrix of the residual autocorrelations from \eqref{eq:asymp_variance_homoskedastic_errors}. It holds that $\sigma^2_\varepsilon = 1$ and that 
					$\mSigma_x = \Var(y_{t-1}) = \Var(y_t) = \frac{1}{1-\phi^2}$. Further, it holds that 
					\begin{align*}
						\vc_h := \E[\vx_t \varepsilon_{t-h}] = \E[y_{t-1} \varepsilon_{t-h}] = \E[ (\phi^h y_{t-h-1} + \sum_{k=0}^{h-1} \phi^k \varepsilon_{t-k-1}) \varepsilon_{t-h}] = \E [ \phi^{h-1} \varepsilon_{t-h}^2] = \phi^{h-1} \, .
					\end{align*}
					Thus,
					$$\mGamma := (\vc_1, \ldots, \vc_H)^\prime = (1, \phi, \phi^2, \ldots, \phi^{H-1}) \, .$$
					Collecting terms yields
					\begin{equation} \label{eq:Sigma_rho_example}
						\mSigma_\rho^{hom} = \mI_H - (1 - \phi^2) \mGamma \mGamma^\prime = \mI_H - (1 - \phi^2) \cdot (\phi^{i+j-2} )_{1 \leq i,j \leq H},
					\end{equation}
					where $( a_{ij} )_{1 \leq i,j \leq H}$ denotes a matrix with entries $a_{ij}$.
					
\begin{figure}
						\noindent \centering{}\includegraphics[scale=0.5]{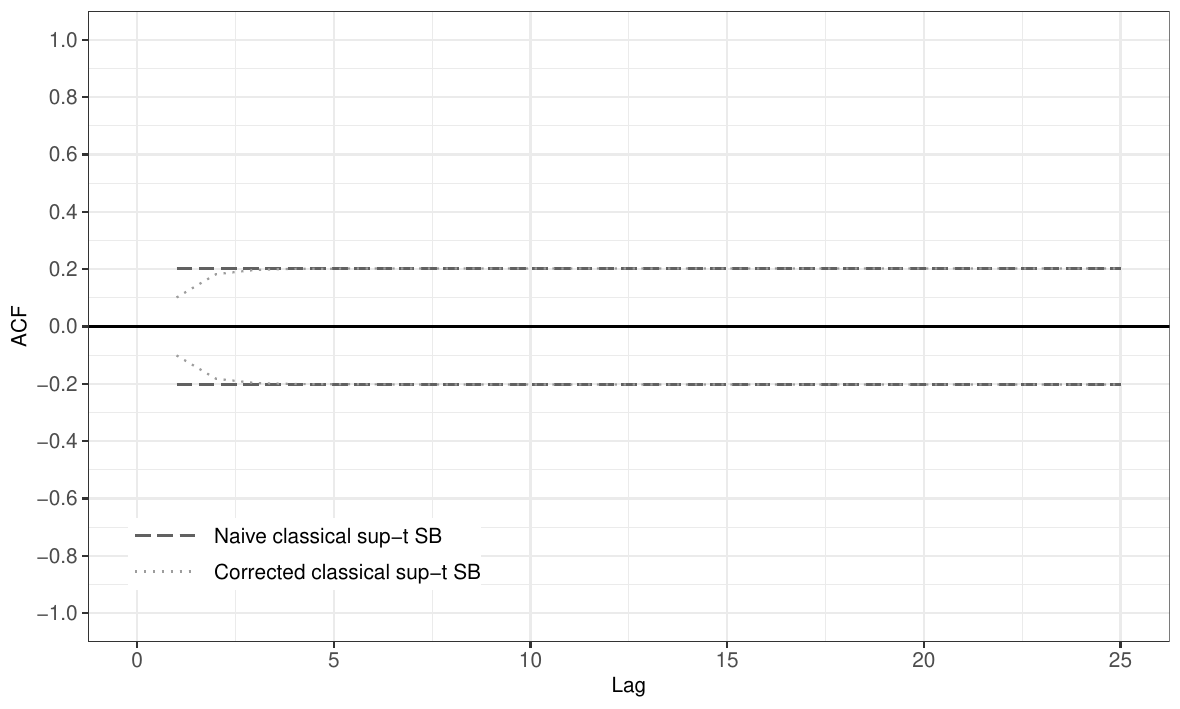}
						\vspace{-0.5cm}
						\caption{$90\%$ corrected versus naive simultaneous significance bands for autocorrelations of regression residuals from dynamic regressions with intercept on one lag with the DGP being an AR(1) process with coefficient 0.5 and sample size $T = 200$.}	
						\label{fig:ar1_residual_sig_bands}
					\end{figure}

					As discussed before Algorithm \ref{alg:shrinkage}, the covariance matrix is substantially different from the identity matrix in its upper left corner but approaches the identity matrix quickly when moving away from that corner. Consequently, the naive significance band ${SB}^{supt}_{\bm{\rho}_e} (\mI_H)$ differs substantially in its width from the corrected band ${SB}^{supt}_{\bm{\rho}_{\varepsilon}} (\mSigma_\rho^{hom})$ for small $h$ due to the asymptotic variances being much smaller than 1 (the equicoordinate quantiles $q_{1-\alpha} (\mSigma_\rho^{hom})$ and $q_{1-\alpha} (\mI_H)$ do not differ substantially), but is virtually identical for moderate and large $h$. Figure \ref{fig:ar1_residual_sig_bands} depicts the two bands for $1-\alpha=0.9$, $\phi=0.5$ and $T = 200$.
					
\begin{figure}
				\noindent \centering{}\includegraphics[scale=0.5]{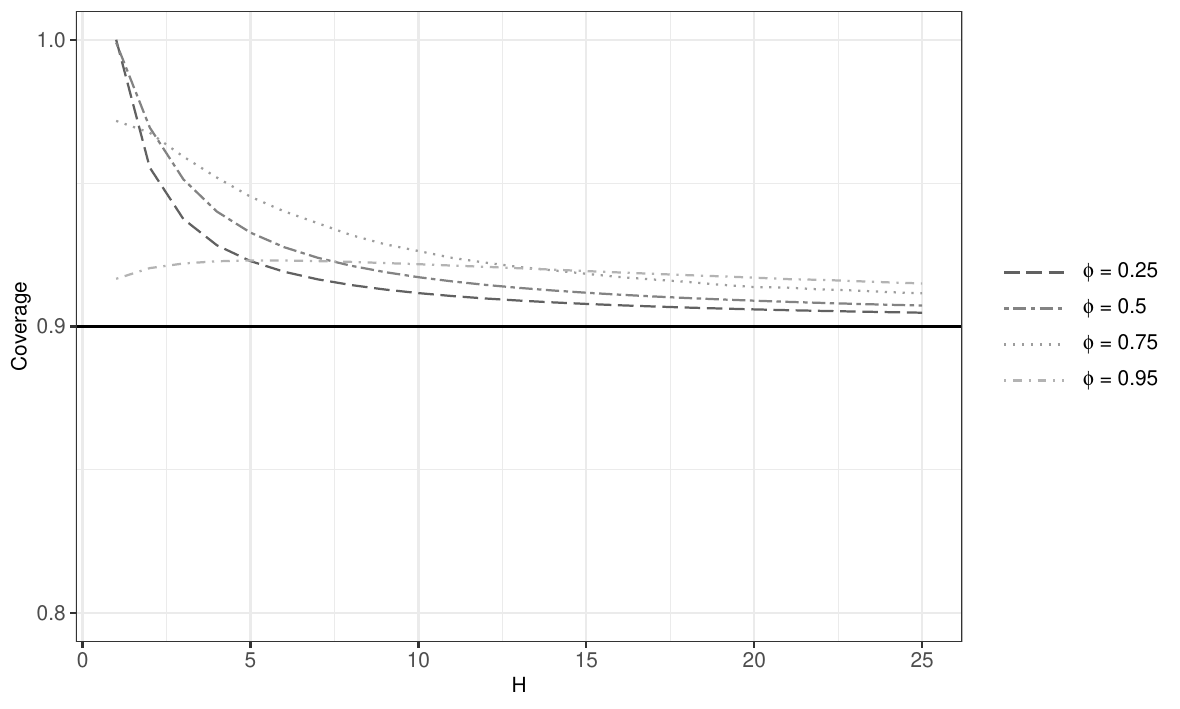}
				\vspace{-0.5cm}
				\caption{Coverage of $90\%$ naive classical sup-t significance bands for autocorrelations of regression residuals from dynamic regressions with intercept on one lag with the DGP being an AR(1) process with coefficient $\phi$.}	
				\label{fig:ar1_residual_cover_naive}
			\end{figure}

					This behaviour of the width carries over to the asymptotic coverage. Short naive bands, i.\,e., with small values of $H$, will have severe overcoverage, while it will be mild for larger values of $H$. Figure \ref{fig:ar1_residual_cover_naive} plots the asymptotic coverage for different values of $\phi$.

\section{Detailed Simulation Results for Time Series}
\label{Sect:sims_ts}

\subsection{Testing White Noise under IID Errors}

\begin{table}[H]
\centering
\caption{Relative frequency of rejections for the null hypothesis of white noise for inference bands (significance level $\alpha = 0.1 $); time series of length $T$ are generated by an AR(1) process with coefficient $\phi$ and iid normal errors, length of the bands is denoted by $H$. The bandwidth for variance estimation is $L = T^{1/2}$ for classical confidence bands and $L = (T-H)^{1/3}$ for robust confidence bands.}
\resizebox{\ifdim\width>\linewidth\linewidth\else\width\fi}{!}{
\begin{tabular}{l|ccc|ccc|cccl|ccc|ccc|cccl|ccc|ccc|cccl|ccc|ccc|cccl|ccc|ccc|cccl|ccc|ccc|cccl|ccc|ccc|cccl|ccc|ccc|cccl|ccc|ccc|cccl|ccc|ccc|ccc}
\toprule
\multicolumn{1}{c}{ } & \multicolumn{3}{c}{$T = 50$} & \multicolumn{3}{c}{$T = 200$} & \multicolumn{3}{c}{$T = 800$} \\
\cmidrule(l{3pt}r{3pt}){2-4} \cmidrule(l{3pt}r{3pt}){5-7} \cmidrule(l{3pt}r{3pt}){8-10}
 & $H = 1$ & $H = 10 $ & $H = 25$ & $H = 1$ & $H = 10 $ & $H = 25$ & $H = 1$ & $H = 10 $ & $H = 25$\\
\midrule
\addlinespace[0.3em]
\hline
\multicolumn{10}{c}{\textbf{size: $\phi = 0$}}\\
\hline
\hspace{1em}Classical sup-t SB & 0.084 & 0.047 & 0.021 & 0.097 & 0.083 & 0.066 & 0.103 & 0.096 & 0.094\\
\hspace{1em}Classical sup-t CB & 0.092 & 0.055 & 0.028 & 0.110 & 0.087 & 0.065 & 0.107 & 0.090 & 0.090\\
\hspace{1em}Classical pointw. SB & 0.084 & 0.509 & 0.678 & 0.097 & 0.605 & 0.887 & 0.103 & 0.668 & 0.935\\
\hspace{1em}Robust Bonf. SB & 0.088 & 0.015 & 0.012 & 0.104 & 0.071 & 0.035 & 0.105 & 0.087 & 0.076\\
\hspace{1em}Robust Bonf. CB & 0.145 & 0.154 & 0.068 & 0.127 & 0.114 & 0.092 & 0.107 & 0.115 & 0.108\\
\addlinespace[0.3em]
\hline
\multicolumn{10}{c}{\textbf{power: $\phi = 0.25$}}\\
\hline
\hspace{1em}Classical sup-t SB & 0.433 & 0.181 & 0.090 & 0.959 & 0.826 & 0.741 & 1.000 & 1.000 & 1.000\\
\hspace{1em}Classical sup-t CB & 0.467 & 0.237 & 0.141 & 0.959 & 0.831 & 0.743 & 1.000 & 1.000 & 1.000\\
\hspace{1em}Classical pointw. SB & 0.433 & 0.701 & 0.817 & 0.959 & 0.985 & 0.995 & 1.000 & 1.000 & 1.000\\
\hspace{1em}Robust Bonf. SB & 0.436 & 0.083 & 0.037 & 0.963 & 0.774 & 0.607 & 1.000 & 1.000 & 1.000\\
\hspace{1em}Robust Bonf. CB & 0.545 & 0.330 & 0.181 & 0.967 & 0.847 & 0.724 & 1.000 & 1.000 & 1.000\\
\addlinespace[0.3em]
\hline
\multicolumn{10}{c}{\textbf{power: $\phi = 0.5$}}\\
\hline
\hspace{1em}Classical sup-t SB & 0.930 & 0.737 & 0.624 & 1.000 & 1.000 & 1.000 & 1.000 & 1.000 & 1.000\\
\hspace{1em}Classical sup-t CB & 0.940 & 0.798 & 0.709 & 1.000 & 1.000 & 1.000 & 1.000 & 1.000 & 1.000\\
\hspace{1em}Classical pointw. SB & 0.930 & 0.965 & 0.979 & 1.000 & 1.000 & 1.000 & 1.000 & 1.000 & 1.000\\
\hspace{1em}Robust Bonf. SB & 0.931 & 0.453 & 0.195 & 1.000 & 1.000 & 1.000 & 1.000 & 1.000 & 1.000\\
\hspace{1em}Robust Bonf. CB & 0.951 & 0.819 & 0.566 & 1.000 & 1.000 & 1.000 & 1.000 & 1.000 & 1.000\\
\addlinespace[0.3em]
\hline
\multicolumn{10}{c}{\textbf{power: $\phi = 0.75$}}\\
\hline
\hspace{1em}Classical sup-t SB & 0.999 & 0.988 & 0.980 & 1.000 & 1.000 & 1.000 & 1.000 & 1.000 & 1.000\\
\hspace{1em}Classical sup-t CB & 0.999 & 0.991 & 0.989 & 1.000 & 1.000 & 1.000 & 1.000 & 1.000 & 1.000\\
\hspace{1em}Classical pointw. SB & 0.999 & 1.000 & 1.000 & 1.000 & 1.000 & 1.000 & 1.000 & 1.000 & 1.000\\
\hspace{1em}Robust Bonf. SB & 0.999 & 0.893 & 0.443 & 1.000 & 1.000 & 1.000 & 1.000 & 1.000 & 1.000\\
\hspace{1em}Robust Bonf. CB & 0.999 & 0.992 & 0.942 & 1.000 & 1.000 & 1.000 & 1.000 & 1.000 & 1.000\\
\addlinespace[0.3em]
\hline
\multicolumn{10}{c}{\textbf{power: $\phi = 0.95$}}\\
\hline
\hspace{1em}Classical sup-t SB & 1.000 & 1.000 & 1.000 & 1.000 & 1.000 & 1.000 & 1.000 & 1.000 & 1.000\\
\hspace{1em}Classical sup-t CB & 1.000 & 1.000 & 1.000 & 1.000 & 1.000 & 1.000 & 1.000 & 1.000 & 1.000\\
\hspace{1em}Classical pointw. SB & 1.000 & 1.000 & 1.000 & 1.000 & 1.000 & 1.000 & 1.000 & 1.000 & 1.000\\
\hspace{1em}Robust Bonf. SB & 1.000 & 0.986 & 0.683 & 1.000 & 1.000 & 1.000 & 1.000 & 1.000 & 1.000\\
\hspace{1em}Robust Bonf. CB & 1.000 & 1.000 & 0.996 & 1.000 & 1.000 & 1.000 & 1.000 & 1.000 & 1.000\\
\bottomrule
\end{tabular}}
\label{tab:significance_bands_coverage}
\end{table}

\begin{table}[H]
\centering
\caption{Relative frequency of rejections for the null hypothesis of white noise of several hypothesis tests (significance level $\alpha = 0.1 $); time series of length $T$ are generated by an AR(1) process with coefficient $\phi$ and iid normal errors, length of the bands is denoted by $H$. NA: missing values due to singularity of the covariance matrix.}
\resizebox{\ifdim\width>\linewidth\linewidth\else\width\fi}{!}{
\begin{tabular}{l|ccc|ccc|cccl|ccc|ccc|cccl|ccc|ccc|cccl|ccc|ccc|cccl|ccc|ccc|cccl|ccc|ccc|cccl|ccc|ccc|cccl|ccc|ccc|cccl|ccc|ccc|cccl|ccc|ccc|ccc}
\toprule
\multicolumn{1}{c}{ } & \multicolumn{3}{c}{$T = 50$} & \multicolumn{3}{c}{$T = 200$} & \multicolumn{3}{c}{$T = 800$} \\
\cmidrule(l{3pt}r{3pt}){2-4} \cmidrule(l{3pt}r{3pt}){5-7} \cmidrule(l{3pt}r{3pt}){8-10}
 & $H = 1$ & $H = 10 $ & $H = 25$ & $H = 1$ & $H = 10 $ & $H = 25$ & $H = 1$ & $H = 10 $ & $H = 25$\\
\midrule
\addlinespace[0.3em]
\hline
\multicolumn{10}{c}{\textbf{size: $\phi = 0$}}\\
\hline
\hspace{1em}Box-Pierce & 0.084 & 0.056 & 0.024 & 0.097 & 0.070 & 0.070 & 0.103 & 0.105 & 0.095\\
\hspace{1em}Ljung-Box & 0.094 & 0.112 & 0.138 & 0.102 & 0.086 & 0.116 & 0.103 & 0.108 & 0.112\\
\hspace{1em}Robust Box-Pierce & 0.088 & 0.030 & 0.978 & 0.104 & 0.043 & 0.008 & 0.105 & 0.096 & 0.062\\
\hspace{1em}Shao & 0.075 & 0.447 & NA & 0.105 & 0.173 & 0.455 & 0.091 & 0.133 & 0.203\\
\addlinespace[0.3em]
\hline
\multicolumn{10}{c}{\textbf{power: $\phi = 0.25$}}\\
\hline
\hspace{1em}Box-Pierce & 0.433 & 0.179 & 0.092 & 0.959 & 0.733 & 0.526 & 1.000 & 1.000 & 0.998\\
\hspace{1em}Ljung-Box & 0.462 & 0.263 & 0.283 & 0.960 & 0.752 & 0.595 & 1.000 & 1.000 & 0.999\\
\hspace{1em}Robust Box-Pierce & 0.436 & 0.065 & 0.974 & 0.963 & 0.596 & 0.143 & 1.000 & 1.000 & 0.997\\
\hspace{1em}Shao & 0.300 & 0.494 & NA & 0.821 & 0.562 & 0.722 & 0.998 & 0.964 & 0.951\\
\addlinespace[0.3em]
\hline
\multicolumn{10}{c}{\textbf{power: $\phi = 0.5$}}\\
\hline
\hspace{1em}Box-Pierce & 0.930 & 0.678 & 0.461 & 1.000 & 1.000 & 0.999 & 1.000 & 1.000 & 1.000\\
\hspace{1em}Ljung-Box & 0.936 & 0.739 & 0.688 & 1.000 & 1.000 & 0.999 & 1.000 & 1.000 & 1.000\\
\hspace{1em}Robust Box-Pierce & 0.931 & 0.206 & 0.986 & 1.000 & 1.000 & 0.875 & 1.000 & 1.000 & 1.000\\
\hspace{1em}Shao & 0.634 & 0.577 & NA & 0.982 & 0.886 & 0.882 & 1.000 & 0.999 & 1.000\\
\addlinespace[0.3em]
\hline
\multicolumn{10}{c}{\textbf{power: $\phi = 0.75$}}\\
\hline
\hspace{1em}Box-Pierce & 0.999 & 0.978 & 0.912 & 1.000 & 1.000 & 1.000 & 1.000 & 1.000 & 1.000\\
\hspace{1em}Ljung-Box & 0.999 & 0.983 & 0.969 & 1.000 & 1.000 & 1.000 & 1.000 & 1.000 & 1.000\\
\hspace{1em}Robust Box-Pierce & 0.999 & 0.467 & 0.985 & 1.000 & 1.000 & 1.000 & 1.000 & 1.000 & 1.000\\
\hspace{1em}Shao & 0.704 & 0.545 & NA & 0.985 & 0.895 & 0.835 & 1.000 & 1.000 & 1.000\\
\addlinespace[0.3em]
\hline
\multicolumn{10}{c}{\textbf{power: $\phi = 0.95$}}\\
\hline
\hspace{1em}Box-Pierce & 1.000 & 1.000 & 0.997 & 1.000 & 1.000 & 1.000 & 1.000 & 1.000 & 1.000\\
\hspace{1em}Ljung-Box & 1.000 & 1.000 & 1.000 & 1.000 & 1.000 & 1.000 & 1.000 & 1.000 & 1.000\\
\hspace{1em}Robust Box-Pierce & 1.000 & 0.751 & 0.990 & 1.000 & 1.000 & 0.999 & 1.000 & 1.000 & 1.000\\
\hspace{1em}Shao & 0.534 & 0.286 & NA & 0.835 & 0.414 & 0.129 & 0.990 & 0.996 & 0.854\\
\bottomrule
\end{tabular}}
\label{tab:tests_iid}
\end{table}

\subsection{Testing White Noise under Normal GARCH Errors}

\begin{table}[H]
\centering
\caption{Relative frequency of rejections for the null hypothesis of white noise for inference bands (significance level $\alpha = 0.1 $); time series of length $T$ are generated by an AR(1) process with coefficient $\phi$ and normal GARCH errors, length of the bands is denoted by $H$. The bandwidth for variance estimation is $L = T^{1/2}$ for classical confidence bands and $L = (T-H)^{1/3}$ for robust confidence bands.}
\resizebox{\ifdim\width>\linewidth\linewidth\else\width\fi}{!}{
\begin{tabular}{l|ccc|ccc|cccl|ccc|ccc|cccl|ccc|ccc|cccl|ccc|ccc|cccl|ccc|ccc|cccl|ccc|ccc|cccl|ccc|ccc|cccl|ccc|ccc|cccl|ccc|ccc|cccl|ccc|ccc|ccc}
\toprule
\multicolumn{1}{c}{ } & \multicolumn{3}{c}{$T = 50$} & \multicolumn{3}{c}{$T = 200$} & \multicolumn{3}{c}{$T = 800$} \\
\cmidrule(l{3pt}r{3pt}){2-4} \cmidrule(l{3pt}r{3pt}){5-7} \cmidrule(l{3pt}r{3pt}){8-10}
 & $H = 1$ & $H = 10 $ & $H = 25$ & $H = 1$ & $H = 10 $ & $H = 25$ & $H = 1$ & $H = 10 $ & $H = 25$\\
\midrule
\addlinespace[0.3em]
\hline
\multicolumn{10}{c}{\textbf{size: $\phi = 0$}}\\
\hline
\hspace{1em}Classical sup-t SB & 0.114 & 0.069 & 0.032 & 0.144 & 0.171 & 0.140 & 0.173 & 0.187 & 0.183\\
\hspace{1em}Classical sup-t CB & 0.120 & 0.082 & 0.040 & 0.143 & 0.173 & 0.134 & 0.169 & 0.183 & 0.178\\
\hspace{1em}Classical pointw. SB & 0.114 & 0.559 & 0.696 & 0.144 & 0.707 & 0.929 & 0.173 & 0.775 & 0.969\\
\hspace{1em}Robust Bonf. SB & 0.097 & 0.069 & 0.123 & 0.090 & 0.118 & 0.186 & 0.104 & 0.111 & 0.152\\
\hspace{1em}Robust Bonf. CB & 0.156 & 0.228 & 0.230 & 0.118 & 0.189 & 0.287 & 0.117 & 0.151 & 0.205\\
\addlinespace[0.3em]
\hline
\multicolumn{10}{c}{\textbf{power: $\phi = 0.25$}}\\
\hline
\hspace{1em}Classical sup-t SB & 0.447 & 0.203 & 0.107 & 0.937 & 0.803 & 0.734 & 1.000 & 1.000 & 1.000\\
\hspace{1em}Classical sup-t CB & 0.465 & 0.264 & 0.168 & 0.936 & 0.812 & 0.744 & 1.000 & 1.000 & 1.000\\
\hspace{1em}Classical pointw. SB & 0.447 & 0.727 & 0.820 & 0.937 & 0.981 & 0.998 & 1.000 & 1.000 & 1.000\\
\hspace{1em}Robust Bonf. SB & 0.399 & 0.183 & 0.213 & 0.906 & 0.730 & 0.672 & 0.998 & 0.997 & 0.999\\
\hspace{1em}Robust Bonf. CB & 0.522 & 0.424 & 0.376 & 0.918 & 0.813 & 0.781 & 0.998 & 0.996 & 1.000\\
\addlinespace[0.3em]
\hline
\multicolumn{10}{c}{\textbf{power: $\phi = 0.5$}}\\
\hline
\hspace{1em}Classical sup-t SB & 0.924 & 0.733 & 0.631 & 1.000 & 0.999 & 0.999 & 1.000 & 1.000 & 1.000\\
\hspace{1em}Classical sup-t CB & 0.927 & 0.785 & 0.701 & 1.000 & 0.999 & 0.998 & 1.000 & 1.000 & 1.000\\
\hspace{1em}Classical pointw. SB & 0.924 & 0.966 & 0.982 & 1.000 & 1.000 & 1.000 & 1.000 & 1.000 & 1.000\\
\hspace{1em}Robust Bonf. SB & 0.909 & 0.529 & 0.385 & 0.998 & 0.998 & 1.000 & 1.000 & 1.000 & 1.000\\
\hspace{1em}Robust Bonf. CB & 0.942 & 0.845 & 0.694 & 0.998 & 0.998 & 1.000 & 1.000 & 1.000 & 1.000\\
\addlinespace[0.3em]
\hline
\multicolumn{10}{c}{\textbf{power: $\phi = 0.75$}}\\
\hline
\hspace{1em}Classical sup-t SB & 0.999 & 0.988 & 0.976 & 1.000 & 1.000 & 1.000 & 1.000 & 1.000 & 1.000\\
\hspace{1em}Classical sup-t CB & 0.999 & 0.992 & 0.987 & 1.000 & 1.000 & 1.000 & 1.000 & 1.000 & 1.000\\
\hspace{1em}Classical pointw. SB & 0.999 & 1.000 & 1.000 & 1.000 & 1.000 & 1.000 & 1.000 & 1.000 & 1.000\\
\hspace{1em}Robust Bonf. SB & 0.996 & 0.917 & 0.618 & 1.000 & 1.000 & 1.000 & 1.000 & 1.000 & 1.000\\
\hspace{1em}Robust Bonf. CB & 0.998 & 0.989 & 0.961 & 1.000 & 1.000 & 1.000 & 1.000 & 1.000 & 1.000\\
\addlinespace[0.3em]
\hline
\multicolumn{10}{c}{\textbf{power: $\phi = 0.95$}}\\
\hline
\hspace{1em}Classical sup-t SB & 1.000 & 1.000 & 1.000 & 1.000 & 1.000 & 1.000 & 1.000 & 1.000 & 1.000\\
\hspace{1em}Classical sup-t CB & 1.000 & 1.000 & 1.000 & 1.000 & 1.000 & 1.000 & 1.000 & 1.000 & 1.000\\
\hspace{1em}Classical pointw. SB & 1.000 & 1.000 & 1.000 & 1.000 & 1.000 & 1.000 & 1.000 & 1.000 & 1.000\\
\hspace{1em}Robust Bonf. SB & 1.000 & 0.989 & 0.799 & 1.000 & 1.000 & 1.000 & 1.000 & 1.000 & 1.000\\
\hspace{1em}Robust Bonf. CB & 1.000 & 1.000 & 1.000 & 1.000 & 1.000 & 1.000 & 1.000 & 1.000 & 1.000\\
\bottomrule
\end{tabular}}
\label{tab:significance_bands_coverage_GARCH_normal}
\end{table}

\begin{table}[H]
\centering
\caption{Relative frequency of rejections for the null hypothesis of white noise of several hypothesis tests (significance level $\alpha = 0.1 $); time series of length $T$ are generated by an AR(1) process with coefficient $\phi$ and normal GARCH errors, length of the bands is denoted by $H$. NA: missing values due to singularity of the covariance matrix.}
\resizebox{\ifdim\width>\linewidth\linewidth\else\width\fi}{!}{
\begin{tabular}{l|ccc|ccc|cccl|ccc|ccc|cccl|ccc|ccc|cccl|ccc|ccc|cccl|ccc|ccc|cccl|ccc|ccc|cccl|ccc|ccc|cccl|ccc|ccc|cccl|ccc|ccc|cccl|ccc|ccc|ccc}
\toprule
\multicolumn{1}{c}{ } & \multicolumn{3}{c}{$T = 50$} & \multicolumn{3}{c}{$T = 200$} & \multicolumn{3}{c}{$T = 800$} \\
\cmidrule(l{3pt}r{3pt}){2-4} \cmidrule(l{3pt}r{3pt}){5-7} \cmidrule(l{3pt}r{3pt}){8-10}
 & $H = 1$ & $H = 10 $ & $H = 25$ & $H = 1$ & $H = 10 $ & $H = 25$ & $H = 1$ & $H = 10 $ & $H = 25$\\
\midrule
\addlinespace[0.3em]
\hline
\multicolumn{10}{c}{\textbf{size: $\phi = 0$}}\\
\hline
\hspace{1em}Box-Pierce & 0.114 & 0.079 & 0.023 & 0.144 & 0.182 & 0.165 & 0.173 & 0.260 & 0.255\\
\hspace{1em}Ljung-Box & 0.125 & 0.131 & 0.142 & 0.146 & 0.204 & 0.228 & 0.174 & 0.265 & 0.272\\
\hspace{1em}Robust Box-Pierce & 0.097 & 0.160 & 0.981 & 0.090 & 0.110 & 0.240 & 0.104 & 0.131 & 0.175\\
\hspace{1em}Shao & 0.066 & 0.468 & NA & 0.067 & 0.111 & 0.433 & 0.089 & 0.094 & 0.119\\
\addlinespace[0.3em]
\hline
\multicolumn{10}{c}{\textbf{power: $\phi = 0.25$}}\\
\hline
\hspace{1em}Box-Pierce & 0.447 & 0.216 & 0.101 & 0.937 & 0.776 & 0.606 & 1.000 & 1.000 & 0.998\\
\hspace{1em}Ljung-Box & 0.467 & 0.286 & 0.272 & 0.938 & 0.793 & 0.672 & 1.000 & 1.000 & 0.998\\
\hspace{1em}Robust Box-Pierce & 0.399 & 0.218 & 0.985 & 0.906 & 0.604 & 0.448 & 0.998 & 0.996 & 0.996\\
\hspace{1em}Shao & 0.297 & 0.531 & NA & 0.637 & 0.386 & 0.634 & 0.948 & 0.894 & 0.830\\
\addlinespace[0.3em]
\hline
\multicolumn{10}{c}{\textbf{power: $\phi = 0.5$}}\\
\hline
\hspace{1em}Box-Pierce & 0.924 & 0.689 & 0.462 & 1.000 & 1.000 & 0.999 & 1.000 & 1.000 & 1.000\\
\hspace{1em}Ljung-Box & 0.932 & 0.752 & 0.688 & 1.000 & 1.000 & 1.000 & 1.000 & 1.000 & 1.000\\
\hspace{1em}Robust Box-Pierce & 0.909 & 0.373 & 0.994 & 0.998 & 0.997 & 0.924 & 1.000 & 1.000 & 1.000\\
\hspace{1em}Shao & 0.596 & 0.608 & NA & 0.840 & 0.701 & 0.796 & 0.953 & 0.980 & 0.988\\
\addlinespace[0.3em]
\hline
\multicolumn{10}{c}{\textbf{power: $\phi = 0.75$}}\\
\hline
\hspace{1em}Box-Pierce & 0.999 & 0.977 & 0.917 & 1.000 & 1.000 & 1.000 & 1.000 & 1.000 & 1.000\\
\hspace{1em}Ljung-Box & 0.999 & 0.987 & 0.968 & 1.000 & 1.000 & 1.000 & 1.000 & 1.000 & 1.000\\
\hspace{1em}Robust Box-Pierce & 0.996 & 0.603 & 0.997 & 1.000 & 1.000 & 0.997 & 1.000 & 1.000 & 1.000\\
\hspace{1em}Shao & 0.717 & 0.578 & NA & 0.843 & 0.686 & 0.655 & 0.950 & 0.986 & 0.989\\
\addlinespace[0.3em]
\hline
\multicolumn{10}{c}{\textbf{power: $\phi = 0.95$}}\\
\hline
\hspace{1em}Box-Pierce & 1.000 & 1.000 & 0.998 & 1.000 & 1.000 & 1.000 & 1.000 & 1.000 & 1.000\\
\hspace{1em}Ljung-Box & 1.000 & 1.000 & 1.000 & 1.000 & 1.000 & 1.000 & 1.000 & 1.000 & 1.000\\
\hspace{1em}Robust Box-Pierce & 1.000 & 0.832 & NA & 1.000 & 1.000 & 0.999 & 1.000 & 1.000 & 1.000\\
\hspace{1em}Shao & 0.584 & 0.333 & NA & 0.782 & 0.277 & 0.085 & 0.936 & 0.911 & 0.508\\
\bottomrule
\end{tabular}}
\label{tab:tests_GARCH_normal}
\end{table}

\subsection{Testing White Noise under Chi-square GARCH Errors}

\begin{table}[H]
\centering
\caption{Relative frequency of rejections for the null hypothesis of white noise for inference bands (significance level $\alpha = 0.1 $); time series of length $T$ are generated by an AR(1) process with coefficient $\phi$ and $\chi^2(3)$ GARCH errors, length of the bands is denoted by $H$. The bandwidth for variance estimation is $L = T^{1/2}$ for classical confidence bands and $L = (T-H)^{1/3}$ for robust confidence bands.}
\resizebox{\ifdim\width>\linewidth\linewidth\else\width\fi}{!}{
\begin{tabular}{l|ccc|ccc|cccl|ccc|ccc|cccl|ccc|ccc|cccl|ccc|ccc|cccl|ccc|ccc|cccl|ccc|ccc|cccl|ccc|ccc|cccl|ccc|ccc|cccl|ccc|ccc|cccl|ccc|ccc|ccc}
\toprule
\multicolumn{1}{c}{ } & \multicolumn{3}{c}{$T = 50$} & \multicolumn{3}{c}{$T = 200$} & \multicolumn{3}{c}{$T = 800$} \\
\cmidrule(l{3pt}r{3pt}){2-4} \cmidrule(l{3pt}r{3pt}){5-7} \cmidrule(l{3pt}r{3pt}){8-10}
 & $H = 1$ & $H = 10 $ & $H = 25$ & $H = 1$ & $H = 10 $ & $H = 25$ & $H = 1$ & $H = 10 $ & $H = 25$\\
\midrule
\addlinespace[0.3em]
\hline
\multicolumn{10}{c}{\textbf{size: $\phi = 0$}}\\
\hline
\hspace{1em}Classical sup-t SB & 0.108 & 0.044 & 0.016 & 0.181 & 0.211 & 0.169 & 0.179 & 0.261 & 0.224\\
\hspace{1em}Classical sup-t CB & 0.129 & 0.051 & 0.021 & 0.176 & 0.212 & 0.156 & 0.179 & 0.252 & 0.218\\
\hspace{1em}Classical pointw. SB & 0.108 & 0.528 & 0.665 & 0.181 & 0.758 & 0.929 & 0.179 & 0.793 & 0.967\\
\hspace{1em}Robust Bonf. SB & 0.102 & 0.076 & 0.167 & 0.106 & 0.129 & 0.184 & 0.106 & 0.125 & 0.129\\
\hspace{1em}Robust Bonf. CB & 0.203 & 0.261 & 0.263 & 0.163 & 0.245 & 0.295 & 0.120 & 0.171 & 0.204\\
\addlinespace[0.3em]
\hline
\multicolumn{10}{c}{\textbf{power: $\phi = 0.25$}}\\
\hline
\hspace{1em}Classical sup-t SB & 0.428 & 0.170 & 0.093 & 0.937 & 0.805 & 0.731 & 1.000 & 1.000 & 1.000\\
\hspace{1em}Classical sup-t CB & 0.447 & 0.223 & 0.148 & 0.934 & 0.820 & 0.727 & 0.999 & 1.000 & 1.000\\
\hspace{1em}Classical pointw. SB & 0.428 & 0.706 & 0.821 & 0.937 & 0.983 & 0.996 & 1.000 & 1.000 & 1.000\\
\hspace{1em}Robust Bonf. SB & 0.410 & 0.164 & 0.220 & 0.911 & 0.677 & 0.557 & 0.997 & 0.994 & 0.999\\
\hspace{1em}Robust Bonf. CB & 0.531 & 0.423 & 0.338 & 0.931 & 0.814 & 0.750 & 0.998 & 1.000 & 1.000\\
\addlinespace[0.3em]
\hline
\multicolumn{10}{c}{\textbf{power: $\phi = 0.5$}}\\
\hline
\hspace{1em}Classical sup-t SB & 0.937 & 0.744 & 0.634 & 1.000 & 0.999 & 0.999 & 1.000 & 1.000 & 1.000\\
\hspace{1em}Classical sup-t CB & 0.944 & 0.798 & 0.703 & 1.000 & 1.000 & 0.999 & 1.000 & 1.000 & 1.000\\
\hspace{1em}Classical pointw. SB & 0.937 & 0.973 & 0.978 & 1.000 & 1.000 & 1.000 & 1.000 & 1.000 & 1.000\\
\hspace{1em}Robust Bonf. SB & 0.911 & 0.499 & 0.400 & 0.998 & 0.989 & 0.989 & 1.000 & 1.000 & 1.000\\
\hspace{1em}Robust Bonf. CB & 0.954 & 0.831 & 0.645 & 0.999 & 1.000 & 0.999 & 1.000 & 1.000 & 1.000\\
\addlinespace[0.3em]
\hline
\multicolumn{10}{c}{\textbf{power: $\phi = 0.75$}}\\
\hline
\hspace{1em}Classical sup-t SB & 0.998 & 0.993 & 0.984 & 1.000 & 1.000 & 1.000 & 1.000 & 1.000 & 1.000\\
\hspace{1em}Classical sup-t CB & 0.997 & 0.995 & 0.990 & 1.000 & 1.000 & 1.000 & 1.000 & 1.000 & 1.000\\
\hspace{1em}Classical pointw. SB & 0.998 & 0.999 & 1.000 & 1.000 & 1.000 & 1.000 & 1.000 & 1.000 & 1.000\\
\hspace{1em}Robust Bonf. SB & 0.997 & 0.865 & 0.588 & 1.000 & 1.000 & 1.000 & 1.000 & 1.000 & 1.000\\
\hspace{1em}Robust Bonf. CB & 0.999 & 0.991 & 0.932 & 1.000 & 1.000 & 1.000 & 1.000 & 1.000 & 1.000\\
\addlinespace[0.3em]
\hline
\multicolumn{10}{c}{\textbf{power: $\phi = 0.95$}}\\
\hline
\hspace{1em}Classical sup-t SB & 1.000 & 1.000 & 1.000 & 1.000 & 1.000 & 1.000 & 1.000 & 1.000 & 1.000\\
\hspace{1em}Classical sup-t CB & 1.000 & 1.000 & 1.000 & 1.000 & 1.000 & 1.000 & 1.000 & 1.000 & 1.000\\
\hspace{1em}Classical pointw. SB & 1.000 & 1.000 & 1.000 & 1.000 & 1.000 & 1.000 & 1.000 & 1.000 & 1.000\\
\hspace{1em}Robust Bonf. SB & 1.000 & 0.989 & 0.749 & 1.000 & 1.000 & 1.000 & 1.000 & 1.000 & 1.000\\
\hspace{1em}Robust Bonf. CB & 1.000 & 1.000 & 0.997 & 1.000 & 1.000 & 1.000 & 1.000 & 1.000 & 1.000\\
\bottomrule
\end{tabular}}
\label{tab:significance_bands_coverage_GARCH_chisqu}
\end{table}

\begin{table}[H]
\centering
\caption{Relative frequency of rejections for the null hypothesis of white noise of several hypothesis tests (significance level $\alpha = 0.1 $); time series of length $T$ are generated by an AR(1) process with coefficient $\phi$ and $\chi^2(3)$ GARCH errors, length of the bands is denoted by $H$. NA: missing values due to singularity of the covariance matrix.}
\resizebox{\ifdim\width>\linewidth\linewidth\else\width\fi}{!}{
\begin{tabular}{l|ccc|ccc|cccl|ccc|ccc|cccl|ccc|ccc|cccl|ccc|ccc|cccl|ccc|ccc|cccl|ccc|ccc|cccl|ccc|ccc|cccl|ccc|ccc|cccl|ccc|ccc|cccl|ccc|ccc|ccc}
\toprule
\multicolumn{1}{c}{ } & \multicolumn{3}{c}{$T = 50$} & \multicolumn{3}{c}{$T = 200$} & \multicolumn{3}{c}{$T = 800$} \\
\cmidrule(l{3pt}r{3pt}){2-4} \cmidrule(l{3pt}r{3pt}){5-7} \cmidrule(l{3pt}r{3pt}){8-10}
 & $H = 1$ & $H = 10 $ & $H = 25$ & $H = 1$ & $H = 10 $ & $H = 25$ & $H = 1$ & $H = 10 $ & $H = 25$\\
\midrule
\addlinespace[0.3em]
\hline
\multicolumn{10}{c}{\textbf{size: $\phi = 0$}}\\
\hline
\hspace{1em}Box-Pierce & 0.108 & 0.076 & 0.022 & 0.181 & 0.244 & 0.186 & 0.179 & 0.299 & 0.307\\
\hspace{1em}Ljung-Box & 0.116 & 0.121 & 0.104 & 0.182 & 0.269 & 0.232 & 0.180 & 0.306 & 0.316\\
\hspace{1em}Robust Box-Pierce & 0.102 & 0.156 & 0.972 & 0.106 & 0.132 & 0.245 & 0.106 & 0.130 & 0.150\\
\hspace{1em}Shao & 0.097 & 0.382 & NA & 0.064 & 0.075 & 0.318 & 0.092 & 0.068 & 0.045\\
\addlinespace[0.3em]
\hline
\multicolumn{10}{c}{\textbf{power: $\phi = 0.25$}}\\
\hline
\hspace{1em}Box-Pierce & 0.428 & 0.188 & 0.077 & 0.937 & 0.761 & 0.602 & 1.000 & 1.000 & 0.999\\
\hspace{1em}Ljung-Box & 0.450 & 0.268 & 0.252 & 0.937 & 0.779 & 0.671 & 1.000 & 1.000 & 0.999\\
\hspace{1em}Robust Box-Pierce & 0.410 & 0.209 & 0.979 & 0.911 & 0.590 & 0.405 & 0.997 & 0.995 & 0.983\\
\hspace{1em}Shao & 0.290 & 0.435 & NA & 0.646 & 0.311 & 0.493 & 0.937 & 0.804 & 0.672\\
\addlinespace[0.3em]
\hline
\multicolumn{10}{c}{\textbf{power: $\phi = 0.5$}}\\
\hline
\hspace{1em}Box-Pierce & 0.937 & 0.685 & 0.448 & 1.000 & 0.999 & 0.997 & 1.000 & 1.000 & 1.000\\
\hspace{1em}Ljung-Box & 0.943 & 0.760 & 0.684 & 1.000 & 0.999 & 0.997 & 1.000 & 1.000 & 1.000\\
\hspace{1em}Robust Box-Pierce & 0.911 & 0.370 & 0.987 & 0.998 & 0.983 & 0.854 & 1.000 & 1.000 & 1.000\\
\hspace{1em}Shao & 0.599 & 0.524 & NA & 0.818 & 0.571 & 0.670 & 0.943 & 0.920 & 0.866\\
\addlinespace[0.3em]
\hline
\multicolumn{10}{c}{\textbf{power: $\phi = 0.75$}}\\
\hline
\hspace{1em}Box-Pierce & 0.998 & 0.978 & 0.915 & 1.000 & 1.000 & 1.000 & 1.000 & 1.000 & 1.000\\
\hspace{1em}Ljung-Box & 0.998 & 0.986 & 0.958 & 1.000 & 1.000 & 1.000 & 1.000 & 1.000 & 1.000\\
\hspace{1em}Robust Box-Pierce & 0.997 & 0.563 & 0.990 & 1.000 & 1.000 & 0.976 & 1.000 & 1.000 & 1.000\\
\hspace{1em}Shao & 0.729 & 0.517 & NA & 0.831 & 0.551 & 0.509 & 0.944 & 0.918 & 0.845\\
\addlinespace[0.3em]
\hline
\multicolumn{10}{c}{\textbf{power: $\phi = 0.95$}}\\
\hline
\hspace{1em}Box-Pierce & 1.000 & 0.997 & 0.993 & 1.000 & 1.000 & 1.000 & 1.000 & 1.000 & 1.000\\
\hspace{1em}Ljung-Box & 1.000 & 0.998 & 0.998 & 1.000 & 1.000 & 1.000 & 1.000 & 1.000 & 1.000\\
\hspace{1em}Robust Box-Pierce & 1.000 & 0.818 & 0.996 & 1.000 & 1.000 & 0.998 & 1.000 & 1.000 & 1.000\\
\hspace{1em}Shao & 0.571 & 0.385 & NA & 0.776 & 0.246 & 0.108 & 0.901 & 0.730 & 0.331\\
\bottomrule
\end{tabular}}
\label{tab:tests_GARCH_chisqu}
\end{table}

\subsection{Confidence Bands under IID Errors}

\begin{table}[H]
\centering
				\caption{Coverage of confidence bands (nominal coverage: $1 - \alpha = 0.9 $); time series of length $T$ are generated by an AR(1) process with coefficient $\phi$ and iid normal errors, length of the bands is denoted by $H$; the bandwidth for variance estimation is $L = T^{1/2}$ for classical bands and  $L = (T-H)^{1/3}$ for robust bands. NA: For 0.1\% of Monte Carlo runs, the equicoordinate quantile could not be computed due to non-positive semi-definiteness of the estimated covariance matrix.}
\resizebox{\ifdim\width>\linewidth\linewidth\else\width\fi}{!}{
\begin{tabular}{l|ccc|ccc|cccl|ccc|ccc|cccl|ccc|ccc|cccl|ccc|ccc|cccl|ccc|ccc|cccl|ccc|ccc|cccl|ccc|ccc|cccl|ccc|ccc|cccl|ccc|ccc|cccl|ccc|ccc|ccc}
\toprule
\multicolumn{1}{c}{ } & \multicolumn{3}{c}{$T = 50$} & \multicolumn{3}{c}{$T = 200$} & \multicolumn{3}{c}{$T = 800$} \\
\cmidrule(l{3pt}r{3pt}){2-4} \cmidrule(l{3pt}r{3pt}){5-7} \cmidrule(l{3pt}r{3pt}){8-10}
 & $H = 1$ & $H = 10 $ & $H = 25$ & $H = 1$ & $H = 10 $ & $H = 25$ & $H = 1$ & $H = 10 $ & $H = 25$\\
\midrule
\addlinespace[0.3em]
\hline
\multicolumn{10}{c}{\textbf{$\phi = 0$}}\\
\hline
\hspace{1em}Classical sup-t CB & 0.908 & 0.945 & 0.972 & 0.890 & 0.913 & 0.935 & 0.893 & 0.910 & 0.910\\
\hspace{1em}Classical Bonf. CB & 0.908 & 0.948 & 0.974 & 0.890 & 0.925 & 0.941 & 0.893 & 0.910 & 0.917\\
\hspace{1em}Classical pointw. CB & 0.908 & 0.483 & 0.325 & 0.890 & 0.388 & 0.117 & 0.893 & 0.334 & 0.071\\
\hspace{1em}Robust sup-t CB & 0.855 & 0.831 & 0.913 & 0.873 & 0.875 & 0.895 & 0.893 & 0.874 & 0.886\\
\hspace{1em}Robust Bonf. CB & 0.855 & 0.846 & 0.932 & 0.873 & 0.886 & 0.908 & 0.893 & 0.885 & 0.892\\
\hspace{1em}Robust pointw. CB & 0.855 & 0.358 & 0.410 & 0.873 & 0.322 & 0.116 & 0.893 & 0.304 & 0.067\\
\addlinespace[0.3em]
\hline
\multicolumn{10}{c}{\textbf{$\phi = 0.25$}}\\
\hline
\hspace{1em}Classical sup-t CB & 0.886 & 0.932 & 0.966 & 0.890 & 0.904 & 0.940 & 0.913 & 0.904 & 0.908\\
\hspace{1em}Classical Bonf. CB & 0.886 & 0.939 & 0.971 & 0.890 & 0.919 & 0.944 & 0.913 & 0.917 & 0.915\\
\hspace{1em}Classical pointw. CB & 0.886 & 0.484 & 0.310 & 0.890 & 0.430 & 0.145 & 0.913 & 0.398 & 0.112\\
\hspace{1em}Robust sup-t CB & 0.824 & 0.798 & 0.887 & 0.865 & 0.861 & 0.871 & 0.901 & 0.871 & 0.879\\
\hspace{1em}Robust Bonf. CB & 0.824 & 0.829 & 0.911 & 0.865 & 0.870 & 0.883 & 0.901 & 0.882 & 0.892\\
\hspace{1em}Robust pointw. CB & 0.824 & 0.359 & 0.348 & 0.865 & 0.372 & 0.129 & 0.901 & 0.378 & 0.093\\
\addlinespace[0.3em]
\hline
\multicolumn{10}{c}{\textbf{$\phi = 0.5$}}\\
\hline
\hspace{1em}Classical sup-t CB & 0.868 & 0.881 & 0.946 & 0.909 & 0.887 & 0.916 & 0.910 & 0.893 & 0.885\\
\hspace{1em}Classical Bonf. CB & 0.868 & 0.902 & 0.953 & 0.909 & 0.911 & 0.943 & 0.910 & 0.919 & 0.923\\
\hspace{1em}Classical pointw. CB & 0.868 & 0.457 & 0.308 & 0.909 & 0.522 & 0.235 & 0.910 & 0.530 & 0.235\\
\hspace{1em}Robust sup-t CB & 0.792 & 0.698 & NA & 0.875 & 0.830 & 0.840 & 0.901 & 0.864 & 0.852\\
\hspace{1em}Robust Bonf. CB & 0.792 & 0.738 & 0.842 & 0.875 & 0.861 & 0.881 & 0.901 & 0.891 & 0.892\\
\hspace{1em}Robust pointw. CB & 0.792 & 0.322 & 0.302 & 0.875 & 0.464 & 0.186 & 0.901 & 0.496 & 0.207\\
\addlinespace[0.3em]
\hline
\multicolumn{10}{c}{\textbf{$\phi = 0.75$}}\\
\hline
\hspace{1em}Classical sup-t CB & 0.843 & 0.770 & 0.849 & 0.910 & 0.868 & 0.856 & 0.900 & 0.895 & 0.882\\
\hspace{1em}Classical Bonf. CB & 0.843 & 0.818 & 0.877 & 0.910 & 0.916 & 0.929 & 0.900 & 0.948 & 0.945\\
\hspace{1em}Classical pointw. CB & 0.843 & 0.501 & 0.327 & 0.910 & 0.625 & 0.375 & 0.900 & 0.638 & 0.415\\
\hspace{1em}Robust sup-t CB & 0.755 & 0.582 & 0.613 & 0.866 & 0.776 & 0.725 & 0.878 & 0.868 & 0.849\\
\hspace{1em}Robust Bonf. CB & 0.755 & 0.648 & 0.675 & 0.866 & 0.840 & 0.819 & 0.878 & 0.930 & 0.912\\
\hspace{1em}Robust pointw. CB & 0.755 & 0.351 & 0.202 & 0.866 & 0.526 & 0.282 & 0.878 & 0.586 & 0.352\\
\addlinespace[0.3em]
\hline
\multicolumn{10}{c}{\textbf{$\phi = 0.95$}}\\
\hline
\hspace{1em}Classical sup-t CB & 0.745 & 0.449 & 0.275 & 0.876 & 0.795 & 0.684 & 0.935 & 0.908 & 0.874\\
\hspace{1em}Classical Bonf. CB & 0.745 & 0.532 & 0.437 & 0.876 & 0.868 & 0.795 & 0.935 & 0.961 & 0.943\\
\hspace{1em}Classical pointw. CB & 0.745 & 0.256 & 0.003 & 0.876 & 0.698 & 0.475 & 0.935 & 0.840 & 0.735\\
\hspace{1em}Robust sup-t CB & 0.523 & 0.192 & 0.008 & 0.774 & 0.674 & 0.519 & 0.850 & 0.843 & 0.781\\
\hspace{1em}Robust Bonf. CB & 0.523 & 0.281 & 0.025 & 0.774 & 0.770 & 0.642 & 0.850 & 0.920 & 0.890\\
\hspace{1em}Robust pointw. CB & 0.523 & 0.095 & 0.000 & 0.774 & 0.530 & 0.328 & 0.850 & 0.725 & 0.577\\
\bottomrule
\end{tabular}}
\label{tab:confidence_bands_coverage}
\end{table}

\begin{table}[H]
\centering
\caption{Coverage of classical sup-t confidence bands (nominal coverage: $1 - \alpha = 0.9 $) for different bandwidth choices for variance estimation; time series of length $T$ are generated by an AR(1) process with coefficient $\phi$ with iid normal errors, length of the bands is denoted by $H$.}
\resizebox{\ifdim\width>\linewidth\linewidth\else\width\fi}{!}{
\begin{tabular}{l|ccc|ccc|cccl|ccc|ccc|cccl|ccc|ccc|cccl|ccc|ccc|cccl|ccc|ccc|cccl|ccc|ccc|cccl|ccc|ccc|cccl|ccc|ccc|cccl|ccc|ccc|cccl|ccc|ccc|ccc}
\toprule
\multicolumn{1}{c}{ } & \multicolumn{3}{c}{$T = 50$} & \multicolumn{3}{c}{$T = 200$} & \multicolumn{3}{c}{$T = 800$} \\
\cmidrule(l{3pt}r{3pt}){2-4} \cmidrule(l{3pt}r{3pt}){5-7} \cmidrule(l{3pt}r{3pt}){8-10}
 & $H = 1$ & $H = 10 $ & $H = 25$ & $H = 1$ & $H = 10 $ & $H = 25$ & $H = 1$ & $H = 10 $ & $H = 25$\\
\midrule
\addlinespace[0.3em]
\hline
\multicolumn{10}{c}{\textbf{$\phi = 0$}}\\
\hline
\hspace{1em}$L = 3T^{1/2}$ & 0.918 & 0.939 & 0.970 & 0.907 & 0.924 & 0.944 & 0.900 & 0.914 & 0.922\\
\hspace{1em}$L = T^{1/2}$ & 0.908 & 0.945 & 0.972 & 0.890 & 0.913 & 0.935 & 0.893 & 0.910 & 0.910\\
\hspace{1em}$L = T^{1/3}$ & 0.911 & 0.950 & 0.975 & 0.892 & 0.918 & 0.934 & 0.893 & 0.903 & 0.907\\
\hspace{1em}$L = 0.75T^{1/3}$ & 0.910 & 0.953 & 0.977 & 0.893 & 0.920 & 0.934 & 0.892 & 0.905 & 0.905\\
\addlinespace[0.3em]
\hline
\multicolumn{10}{c}{\textbf{$\phi = 0.25$}}\\
\hline
\hspace{1em}$L = 3T^{1/2}$ & 0.896 & 0.929 & 0.966 & 0.899 & 0.918 & 0.947 & 0.920 & 0.913 & 0.917\\
\hspace{1em}$L = T^{1/2}$ & 0.886 & 0.932 & 0.966 & 0.890 & 0.904 & 0.940 & 0.913 & 0.904 & 0.908\\
\hspace{1em}$L = T^{1/3}$ & 0.895 & 0.934 & 0.968 & 0.890 & 0.903 & 0.926 & 0.914 & 0.899 & 0.899\\
\hspace{1em}$L = 0.75T^{1/3}$ & 0.898 & 0.927 & 0.966 & 0.894 & 0.900 & 0.925 & 0.915 & 0.897 & 0.896\\
\addlinespace[0.3em]
\hline
\multicolumn{10}{c}{\textbf{$\phi = 0.5$}}\\
\hline
\hspace{1em}$L = 3T^{1/2}$ & 0.880 & 0.881 & 0.934 & 0.916 & 0.891 & 0.923 & 0.910 & 0.902 & 0.903\\
\hspace{1em}$L = T^{1/2}$ & 0.868 & 0.881 & 0.946 & 0.909 & 0.887 & 0.916 & 0.910 & 0.893 & 0.885\\
\hspace{1em}$L = T^{1/3}$ & 0.880 & 0.874 & 0.929 & 0.916 & 0.879 & 0.885 & 0.911 & 0.888 & 0.863\\
\hspace{1em}$L = 0.75T^{1/3}$ & 0.886 & 0.860 & 0.913 & 0.914 & 0.871 & 0.873 & 0.914 & 0.876 & 0.841\\
\addlinespace[0.3em]
\hline
\multicolumn{10}{c}{\textbf{$\phi = 0.75$}}\\
\hline
\hspace{1em}$L = 3T^{1/2}$ & 0.837 & 0.776 & 0.844 & 0.903 & 0.873 & 0.870 & 0.897 & 0.908 & 0.890\\
\hspace{1em}$L = T^{1/2}$ & 0.843 & 0.770 & 0.849 & 0.910 & 0.868 & 0.856 & 0.900 & 0.895 & 0.882\\
\hspace{1em}$L = T^{1/3}$ & 0.865 & 0.694 & 0.771 & 0.928 & 0.807 & 0.779 & 0.916 & 0.875 & 0.840\\
\hspace{1em}$L = 0.75T^{1/3}$ & 0.872 & 0.652 & 0.707 & 0.931 & 0.791 & 0.741 & 0.928 & 0.859 & 0.790\\
\addlinespace[0.3em]
\hline
\multicolumn{10}{c}{\textbf{$\phi = 0.95$}}\\
\hline
\hspace{1em}$L = 3T^{1/2}$ & 0.673 & 0.534 & 0.496 & 0.836 & 0.796 & 0.762 & 0.898 & 0.892 & 0.881\\
\hspace{1em}$L = T^{1/2}$ & 0.745 & 0.449 & 0.275 & 0.876 & 0.795 & 0.684 & 0.935 & 0.908 & 0.874\\
\hspace{1em}$L = T^{1/3}$ & 0.791 & 0.309 & 0.066 & 0.918 & 0.734 & 0.542 & 0.959 & 0.906 & 0.794\\
\hspace{1em}$L = 0.75T^{1/3}$ & 0.822 & 0.250 & 0.023 & 0.931 & 0.707 & 0.505 & 0.970 & 0.884 & 0.749\\
\bottomrule
\end{tabular}}
\label{tab:confidence_bands_coverage_bandwidth_classical_iid}
\end{table}

\begin{table}[H]
\centering
\caption{Coverage of robust Bonferroni confidence bands (nominal coverage: $1 - \alpha = 0.9 $) for different bandwidth choices for variance estimation; time series of length $T$ are generated by an AR(1) process with coefficient $\phi$ with iid normal errors, length of the bands is denoted by $H$, and $N = T-H$.}
\resizebox{\ifdim\width>\linewidth\linewidth\else\width\fi}{!}{
\begin{tabular}{l|ccc|ccc|cccl|ccc|ccc|cccl|ccc|ccc|cccl|ccc|ccc|cccl|ccc|ccc|cccl|ccc|ccc|cccl|ccc|ccc|cccl|ccc|ccc|cccl|ccc|ccc|cccl|ccc|ccc|ccc}
\toprule
\multicolumn{1}{c}{ } & \multicolumn{3}{c}{$T = 50$} & \multicolumn{3}{c}{$T = 200$} & \multicolumn{3}{c}{$T = 800$} \\
\cmidrule(l{3pt}r{3pt}){2-4} \cmidrule(l{3pt}r{3pt}){5-7} \cmidrule(l{3pt}r{3pt}){8-10}
 & $H = 1$ & $H = 10 $ & $H = 25$ & $H = 1$ & $H = 10 $ & $H = 25$ & $H = 1$ & $H = 10 $ & $H = 25$\\
\midrule
\addlinespace[0.3em]
\hline
\multicolumn{10}{c}{\textbf{$\phi = 0$}}\\
\hline
\hspace{1em}$L = 3N^{1/2}$ & 0.726 & 0.417 & 0.431 & 0.812 & 0.613 & 0.546 & 0.867 & 0.739 & 0.657\\
\hspace{1em}$L = N^{1/2}$ & 0.820 & 0.750 & 0.852 & 0.852 & 0.826 & 0.825 & 0.888 & 0.830 & 0.837\\
\hspace{1em}$L = N^{1/3}$ & 0.855 & 0.846 & 0.932 & 0.873 & 0.886 & 0.908 & 0.893 & 0.885 & 0.892\\
\hspace{1em}$L = 0.75N^{1/3}$ & 0.865 & 0.876 & 0.932 & 0.877 & 0.887 & 0.915 & 0.891 & 0.892 & 0.903\\
\addlinespace[0.3em]
\hline
\multicolumn{10}{c}{\textbf{$\phi = 0.25$}}\\
\hline
\hspace{1em}$L = 3N^{1/2}$ & 0.671 & 0.393 & 0.387 & 0.808 & 0.630 & 0.534 & 0.853 & 0.743 & 0.684\\
\hspace{1em}$L = N^{1/2}$ & 0.777 & 0.742 & 0.801 & 0.849 & 0.822 & 0.815 & 0.888 & 0.850 & 0.839\\
\hspace{1em}$L = N^{1/3}$ & 0.824 & 0.829 & 0.911 & 0.865 & 0.870 & 0.883 & 0.901 & 0.882 & 0.892\\
\hspace{1em}$L = 0.75N^{1/3}$ & 0.846 & 0.850 & 0.911 & 0.870 & 0.876 & 0.896 & 0.904 & 0.887 & 0.897\\
\addlinespace[0.3em]
\hline
\multicolumn{10}{c}{\textbf{$\phi = 0.5$}}\\
\hline
\hspace{1em}$L = 3N^{1/2}$ & 0.653 & 0.369 & 0.341 & 0.804 & 0.674 & 0.562 & 0.870 & 0.783 & 0.736\\
\hspace{1em}$L = N^{1/2}$ & 0.757 & 0.655 & 0.721 & 0.859 & 0.808 & 0.815 & 0.885 & 0.870 & 0.865\\
\hspace{1em}$L = N^{1/3}$ & 0.792 & 0.738 & 0.842 & 0.875 & 0.861 & 0.881 & 0.901 & 0.891 & 0.892\\
\hspace{1em}$L = 0.75N^{1/3}$ & 0.811 & 0.753 & 0.842 & 0.880 & 0.865 & 0.882 & 0.900 & 0.896 & 0.892\\
\addlinespace[0.3em]
\hline
\multicolumn{10}{c}{\textbf{$\phi = 0.75$}}\\
\hline
\hspace{1em}$L = 3N^{1/2}$ & 0.602 & 0.394 & 0.310 & 0.788 & 0.694 & 0.599 & 0.843 & 0.879 & 0.842\\
\hspace{1em}$L = N^{1/2}$ & 0.704 & 0.598 & 0.591 & 0.837 & 0.810 & 0.780 & 0.862 & 0.924 & 0.907\\
\hspace{1em}$L = N^{1/3}$ & 0.755 & 0.648 & 0.675 & 0.866 & 0.840 & 0.819 & 0.878 & 0.930 & 0.912\\
\hspace{1em}$L = 0.75N^{1/3}$ & 0.776 & 0.642 & 0.675 & 0.874 & 0.833 & 0.813 & 0.881 & 0.918 & 0.899\\
\addlinespace[0.3em]
\hline
\multicolumn{10}{c}{\textbf{$\phi = 0.95$}}\\
\hline
\hspace{1em}$L = 3N^{1/2}$ & 0.330 & 0.171 & 0.030 & 0.665 & 0.711 & 0.610 & 0.799 & 0.902 & 0.888\\
\hspace{1em}$L = N^{1/2}$ & 0.471 & 0.322 & 0.064 & 0.745 & 0.800 & 0.718 & 0.839 & 0.930 & 0.922\\
\hspace{1em}$L = N^{1/3}$ & 0.523 & 0.281 & 0.025 & 0.774 & 0.770 & 0.642 & 0.850 & 0.920 & 0.890\\
\hspace{1em}$L = 0.75N^{1/3}$ & 0.538 & 0.234 & 0.025 & 0.774 & 0.748 & 0.602 & 0.852 & 0.907 & 0.849\\
\bottomrule
\end{tabular}}
\label{tab:confidence_bands_coverage_bandwidth_robust_iid}
\end{table}

\subsection{Confidence Bands under Normal GARCH Errors}		

\begin{table}[H]
\centering
\caption{Coverage of confidence bands (nominal coverage: $1 - \alpha = 0.9 $); time series of length $T$ are generated by an AR(1) process with coefficient $\phi$ and normal GARCH errors, length of the bands is denoted by $H$; the bandwidth for variance estimation is $L = T^{1/2}$ for classical bands and  $L = (T-H)^{1/3}$ for robust bands. NA: For 0.1\% of Monte Carlo runs, the equicoordinate quantile could not be computed due to non-positive semi-definiteness of the estimated covariance matrix.}
\resizebox{\ifdim\width>\linewidth\linewidth\else\width\fi}{!}{
\begin{tabular}{l|ccc|ccc|cccl|ccc|ccc|cccl|ccc|ccc|cccl|ccc|ccc|cccl|ccc|ccc|cccl|ccc|ccc|cccl|ccc|ccc|cccl|ccc|ccc|cccl|ccc|ccc|cccl|ccc|ccc|ccc}
\toprule
\multicolumn{1}{c}{ } & \multicolumn{3}{c}{$T = 50$} & \multicolumn{3}{c}{$T = 200$} & \multicolumn{3}{c}{$T = 800$} \\
\cmidrule(l{3pt}r{3pt}){2-4} \cmidrule(l{3pt}r{3pt}){5-7} \cmidrule(l{3pt}r{3pt}){8-10}
 & $H = 1$ & $H = 10 $ & $H = 25$ & $H = 1$ & $H = 10 $ & $H = 25$ & $H = 1$ & $H = 10 $ & $H = 25$\\
\midrule
\addlinespace[0.3em]
\hline
\multicolumn{10}{c}{\textbf{$\phi = 0$}}\\
\hline
\hspace{1em}Classical sup-t CB & 0.880 & 0.918 & 0.960 & 0.857 & 0.827 & 0.866 & 0.831 & 0.817 & 0.822\\
\hspace{1em}Classical Bonf. CB & 0.880 & 0.925 & 0.964 & 0.857 & 0.836 & 0.872 & 0.831 & 0.825 & 0.827\\
\hspace{1em}Classical pointw. CB & 0.880 & 0.451 & 0.333 & 0.857 & 0.303 & 0.084 & 0.831 & 0.231 & 0.035\\
\hspace{1em}Robust sup-t CB & 0.844 & 0.744 & 0.747 & 0.882 & 0.791 & 0.695 & 0.883 & 0.841 & 0.780\\
\hspace{1em}Robust Bonf. CB & 0.844 & 0.772 & 0.770 & 0.882 & 0.811 & 0.713 & 0.883 & 0.849 & 0.795\\
\hspace{1em}Robust pointw. CB & 0.844 & 0.270 & 0.252 & 0.882 & 0.267 & 0.070 & 0.883 & 0.282 & 0.048\\
\addlinespace[0.3em]
\hline
\multicolumn{10}{c}{\textbf{$\phi = 0.25$}}\\
\hline
\hspace{1em}Classical sup-t CB & 0.869 & 0.914 & 0.959 & 0.837 & 0.814 & 0.850 & 0.847 & 0.795 & 0.826\\
\hspace{1em}Classical Bonf. CB & 0.869 & 0.923 & 0.964 & 0.837 & 0.836 & 0.865 & 0.847 & 0.813 & 0.839\\
\hspace{1em}Classical pointw. CB & 0.869 & 0.440 & 0.303 & 0.837 & 0.315 & 0.107 & 0.847 & 0.303 & 0.067\\
\hspace{1em}Robust sup-t CB & 0.827 & 0.721 & 0.692 & 0.860 & 0.776 & 0.695 & 0.889 & 0.837 & 0.789\\
\hspace{1em}Robust Bonf. CB & 0.827 & 0.746 & 0.729 & 0.860 & 0.793 & 0.713 & 0.889 & 0.855 & 0.806\\
\hspace{1em}Robust pointw. CB & 0.827 & 0.292 & 0.227 & 0.860 & 0.292 & 0.082 & 0.889 & 0.340 & 0.074\\
\addlinespace[0.3em]
\hline
\multicolumn{10}{c}{\textbf{$\phi = 0.5$}}\\
\hline
\hspace{1em}Classical sup-t CB & 0.850 & 0.857 & 0.929 & 0.866 & 0.822 & 0.854 & 0.864 & 0.799 & 0.825\\
\hspace{1em}Classical Bonf. CB & 0.850 & 0.878 & 0.938 & 0.866 & 0.862 & 0.885 & 0.864 & 0.838 & 0.862\\
\hspace{1em}Classical pointw. CB & 0.850 & 0.455 & 0.321 & 0.866 & 0.425 & 0.185 & 0.864 & 0.411 & 0.175\\
\hspace{1em}Robust sup-t CB & 0.782 & 0.637 & 0.623 & 0.863 & 0.759 & 0.696 & 0.890 & 0.822 & 0.768\\
\hspace{1em}Robust Bonf. CB & 0.782 & 0.681 & 0.671 & 0.863 & 0.798 & 0.749 & 0.890 & 0.874 & 0.820\\
\hspace{1em}Robust pointw. CB & 0.782 & 0.265 & 0.209 & 0.863 & 0.389 & 0.134 & 0.890 & 0.434 & 0.160\\
\addlinespace[0.3em]
\hline
\multicolumn{10}{c}{\textbf{$\phi = 0.75$}}\\
\hline
\hspace{1em}Classical sup-t CB & 0.837 & 0.771 & 0.851 & 0.880 & 0.805 & 0.820 & 0.856 & 0.824 & 0.832\\
\hspace{1em}Classical Bonf. CB & 0.837 & 0.818 & 0.878 & 0.880 & 0.869 & 0.894 & 0.856 & 0.908 & 0.904\\
\hspace{1em}Classical pointw. CB & 0.837 & 0.488 & 0.336 & 0.880 & 0.572 & 0.352 & 0.856 & 0.582 & 0.368\\
\hspace{1em}Robust sup-t CB & 0.769 & 0.556 & NA & 0.862 & 0.725 & 0.612 & 0.871 & 0.835 & 0.774\\
\hspace{1em}Robust Bonf. CB & 0.769 & 0.618 & 0.580 & 0.862 & 0.798 & 0.697 & 0.871 & 0.913 & 0.858\\
\hspace{1em}Robust pointw. CB & 0.769 & 0.327 & 0.171 & 0.862 & 0.482 & 0.226 & 0.871 & 0.553 & 0.295\\
\addlinespace[0.3em]
\hline
\multicolumn{10}{c}{\textbf{$\phi = 0.95$}}\\
\hline
\hspace{1em}Classical sup-t CB & 0.757 & 0.487 & 0.346 & 0.865 & 0.788 & 0.697 & 0.912 & 0.884 & 0.859\\
\hspace{1em}Classical Bonf. CB & 0.757 & 0.582 & 0.498 & 0.865 & 0.875 & 0.804 & 0.912 & 0.949 & 0.931\\
\hspace{1em}Classical pointw. CB & 0.757 & 0.310 & 0.006 & 0.865 & 0.696 & 0.502 & 0.912 & 0.816 & 0.717\\
\hspace{1em}Robust sup-t CB & 0.602 & 0.205 & 0.007 & 0.828 & 0.668 & 0.466 & 0.875 & 0.837 & 0.753\\
\hspace{1em}Robust Bonf. CB & 0.602 & 0.303 & 0.026 & 0.828 & 0.759 & 0.583 & 0.875 & 0.923 & 0.862\\
\hspace{1em}Robust pointw. CB & 0.602 & 0.104 & 0.000 & 0.828 & 0.534 & 0.285 & 0.875 & 0.735 & 0.553\\
\bottomrule
\end{tabular}}
\label{tab:confidence_bands_coverage_GARCH_normal}
\end{table}

\begin{table}[H]
\centering
\caption{Coverage of robust Bonferroni confidence bands (nominal coverage: $1 - \alpha = 0.9 $) for different bandwidth choices for variance estimation; time series of length $T$ are generated by an AR(1) process with coefficient $\phi$ with normal GARCH errors, length of the bands is denoted by $H$, and $N = T-H$.}
\resizebox{\ifdim\width>\linewidth\linewidth\else\width\fi}{!}{
\begin{tabular}{l|ccc|ccc|cccl|ccc|ccc|cccl|ccc|ccc|cccl|ccc|ccc|cccl|ccc|ccc|cccl|ccc|ccc|cccl|ccc|ccc|cccl|ccc|ccc|cccl|ccc|ccc|cccl|ccc|ccc|ccc}
\toprule
\multicolumn{1}{c}{ } & \multicolumn{3}{c}{$T = 50$} & \multicolumn{3}{c}{$T = 200$} & \multicolumn{3}{c}{$T = 800$} \\
\cmidrule(l{3pt}r{3pt}){2-4} \cmidrule(l{3pt}r{3pt}){5-7} \cmidrule(l{3pt}r{3pt}){8-10}
 & $H = 1$ & $H = 10 $ & $H = 25$ & $H = 1$ & $H = 10 $ & $H = 25$ & $H = 1$ & $H = 10 $ & $H = 25$\\
\midrule
\addlinespace[0.3em]
\hline
\multicolumn{10}{c}{\textbf{$\phi = 0$}}\\
\hline
\hspace{1em}$L = 3N^{1/2}$ & 0.701 & 0.363 & 0.290 & 0.787 & 0.514 & 0.395 & 0.832 & 0.683 & 0.552\\
\hspace{1em}$L = N^{1/2}$ & 0.798 & 0.661 & 0.660 & 0.852 & 0.732 & 0.637 & 0.872 & 0.800 & 0.737\\
\hspace{1em}$L = N^{1/3}$ & 0.844 & 0.772 & 0.770 & 0.882 & 0.811 & 0.713 & 0.883 & 0.849 & 0.795\\
\hspace{1em}$L = 0.75N^{1/3}$ & 0.855 & 0.799 & 0.770 & 0.884 & 0.815 & 0.726 & 0.883 & 0.864 & 0.802\\
\addlinespace[0.3em]
\hline
\multicolumn{10}{c}{\textbf{$\phi = 0.25$}}\\
\hline
\hspace{1em}$L = 3N^{1/2}$ & 0.657 & 0.336 & 0.254 & 0.771 & 0.518 & 0.380 & 0.835 & 0.700 & 0.586\\
\hspace{1em}$L = N^{1/2}$ & 0.765 & 0.655 & 0.598 & 0.833 & 0.728 & 0.611 & 0.869 & 0.819 & 0.750\\
\hspace{1em}$L = N^{1/3}$ & 0.827 & 0.746 & 0.729 & 0.860 & 0.793 & 0.713 & 0.889 & 0.855 & 0.806\\
\hspace{1em}$L = 0.75N^{1/3}$ & 0.835 & 0.775 & 0.729 & 0.862 & 0.804 & 0.726 & 0.896 & 0.859 & 0.819\\
\addlinespace[0.3em]
\hline
\multicolumn{10}{c}{\textbf{$\phi = 0.5$}}\\
\hline
\hspace{1em}$L = 3N^{1/2}$ & 0.634 & 0.323 & 0.250 & 0.785 & 0.589 & 0.432 & 0.853 & 0.745 & 0.656\\
\hspace{1em}$L = N^{1/2}$ & 0.749 & 0.591 & 0.560 & 0.841 & 0.735 & 0.680 & 0.878 & 0.829 & 0.796\\
\hspace{1em}$L = N^{1/3}$ & 0.782 & 0.681 & 0.671 & 0.863 & 0.798 & 0.749 & 0.890 & 0.874 & 0.820\\
\hspace{1em}$L = 0.75N^{1/3}$ & 0.800 & 0.695 & 0.671 & 0.867 & 0.805 & 0.751 & 0.890 & 0.877 & 0.821\\
\addlinespace[0.3em]
\hline
\multicolumn{10}{c}{\textbf{$\phi = 0.75$}}\\
\hline
\hspace{1em}$L = 3N^{1/2}$ & 0.621 & 0.362 & 0.254 & 0.762 & 0.638 & 0.516 & 0.832 & 0.846 & 0.755\\
\hspace{1em}$L = N^{1/2}$ & 0.728 & 0.583 & 0.514 & 0.829 & 0.762 & 0.672 & 0.856 & 0.891 & 0.848\\
\hspace{1em}$L = N^{1/3}$ & 0.769 & 0.618 & 0.580 & 0.862 & 0.798 & 0.697 & 0.871 & 0.913 & 0.858\\
\hspace{1em}$L = 0.75N^{1/3}$ & 0.776 & 0.611 & 0.580 & 0.864 & 0.794 & 0.691 & 0.873 & 0.910 & 0.848\\
\addlinespace[0.3em]
\hline
\multicolumn{10}{c}{\textbf{$\phi = 0.95$}}\\
\hline
\hspace{1em}$L = 3N^{1/2}$ & 0.416 & 0.191 & 0.038 & 0.732 & 0.717 & 0.590 & 0.825 & 0.897 & 0.866\\
\hspace{1em}$L = N^{1/2}$ & 0.546 & 0.342 & 0.073 & 0.788 & 0.785 & 0.668 & 0.865 & 0.929 & 0.896\\
\hspace{1em}$L = N^{1/3}$ & 0.602 & 0.303 & 0.026 & 0.828 & 0.759 & 0.583 & 0.875 & 0.923 & 0.862\\
\hspace{1em}$L = 0.75N^{1/3}$ & 0.610 & 0.257 & 0.026 & 0.829 & 0.740 & 0.544 & 0.873 & 0.907 & 0.823\\
\bottomrule
\end{tabular}}
\label{tab:confidence_bands_coverage_bandwidth_robust_GARCH_normal}
\end{table}

\subsection{Confidence Bands under Chi-square GARCH Errors}

\begin{table}[H]
\centering
\caption{Coverage of confidence bands (nominal coverage: $1 - \alpha = 0.9 $); time series of length $T$ are generated by an AR(1) process with coefficient $\phi$ and $\chi^2(3)$ GARCH errors, length of the bands is denoted by $H$; the bandwidth for variance estimation is $L = T^{1/2}$ for classical bands and  $L = (T-H)^{1/3}$ for robust bands. NA: For 0.1\% of Monte Carlo runs, the equicoordinate quantile could not be computed due to non-positive semi-definiteness of the estimated covariance matrix.}
\resizebox{\ifdim\width>\linewidth\linewidth\else\width\fi}{!}{
\begin{tabular}{l|ccc|ccc|cccl|ccc|ccc|cccl|ccc|ccc|cccl|ccc|ccc|cccl|ccc|ccc|cccl|ccc|ccc|cccl|ccc|ccc|cccl|ccc|ccc|cccl|ccc|ccc|cccl|ccc|ccc|ccc}
\toprule
\multicolumn{1}{c}{ } & \multicolumn{3}{c}{$T = 50$} & \multicolumn{3}{c}{$T = 200$} & \multicolumn{3}{c}{$T = 800$} \\
\cmidrule(l{3pt}r{3pt}){2-4} \cmidrule(l{3pt}r{3pt}){5-7} \cmidrule(l{3pt}r{3pt}){8-10}
 & $H = 1$ & $H = 10 $ & $H = 25$ & $H = 1$ & $H = 10 $ & $H = 25$ & $H = 1$ & $H = 10 $ & $H = 25$\\
\midrule
\addlinespace[0.3em]
\hline
\multicolumn{10}{c}{\textbf{$\phi = 0$}}\\
\hline
\hspace{1em}Classical sup-t CB & 0.871 & 0.949 & 0.979 & 0.824 & 0.788 & 0.844 & 0.821 & 0.748 & 0.782\\
\hspace{1em}Classical Bonf. CB & 0.871 & 0.955 & 0.982 & 0.824 & 0.802 & 0.852 & 0.821 & 0.759 & 0.795\\
\hspace{1em}Classical pointw. CB & 0.871 & 0.475 & 0.354 & 0.824 & 0.249 & 0.081 & 0.821 & 0.207 & 0.035\\
\hspace{1em}Robust sup-t CB & 0.797 & 0.708 & 0.703 & 0.837 & 0.740 & 0.678 & 0.880 & 0.817 & 0.781\\
\hspace{1em}Robust Bonf. CB & 0.797 & 0.739 & 0.737 & 0.837 & 0.755 & 0.705 & 0.880 & 0.829 & 0.796\\
\hspace{1em}Robust pointw. CB & 0.797 & 0.245 & 0.202 & 0.837 & 0.233 & 0.049 & 0.880 & 0.271 & 0.049\\
\addlinespace[0.3em]
\hline
\multicolumn{10}{c}{\textbf{$\phi = 0.25$}}\\
\hline
\hspace{1em}Classical sup-t CB & 0.875 & 0.942 & 0.975 & 0.851 & 0.805 & 0.837 & 0.829 & 0.766 & 0.777\\
\hspace{1em}Classical Bonf. CB & 0.875 & 0.951 & 0.977 & 0.851 & 0.817 & 0.850 & 0.829 & 0.781 & 0.794\\
\hspace{1em}Classical pointw. CB & 0.875 & 0.476 & 0.351 & 0.851 & 0.281 & 0.101 & 0.829 & 0.264 & 0.072\\
\hspace{1em}Robust sup-t CB & 0.820 & 0.658 & 0.684 & 0.849 & 0.755 & 0.699 & 0.876 & 0.811 & 0.771\\
\hspace{1em}Robust Bonf. CB & 0.820 & 0.695 & 0.727 & 0.849 & 0.778 & 0.729 & 0.876 & 0.833 & 0.790\\
\hspace{1em}Robust pointw. CB & 0.820 & 0.269 & 0.210 & 0.849 & 0.248 & 0.068 & 0.876 & 0.311 & 0.074\\
\addlinespace[0.3em]
\hline
\multicolumn{10}{c}{\textbf{$\phi = 0.5$}}\\
\hline
\hspace{1em}Classical sup-t CB & 0.847 & 0.888 & 0.941 & 0.861 & 0.808 & 0.858 & 0.842 & 0.775 & 0.767\\
\hspace{1em}Classical Bonf. CB & 0.847 & 0.904 & 0.953 & 0.861 & 0.843 & 0.883 & 0.842 & 0.813 & 0.819\\
\hspace{1em}Classical pointw. CB & 0.847 & 0.493 & 0.384 & 0.861 & 0.396 & 0.205 & 0.842 & 0.376 & 0.137\\
\hspace{1em}Robust sup-t CB & 0.813 & 0.642 & 0.653 & 0.881 & 0.747 & 0.719 & 0.893 & 0.796 & 0.776\\
\hspace{1em}Robust Bonf. CB & 0.813 & 0.687 & 0.688 & 0.881 & 0.781 & 0.765 & 0.893 & 0.849 & 0.816\\
\hspace{1em}Robust pointw. CB & 0.813 & 0.290 & 0.221 & 0.881 & 0.364 & 0.137 & 0.893 & 0.437 & 0.144\\
\addlinespace[0.3em]
\hline
\multicolumn{10}{c}{\textbf{$\phi = 0.75$}}\\
\hline
\hspace{1em}Classical sup-t CB & 0.848 & 0.813 & 0.879 & 0.858 & 0.829 & 0.834 & 0.863 & 0.800 & 0.798\\
\hspace{1em}Classical Bonf. CB & 0.848 & 0.847 & 0.906 & 0.858 & 0.880 & 0.898 & 0.863 & 0.883 & 0.883\\
\hspace{1em}Classical pointw. CB & 0.848 & 0.530 & 0.385 & 0.858 & 0.532 & 0.332 & 0.863 & 0.556 & 0.343\\
\hspace{1em}Robust sup-t CB & 0.819 & 0.566 & 0.527 & 0.885 & 0.730 & 0.638 & 0.913 & 0.834 & 0.768\\
\hspace{1em}Robust Bonf. CB & 0.819 & 0.623 & 0.600 & 0.885 & 0.798 & 0.717 & 0.913 & 0.897 & 0.851\\
\hspace{1em}Robust pointw. CB & 0.819 & 0.336 & 0.154 & 0.885 & 0.477 & 0.231 & 0.913 & 0.600 & 0.338\\
\addlinespace[0.3em]
\hline
\multicolumn{10}{c}{\textbf{$\phi = 0.95$}}\\
\hline
\hspace{1em}Classical sup-t CB & 0.770 & 0.469 & 0.337 & 0.868 & 0.800 & 0.699 & 0.889 & 0.851 & 0.813\\
\hspace{1em}Classical Bonf. CB & 0.770 & 0.569 & 0.485 & 0.868 & 0.875 & 0.810 & 0.889 & 0.950 & 0.929\\
\hspace{1em}Classical pointw. CB & 0.770 & 0.273 & 0.003 & 0.868 & 0.676 & 0.497 & 0.889 & 0.751 & 0.647\\
\hspace{1em}Robust sup-t CB & 0.645 & 0.202 & NA & 0.841 & 0.674 & 0.472 & 0.906 & 0.846 & 0.725\\
\hspace{1em}Robust Bonf. CB & 0.645 & 0.282 & 0.033 & 0.841 & 0.766 & 0.579 & 0.906 & 0.925 & 0.836\\
\hspace{1em}Robust pointw. CB & 0.645 & 0.085 & 0.000 & 0.841 & 0.544 & 0.292 & 0.906 & 0.736 & 0.530\\
\bottomrule
\end{tabular}}
\label{tab:confidence_bands_coverage_GARCH_chisqu}
\end{table}

\begin{table}[H]
\centering
\caption{Coverage of robust Bonferroni confidence bands (nominal coverage: $1 - \alpha = 0.9 $) for different bandwidth choices for variance estimation; time series of length $T$ are generated by an AR(1) process with coefficient $\phi$ with $\chi^2(3)$ GARCH errors, length of the bands is denoted by $H$, and $N = T-H$.}
\resizebox{\ifdim\width>\linewidth\linewidth\else\width\fi}{!}{
\begin{tabular}{l|ccc|ccc|cccl|ccc|ccc|cccl|ccc|ccc|cccl|ccc|ccc|cccl|ccc|ccc|cccl|ccc|ccc|cccl|ccc|ccc|cccl|ccc|ccc|cccl|ccc|ccc|cccl|ccc|ccc|ccc}
\toprule
\multicolumn{1}{c}{ } & \multicolumn{3}{c}{$T = 50$} & \multicolumn{3}{c}{$T = 200$} & \multicolumn{3}{c}{$T = 800$} \\
\cmidrule(l{3pt}r{3pt}){2-4} \cmidrule(l{3pt}r{3pt}){5-7} \cmidrule(l{3pt}r{3pt}){8-10}
 & $H = 1$ & $H = 10 $ & $H = 25$ & $H = 1$ & $H = 10 $ & $H = 25$ & $H = 1$ & $H = 10 $ & $H = 25$\\
\midrule
\addlinespace[0.3em]
\hline
\multicolumn{10}{c}{\textbf{$\phi = 0$}}\\
\hline
\hspace{1em}$L = 3N^{1/2}$ & 0.649 & 0.280 & 0.271 & 0.732 & 0.444 & 0.347 & 0.821 & 0.649 & 0.557\\
\hspace{1em}$L = N^{1/2}$ & 0.755 & 0.643 & 0.616 & 0.799 & 0.671 & 0.617 & 0.868 & 0.769 & 0.733\\
\hspace{1em}$L = N^{1/3}$ & 0.797 & 0.739 & 0.737 & 0.837 & 0.755 & 0.705 & 0.880 & 0.829 & 0.796\\
\hspace{1em}$L = 0.75N^{1/3}$ & 0.819 & 0.773 & 0.737 & 0.841 & 0.761 & 0.712 & 0.882 & 0.837 & 0.813\\
\addlinespace[0.3em]
\hline
\multicolumn{10}{c}{\textbf{$\phi = 0.25$}}\\
\hline
\hspace{1em}$L = 3N^{1/2}$ & 0.661 & 0.318 & 0.290 & 0.739 & 0.445 & 0.365 & 0.810 & 0.645 & 0.547\\
\hspace{1em}$L = N^{1/2}$ & 0.778 & 0.600 & 0.607 & 0.814 & 0.689 & 0.652 & 0.860 & 0.770 & 0.726\\
\hspace{1em}$L = N^{1/3}$ & 0.820 & 0.695 & 0.727 & 0.849 & 0.778 & 0.729 & 0.876 & 0.833 & 0.790\\
\hspace{1em}$L = 0.75N^{1/3}$ & 0.834 & 0.734 & 0.727 & 0.855 & 0.783 & 0.740 & 0.884 & 0.840 & 0.804\\
\addlinespace[0.3em]
\hline
\multicolumn{10}{c}{\textbf{$\phi = 0.5$}}\\
\hline
\hspace{1em}$L = 3N^{1/2}$ & 0.642 & 0.356 & 0.287 & 0.771 & 0.526 & 0.446 & 0.825 & 0.709 & 0.591\\
\hspace{1em}$L = N^{1/2}$ & 0.757 & 0.602 & 0.586 & 0.845 & 0.719 & 0.675 & 0.868 & 0.803 & 0.767\\
\hspace{1em}$L = N^{1/3}$ & 0.813 & 0.687 & 0.688 & 0.881 & 0.781 & 0.765 & 0.893 & 0.849 & 0.816\\
\hspace{1em}$L = 0.75N^{1/3}$ & 0.822 & 0.716 & 0.688 & 0.888 & 0.789 & 0.766 & 0.900 & 0.846 & 0.819\\
\addlinespace[0.3em]
\hline
\multicolumn{10}{c}{\textbf{$\phi = 0.75$}}\\
\hline
\hspace{1em}$L = 3N^{1/2}$ & 0.641 & 0.410 & 0.240 & 0.770 & 0.641 & 0.515 & 0.835 & 0.829 & 0.749\\
\hspace{1em}$L = N^{1/2}$ & 0.769 & 0.596 & 0.524 & 0.852 & 0.763 & 0.686 & 0.880 & 0.884 & 0.836\\
\hspace{1em}$L = N^{1/3}$ & 0.819 & 0.623 & 0.600 & 0.885 & 0.798 & 0.717 & 0.913 & 0.897 & 0.851\\
\hspace{1em}$L = 0.75N^{1/3}$ & 0.829 & 0.626 & 0.600 & 0.899 & 0.792 & 0.710 & 0.929 & 0.894 & 0.835\\
\addlinespace[0.3em]
\hline
\multicolumn{10}{c}{\textbf{$\phi = 0.95$}}\\
\hline
\hspace{1em}$L = 3N^{1/2}$ & 0.412 & 0.176 & 0.041 & 0.734 & 0.709 & 0.578 & 0.819 & 0.889 & 0.840\\
\hspace{1em}$L = N^{1/2}$ & 0.582 & 0.316 & 0.074 & 0.797 & 0.786 & 0.656 & 0.859 & 0.919 & 0.881\\
\hspace{1em}$L = N^{1/3}$ & 0.645 & 0.282 & 0.033 & 0.841 & 0.766 & 0.579 & 0.906 & 0.925 & 0.836\\
\hspace{1em}$L = 0.75N^{1/3}$ & 0.658 & 0.242 & 0.033 & 0.848 & 0.753 & 0.551 & 0.915 & 0.910 & 0.797\\
\bottomrule
\end{tabular}}
\label{tab:confidence_bands_coverage_bandwidth_robust_GARCH_chisqu}
\end{table}

\subsection{Average Width of Inference Bands}
\label{appsubsec:width_inference_bands}

\begin{table}[H]
\centering
\caption{Average width of inference bands (significance level $\alpha = 0.1 $) -- averaged over the lag $h$ as well as over simulation runs; time series of length $T$ are generated by an AR(1) process with coefficient $\phi$ and iid normal errors, length of the bands is denoted by $H$; the bandwidth for variance estimation is $L = T^{1/2}$ for classical confidence bands and $L = (T-H)^{1/3}$ for robust confidence bands.}
\resizebox{\ifdim\width>\linewidth\linewidth\else\width\fi}{!}{
\begin{tabular}{l|ccc|ccc|cccl|ccc|ccc|cccl|ccc|ccc|cccl|ccc|ccc|cccl|ccc|ccc|cccl|ccc|ccc|cccl|ccc|ccc|cccl|ccc|ccc|cccl|ccc|ccc|cccl|ccc|ccc|ccc}
\toprule
\multicolumn{1}{c}{ } & \multicolumn{3}{c}{$T = 50$} & \multicolumn{3}{c}{$T = 200$} & \multicolumn{3}{c}{$T = 800$} \\
\cmidrule(l{3pt}r{3pt}){2-4} \cmidrule(l{3pt}r{3pt}){5-7} \cmidrule(l{3pt}r{3pt}){8-10}
 & $H = 1$ & $H = 10 $ & $H = 25$ & $H = 1$ & $H = 10 $ & $H = 25$ & $H = 1$ & $H = 10 $ & $H = 25$\\
\midrule
\addlinespace[0.3em]
\hline
\multicolumn{10}{c}{\textbf{$\phi = 0$}}\\
\hline
\hspace{1em}Classical sup-t SB & 0.465 & 0.724 & 0.810 & 0.233 & 0.362 & 0.405 & 0.116 & 0.181 & \vphantom{4} 0.202\\
\hspace{1em}Classical sup-t CB & 0.455 & 0.736 & 0.830 & 0.234 & 0.366 & 0.411 & 0.117 & 0.182 & 0.204\\
\hspace{1em}Classical pointw. SB & 0.465 & 0.465 & 0.465 & 0.233 & 0.233 & 0.233 & 0.116 & 0.116 & \vphantom{4} 0.116\\
\hspace{1em}Robust Bonf. SB & 0.455 & 0.787 & 1.101 & 0.232 & 0.371 & 0.431 & 0.116 & 0.183 & 0.206\\
\hspace{1em}Robust Bonf. CB & 0.407 & 0.723 & 1.003 & 0.224 & 0.360 & 0.419 & 0.115 & 0.181 & 0.204\\
\addlinespace[0.3em]
\hline
\multicolumn{10}{c}{\textbf{$\phi = 0.25$}}\\
\hline
\hspace{1em}Classical sup-t SB & 0.465 & 0.724 & 0.810 & 0.233 & 0.362 & 0.405 & 0.116 & 0.181 & \vphantom{3} 0.202\\
\hspace{1em}Classical sup-t CB & 0.446 & 0.749 & 0.850 & 0.227 & 0.377 & 0.427 & 0.114 & 0.189 & 0.214\\
\hspace{1em}Classical pointw. SB & 0.465 & 0.465 & 0.465 & 0.233 & 0.233 & 0.233 & 0.116 & 0.116 & \vphantom{3} 0.116\\
\hspace{1em}Robust Bonf. SB & 0.475 & 0.788 & 1.096 & 0.244 & 0.372 & 0.431 & 0.123 & 0.184 & 0.207\\
\hspace{1em}Robust Bonf. CB & 0.396 & 0.735 & 1.013 & 0.215 & 0.372 & 0.435 & 0.112 & 0.189 & 0.215\\
\addlinespace[0.3em]
\hline
\multicolumn{10}{c}{\textbf{$\phi = 0.5$}}\\
\hline
\hspace{1em}Classical sup-t SB & 0.465 & 0.724 & 0.810 & 0.233 & 0.362 & 0.405 & 0.116 & 0.181 & \vphantom{2} 0.202\\
\hspace{1em}Classical sup-t CB & 0.415 & 0.793 & 0.917 & 0.207 & 0.409 & 0.477 & 0.102 & 0.208 & 0.243\\
\hspace{1em}Classical pointw. SB & 0.465 & 0.465 & 0.465 & 0.233 & 0.233 & 0.233 & 0.116 & 0.116 & \vphantom{2} 0.116\\
\hspace{1em}Robust Bonf. SB & 0.530 & 0.792 & 1.094 & 0.278 & 0.380 & 0.436 & 0.142 & 0.188 & 0.209\\
\hspace{1em}Robust Bonf. CB & 0.359 & 0.772 & 1.067 & 0.195 & 0.414 & 0.494 & 0.099 & 0.214 & 0.249\\
\addlinespace[0.3em]
\hline
\multicolumn{10}{c}{\textbf{$\phi = 0.75$}}\\
\hline
\hspace{1em}Classical sup-t SB & 0.465 & 0.724 & 0.810 & 0.233 & 0.362 & 0.405 & 0.116 & 0.181 & \vphantom{1} 0.202\\
\hspace{1em}Classical sup-t CB & 0.359 & 0.862 & 1.032 & 0.168 & 0.457 & 0.569 & 0.081 & 0.235 & 0.298\\
\hspace{1em}Classical pointw. SB & 0.465 & 0.465 & 0.465 & 0.233 & 0.233 & 0.233 & 0.116 & 0.116 & \vphantom{1} 0.116\\
\hspace{1em}Robust Bonf. SB & 0.602 & 0.796 & 1.089 & 0.327 & 0.400 & 0.441 & 0.168 & 0.202 & 0.214\\
\hspace{1em}Robust Bonf. CB & 0.299 & 0.827 & 1.138 & 0.151 & 0.475 & 0.596 & 0.076 & 0.258 & 0.323\\
\addlinespace[0.3em]
\hline
\multicolumn{10}{c}{\textbf{$\phi = 0.95$}}\\
\hline
\hspace{1em}Classical sup-t SB & 0.465 & 0.724 & 0.810 & 0.233 & 0.362 & 0.405 & 0.116 & 0.181 & 0.202\\
\hspace{1em}Classical sup-t CB & 0.293 & 0.923 & 1.163 & 0.109 & 0.459 & 0.689 & 0.046 & 0.212 & 0.362\\
\hspace{1em}Classical pointw. SB & 0.465 & 0.465 & 0.465 & 0.233 & 0.233 & 0.233 & 0.116 & 0.116 & 0.116\\
\hspace{1em}Robust Bonf. SB & 0.640 & 0.781 & 1.023 & 0.357 & 0.470 & 0.464 & 0.189 & 0.257 & 0.251\\
\hspace{1em}Robust Bonf. CB & 0.211 & 0.770 & 1.061 & 0.082 & 0.428 & 0.644 & 0.037 & 0.227 & 0.383\\
\bottomrule
\end{tabular}}
\label{tab:width_significance_bands_iid}
\end{table}

\begin{table}[H]
\centering
\caption{Average width of inference bands (significance level $\alpha = 0.1 $) -- averaged over the lag $h$ as well as over simulation runs; time series of length $T$ are generated by an AR(1) process with coefficient $\phi$ and normal GARCH errors, length of the bands is denoted by $H$; the bandwidth for variance estimation is $L = T^{1/2}$ for classical confidence bands and $L = (T-H)^{1/3}$ for robust confidence bands.}
\resizebox{\ifdim\width>\linewidth\linewidth\else\width\fi}{!}{
\begin{tabular}{l|ccc|ccc|cccl|ccc|ccc|cccl|ccc|ccc|cccl|ccc|ccc|cccl|ccc|ccc|cccl|ccc|ccc|cccl|ccc|ccc|cccl|ccc|ccc|cccl|ccc|ccc|cccl|ccc|ccc|ccc}
\toprule
\multicolumn{1}{c}{ } & \multicolumn{3}{c}{$T = 50$} & \multicolumn{3}{c}{$T = 200$} & \multicolumn{3}{c}{$T = 800$} \\
\cmidrule(l{3pt}r{3pt}){2-4} \cmidrule(l{3pt}r{3pt}){5-7} \cmidrule(l{3pt}r{3pt}){8-10}
 & $H = 1$ & $H = 10 $ & $H = 25$ & $H = 1$ & $H = 10 $ & $H = 25$ & $H = 1$ & $H = 10 $ & $H = 25$\\
\midrule
\addlinespace[0.3em]
\hline
\multicolumn{10}{c}{\textbf{$\phi = 0$}}\\
\hline
\hspace{1em}Classical sup-t SB & 0.465 & 0.724 & 0.810 & 0.233 & 0.362 & 0.405 & 0.116 & 0.181 & \vphantom{4} 0.202\\
\hspace{1em}Classical sup-t CB & 0.454 & 0.737 & 0.833 & 0.235 & 0.367 & 0.413 & 0.117 & 0.183 & 0.205\\
\hspace{1em}Classical pointw. SB & 0.465 & 0.465 & 0.465 & 0.233 & 0.233 & 0.233 & 0.116 & 0.116 & \vphantom{4} 0.116\\
\hspace{1em}Robust Bonf. SB & 0.483 & 0.761 & 0.969 & 0.269 & 0.392 & 0.411 & 0.135 & 0.200 & 0.211\\
\hspace{1em}Robust Bonf. CB & 0.423 & 0.696 & 0.885 & 0.252 & 0.374 & 0.397 & 0.131 & 0.195 & 0.208\\
\addlinespace[0.3em]
\hline
\multicolumn{10}{c}{\textbf{$\phi = 0.25$}}\\
\hline
\hspace{1em}Classical sup-t SB & 0.465 & 0.724 & 0.810 & 0.233 & 0.362 & 0.405 & 0.116 & 0.181 & \vphantom{3} 0.202\\
\hspace{1em}Classical sup-t CB & 0.444 & 0.751 & 0.852 & 0.227 & 0.378 & 0.429 & 0.114 & 0.189 & 0.214\\
\hspace{1em}Classical pointw. SB & 0.465 & 0.465 & 0.465 & 0.233 & 0.233 & 0.233 & 0.116 & 0.116 & \vphantom{3} 0.116\\
\hspace{1em}Robust Bonf. SB & 0.501 & 0.762 & 0.956 & 0.282 & 0.393 & 0.412 & 0.142 & 0.201 & 0.211\\
\hspace{1em}Robust Bonf. CB & 0.409 & 0.708 & 0.890 & 0.240 & 0.386 & 0.412 & 0.127 & 0.204 & 0.218\\
\addlinespace[0.3em]
\hline
\multicolumn{10}{c}{\textbf{$\phi = 0.5$}}\\
\hline
\hspace{1em}Classical sup-t SB & 0.465 & 0.724 & 0.810 & 0.233 & 0.362 & 0.405 & 0.116 & 0.181 & \vphantom{2} 0.202\\
\hspace{1em}Classical sup-t CB & 0.414 & 0.795 & 0.919 & 0.208 & 0.410 & 0.478 & 0.103 & 0.208 & 0.243\\
\hspace{1em}Classical pointw. SB & 0.465 & 0.465 & 0.465 & 0.233 & 0.233 & 0.233 & 0.116 & 0.116 & \vphantom{2} 0.116\\
\hspace{1em}Robust Bonf. SB & 0.550 & 0.766 & 0.969 & 0.313 & 0.395 & 0.413 & 0.162 & 0.205 & 0.212\\
\hspace{1em}Robust Bonf. CB & 0.368 & 0.744 & 0.953 & 0.216 & 0.423 & 0.464 & 0.112 & 0.230 & 0.252\\
\addlinespace[0.3em]
\hline
\multicolumn{10}{c}{\textbf{$\phi = 0.75$}}\\
\hline
\hspace{1em}Classical sup-t SB & 0.465 & 0.724 & 0.810 & 0.233 & 0.362 & 0.405 & 0.116 & 0.181 & \vphantom{1} 0.202\\
\hspace{1em}Classical sup-t CB & 0.359 & 0.863 & 1.034 & 0.168 & 0.457 & 0.569 & 0.080 & 0.235 & 0.298\\
\hspace{1em}Classical pointw. SB & 0.465 & 0.465 & 0.465 & 0.233 & 0.233 & 0.233 & 0.116 & 0.116 & \vphantom{1} 0.116\\
\hspace{1em}Robust Bonf. SB & 0.618 & 0.763 & 0.971 & 0.365 & 0.415 & 0.417 & 0.193 & 0.220 & 0.218\\
\hspace{1em}Robust Bonf. CB & 0.306 & 0.794 & 1.026 & 0.167 & 0.486 & 0.561 & 0.086 & 0.277 & 0.326\\
\addlinespace[0.3em]
\hline
\multicolumn{10}{c}{\textbf{$\phi = 0.95$}}\\
\hline
\hspace{1em}Classical sup-t SB & 0.465 & 0.724 & 0.810 & 0.233 & 0.362 & 0.405 & 0.116 & 0.181 & 0.202\\
\hspace{1em}Classical sup-t CB & 0.290 & 0.925 & 1.169 & 0.109 & 0.458 & 0.690 & 0.046 & 0.212 & 0.362\\
\hspace{1em}Classical pointw. SB & 0.465 & 0.465 & 0.465 & 0.233 & 0.233 & 0.233 & 0.116 & 0.116 & 0.116\\
\hspace{1em}Robust Bonf. SB & 0.656 & 0.740 & 0.921 & 0.393 & 0.474 & 0.429 & 0.214 & 0.274 & 0.251\\
\hspace{1em}Robust Bonf. CB & 0.217 & 0.721 & 0.961 & 0.092 & 0.431 & 0.591 & 0.044 & 0.246 & 0.384\\
\bottomrule
\end{tabular}}
\end{table}

\begin{table}[H]
\centering
\caption{Average width of inference bands (significance level $\alpha = 0.1 $) -- averaged over the lag $h$ as well as over simulation runs; time series of length $T$ are generated by an AR(1) process with coefficient $\phi$ and $\chi^2(3)$ GARCH errors, length of the bands is denoted by $H$; the bandwidth for variance estimation is $L = T^{1/2}$ for classical confidence bands and $L = (T-H)^{1/3}$ for robust confidence bands.}
\resizebox{\ifdim\width>\linewidth\linewidth\else\width\fi}{!}{
\begin{tabular}{l|ccc|ccc|cccl|ccc|ccc|cccl|ccc|ccc|cccl|ccc|ccc|cccl|ccc|ccc|cccl|ccc|ccc|cccl|ccc|ccc|cccl|ccc|ccc|cccl|ccc|ccc|cccl|ccc|ccc|ccc}
\toprule
\multicolumn{1}{c}{ } & \multicolumn{3}{c}{$T = 50$} & \multicolumn{3}{c}{$T = 200$} & \multicolumn{3}{c}{$T = 800$} \\
\cmidrule(l{3pt}r{3pt}){2-4} \cmidrule(l{3pt}r{3pt}){5-7} \cmidrule(l{3pt}r{3pt}){8-10}
 & $H = 1$ & $H = 10 $ & $H = 25$ & $H = 1$ & $H = 10 $ & $H = 25$ & $H = 1$ & $H = 10 $ & $H = 25$\\
\midrule
\addlinespace[0.3em]
\hline
\multicolumn{10}{c}{\textbf{$\phi = 0$}}\\
\hline
\hspace{1em}Classical sup-t SB & 0.465 & 0.724 & 0.810 & 0.233 & 0.362 & 0.405 & 0.116 & 0.181 & \vphantom{4} 0.202\\
\hspace{1em}Classical sup-t CB & 0.458 & 0.738 & 0.833 & 0.233 & 0.367 & 0.414 & 0.117 & 0.183 & 0.205\\
\hspace{1em}Classical pointw. SB & 0.465 & 0.465 & 0.465 & 0.233 & 0.233 & 0.233 & 0.116 & 0.116 & \vphantom{4} 0.116\\
\hspace{1em}Robust Bonf. SB & 0.465 & 0.748 & 0.946 & 0.272 & 0.398 & 0.416 & 0.141 & 0.210 & 0.221\\
\hspace{1em}Robust Bonf. CB & 0.401 & 0.675 & 0.855 & 0.246 & 0.376 & 0.399 & 0.135 & 0.203 & 0.216\\
\addlinespace[0.3em]
\hline
\multicolumn{10}{c}{\textbf{$\phi = 0.25$}}\\
\hline
\hspace{1em}Classical sup-t SB & 0.465 & 0.724 & 0.810 & 0.233 & 0.362 & 0.405 & 0.116 & 0.181 & \vphantom{3} 0.202\\
\hspace{1em}Classical sup-t CB & 0.449 & 0.750 & 0.851 & 0.228 & 0.377 & 0.429 & 0.114 & 0.189 & 0.214\\
\hspace{1em}Classical pointw. SB & 0.465 & 0.465 & 0.465 & 0.233 & 0.233 & 0.233 & 0.116 & 0.116 & \vphantom{3} 0.116\\
\hspace{1em}Robust Bonf. SB & 0.483 & 0.748 & 0.955 & 0.286 & 0.399 & 0.421 & 0.154 & 0.212 & 0.221\\
\hspace{1em}Robust Bonf. CB & 0.395 & 0.687 & 0.879 & 0.239 & 0.387 & 0.418 & 0.129 & 0.210 & 0.225\\
\addlinespace[0.3em]
\hline
\multicolumn{10}{c}{\textbf{$\phi = 0.5$}}\\
\hline
\hspace{1em}Classical sup-t SB & 0.465 & 0.724 & 0.810 & 0.233 & 0.362 & 0.405 & 0.116 & 0.181 & \vphantom{2} 0.202\\
\hspace{1em}Classical sup-t CB & 0.415 & 0.796 & 0.921 & 0.206 & 0.409 & 0.477 & 0.103 & 0.208 & 0.243\\
\hspace{1em}Classical pointw. SB & 0.465 & 0.465 & 0.465 & 0.233 & 0.233 & 0.233 & 0.116 & 0.116 & \vphantom{2} 0.116\\
\hspace{1em}Robust Bonf. SB & 0.569 & 0.759 & 0.960 & 0.345 & 0.405 & 0.423 & 0.193 & 0.222 & 0.225\\
\hspace{1em}Robust Bonf. CB & 0.382 & 0.740 & 0.941 & 0.226 & 0.425 & 0.469 & 0.121 & 0.241 & 0.262\\
\addlinespace[0.3em]
\hline
\multicolumn{10}{c}{\textbf{$\phi = 0.75$}}\\
\hline
\hspace{1em}Classical sup-t SB & 0.465 & 0.724 & 0.810 & 0.233 & 0.362 & 0.405 & 0.116 & 0.181 & \vphantom{1} 0.202\\
\hspace{1em}Classical sup-t CB & 0.358 & 0.866 & 1.037 & 0.167 & 0.457 & 0.569 & 0.081 & 0.235 & 0.298\\
\hspace{1em}Classical pointw. SB & 0.465 & 0.465 & 0.465 & 0.233 & 0.233 & 0.233 & 0.116 & 0.116 & \vphantom{1} 0.116\\
\hspace{1em}Robust Bonf. SB & 0.645 & 0.773 & 0.970 & 0.397 & 0.425 & 0.430 & 0.223 & 0.238 & 0.233\\
\hspace{1em}Robust Bonf. CB & 0.335 & 0.807 & 1.030 & 0.186 & 0.495 & 0.577 & 0.097 & 0.287 & 0.339\\
\addlinespace[0.3em]
\hline
\multicolumn{10}{c}{\textbf{$\phi = 0.95$}}\\
\hline
\hspace{1em}Classical sup-t SB & 0.465 & 0.724 & 0.810 & 0.233 & 0.362 & 0.405 & 0.116 & 0.181 & 0.202\\
\hspace{1em}Classical sup-t CB & 0.290 & 0.924 & 1.168 & 0.110 & 0.459 & 0.689 & 0.046 & 0.213 & 0.362\\
\hspace{1em}Classical pointw. SB & 0.465 & 0.465 & 0.465 & 0.233 & 0.233 & 0.233 & 0.116 & 0.116 & 0.116\\
\hspace{1em}Robust Bonf. SB & 0.667 & 0.743 & 0.930 & 0.404 & 0.481 & 0.436 & 0.231 & 0.293 & 0.263\\
\hspace{1em}Robust Bonf. CB & 0.233 & 0.740 & 0.979 & 0.104 & 0.457 & 0.607 & 0.051 & 0.280 & 0.413\\
\bottomrule
\end{tabular}}
\end{table}

\begin{table}[H]
\centering
			\caption{Average width of confidence bands (nominal coverage: $1 - \alpha = 0.9 $) -- averaged over the lag $h$ as well as over simulation runs; time series of length $T$ are generated by an AR(1) process with coefficient $\phi$ and iid normal errors, length of the bands is denoted by $H$; the bandwidth for variance estimation is $L = T^{1/2}$ for classical bands and $L = (T-H)^{1/3}$ for robust bands. NA: For 0.1\% of Monte Carlo runs, the equicoordinate quantile could not be computed due to non-positive semi-definiteness of the estimated covariance matrix.}
\resizebox{\ifdim\width>\linewidth\linewidth\else\width\fi}{!}{
\begin{tabular}{l|ccc|ccc|cccl|ccc|ccc|cccl|ccc|ccc|cccl|ccc|ccc|cccl|ccc|ccc|cccl|ccc|ccc|cccl|ccc|ccc|cccl|ccc|ccc|cccl|ccc|ccc|cccl|ccc|ccc|ccc}
\toprule
\multicolumn{1}{c}{ } & \multicolumn{3}{c}{$T = 50$} & \multicolumn{3}{c}{$T = 200$} & \multicolumn{3}{c}{$T = 800$} \\
\cmidrule(l{3pt}r{3pt}){2-4} \cmidrule(l{3pt}r{3pt}){5-7} \cmidrule(l{3pt}r{3pt}){8-10}
 & $H = 1$ & $H = 10 $ & $H = 25$ & $H = 1$ & $H = 10 $ & $H = 25$ & $H = 1$ & $H = 10 $ & $H = 25$\\
\midrule
\addlinespace[0.3em]
\hline
\multicolumn{10}{c}{\textbf{$\phi = 0$}}\\
\hline
\hspace{1em}Classical sup-t CB & 0.455 & 0.736 & 0.830 & 0.234 & 0.366 & 0.411 & 0.117 & 0.182 & 0.204\\
\hspace{1em}Classical Bonf. CB & 0.455 & 0.744 & 0.837 & 0.234 & 0.369 & 0.414 & 0.117 & 0.184 & 0.205\\
\hspace{1em}Classical pointw. CB & 0.455 & 0.475 & 0.479 & 0.234 & 0.236 & 0.237 & 0.117 & 0.117 & 0.117\\
\hspace{1em}Robust sup-t CB & 0.407 & 0.703 & 0.958 & 0.224 & 0.355 & 0.413 & 0.115 & 0.179 & 0.203\\
\hspace{1em}Robust Bonf. CB & 0.407 & 0.723 & 1.003 & 0.224 & 0.360 & 0.419 & 0.115 & 0.181 & 0.204\\
\hspace{1em}Robust pointw. CB & 0.407 & 0.462 & 0.573 & 0.224 & 0.230 & 0.240 & 0.115 & 0.116 & 0.117\\
\addlinespace[0.3em]
\hline
\multicolumn{10}{c}{\textbf{$\phi = 0.25$}}\\
\hline
\hspace{1em}Classical sup-t CB & 0.446 & 0.749 & 0.850 & 0.227 & 0.377 & 0.427 & 0.114 & 0.189 & 0.214\\
\hspace{1em}Classical Bonf. CB & 0.446 & 0.761 & 0.860 & 0.227 & 0.383 & 0.432 & 0.114 & 0.192 & 0.216\\
\hspace{1em}Classical pointw. CB & 0.446 & 0.486 & 0.491 & 0.227 & 0.245 & 0.247 & 0.114 & 0.123 & 0.124\\
\hspace{1em}Robust sup-t CB & 0.396 & 0.711 & 0.964 & 0.215 & 0.364 & 0.426 & 0.112 & 0.186 & 0.211\\
\hspace{1em}Robust Bonf. CB & 0.396 & 0.735 & 1.013 & 0.215 & 0.372 & 0.435 & 0.112 & 0.189 & 0.215\\
\hspace{1em}Robust pointw. CB & 0.396 & 0.469 & 0.579 & 0.215 & 0.238 & 0.249 & 0.112 & 0.121 & 0.123\\
\addlinespace[0.3em]
\hline
\multicolumn{10}{c}{\textbf{$\phi = 0.5$}}\\
\hline
\hspace{1em}Classical sup-t CB & 0.415 & 0.793 & 0.917 & 0.207 & 0.409 & 0.477 & 0.102 & 0.208 & 0.243\\
\hspace{1em}Classical Bonf. CB & 0.415 & 0.824 & 0.943 & 0.207 & 0.430 & 0.496 & 0.102 & 0.220 & 0.253\\
\hspace{1em}Classical pointw. CB & 0.415 & 0.526 & 0.539 & 0.207 & 0.275 & 0.283 & 0.102 & 0.140 & 0.145\\
\hspace{1em}Robust sup-t CB & 0.359 & 0.734 & NA & 0.195 & 0.393 & 0.473 & 0.099 & 0.203 & 0.238\\
\hspace{1em}Robust Bonf. CB & 0.359 & 0.772 & 1.067 & 0.195 & 0.414 & 0.494 & 0.099 & 0.214 & 0.249\\
\hspace{1em}Robust pointw. CB & 0.359 & 0.493 & 0.610 & 0.195 & 0.264 & 0.283 & 0.099 & 0.137 & 0.142\\
\addlinespace[0.3em]
\hline
\multicolumn{10}{c}{\textbf{$\phi = 0.75$}}\\
\hline
\hspace{1em}Classical sup-t CB & 0.359 & 0.862 & 1.032 & 0.168 & 0.457 & 0.569 & 0.081 & 0.235 & 0.298\\
\hspace{1em}Classical Bonf. CB & 0.359 & 0.933 & 1.099 & 0.168 & 0.516 & 0.630 & 0.081 & 0.271 & 0.337\\
\hspace{1em}Classical pointw. CB & 0.359 & 0.596 & 0.628 & 0.168 & 0.329 & 0.360 & 0.081 & 0.173 & 0.193\\
\hspace{1em}Robust sup-t CB & 0.299 & 0.759 & 1.045 & 0.151 & 0.424 & 0.540 & 0.076 & 0.225 & 0.287\\
\hspace{1em}Robust Bonf. CB & 0.299 & 0.827 & 1.138 & 0.151 & 0.475 & 0.596 & 0.076 & 0.258 & 0.323\\
\hspace{1em}Robust pointw. CB & 0.299 & 0.528 & 0.650 & 0.151 & 0.303 & 0.341 & 0.076 & 0.165 & 0.184\\
\addlinespace[0.3em]
\hline
\multicolumn{10}{c}{\textbf{$\phi = 0.95$}}\\
\hline
\hspace{1em}Classical sup-t CB & 0.293 & 0.923 & 1.163 & 0.109 & 0.459 & 0.689 & 0.046 & 0.212 & 0.362\\
\hspace{1em}Classical Bonf. CB & 0.293 & 1.047 & 1.294 & 0.109 & 0.565 & 0.840 & 0.046 & 0.275 & 0.472\\
\hspace{1em}Classical pointw. CB & 0.293 & 0.669 & 0.739 & 0.109 & 0.361 & 0.480 & 0.046 & 0.176 & 0.270\\
\hspace{1em}Robust sup-t CB & 0.211 & 0.673 & 0.939 & 0.082 & 0.353 & 0.532 & 0.037 & 0.180 & 0.301\\
\hspace{1em}Robust Bonf. CB & 0.211 & 0.770 & 1.061 & 0.082 & 0.428 & 0.644 & 0.037 & 0.227 & 0.383\\
\hspace{1em}Robust pointw. CB & 0.211 & 0.492 & 0.607 & 0.082 & 0.273 & 0.368 & 0.037 & 0.145 & 0.219\\
\bottomrule
\end{tabular}}
\end{table}

\begin{table}[H]
\centering
\caption{Average width of confidence bands (nominal coverage: $1 - \alpha = 0.9 $) -- averaged over the lag $h$ as well as over simulation runs; time series of length $T$ are generated by an AR(1) process with coefficient $\phi$ and normal GARCH errors, length of the bands is denoted by $H$; the bandwidth for variance estimation is $L = T^{1/2}$ for classical bands and $L = (T-H)^{1/3}$ for robust bands. NA: For 0.1\% of Monte Carlo runs, the equicoordinate quantile could not be computed due to non-positive semi-definiteness of the estimated covariance matrix.}
\resizebox{\ifdim\width>\linewidth\linewidth\else\width\fi}{!}{
\begin{tabular}{l|ccc|ccc|cccl|ccc|ccc|cccl|ccc|ccc|cccl|ccc|ccc|cccl|ccc|ccc|cccl|ccc|ccc|cccl|ccc|ccc|cccl|ccc|ccc|cccl|ccc|ccc|cccl|ccc|ccc|ccc}
\toprule
\multicolumn{1}{c}{ } & \multicolumn{3}{c}{$T = 50$} & \multicolumn{3}{c}{$T = 200$} & \multicolumn{3}{c}{$T = 800$} \\
\cmidrule(l{3pt}r{3pt}){2-4} \cmidrule(l{3pt}r{3pt}){5-7} \cmidrule(l{3pt}r{3pt}){8-10}
 & $H = 1$ & $H = 10 $ & $H = 25$ & $H = 1$ & $H = 10 $ & $H = 25$ & $H = 1$ & $H = 10 $ & $H = 25$\\
\midrule
\addlinespace[0.3em]
\hline
\multicolumn{10}{c}{\textbf{$\phi = 0$}}\\
\hline
\hspace{1em}Classical sup-t CB & 0.454 & 0.737 & 0.833 & 0.235 & 0.367 & 0.413 & 0.117 & 0.183 & 0.205\\
\hspace{1em}Classical Bonf. CB & 0.454 & 0.746 & 0.840 & 0.235 & 0.371 & 0.416 & 0.117 & 0.184 & 0.206\\
\hspace{1em}Classical pointw. CB & 0.454 & 0.476 & 0.480 & 0.235 & 0.237 & 0.238 & 0.117 & 0.118 & 0.118\\
\hspace{1em}Robust sup-t CB & 0.423 & 0.676 & 0.844 & 0.252 & 0.368 & 0.390 & 0.131 & 0.193 & 0.206\\
\hspace{1em}Robust Bonf. CB & 0.423 & 0.696 & 0.885 & 0.252 & 0.374 & 0.397 & 0.131 & 0.195 & 0.208\\
\hspace{1em}Robust pointw. CB & 0.423 & 0.445 & 0.506 & 0.252 & 0.239 & 0.227 & 0.131 & 0.125 & 0.119\\
\addlinespace[0.3em]
\hline
\multicolumn{10}{c}{\textbf{$\phi = 0.25$}}\\
\hline
\hspace{1em}Classical sup-t CB & 0.444 & 0.751 & 0.852 & 0.227 & 0.378 & 0.429 & 0.114 & 0.189 & 0.214\\
\hspace{1em}Classical Bonf. CB & 0.444 & 0.763 & 0.862 & 0.227 & 0.385 & 0.435 & 0.114 & 0.193 & 0.217\\
\hspace{1em}Classical pointw. CB & 0.444 & 0.487 & 0.493 & 0.227 & 0.246 & 0.249 & 0.114 & 0.123 & 0.124\\
\hspace{1em}Robust sup-t CB & 0.409 & 0.683 & 0.846 & 0.240 & 0.376 & 0.402 & 0.127 & 0.199 & 0.214\\
\hspace{1em}Robust Bonf. CB & 0.409 & 0.708 & 0.890 & 0.240 & 0.386 & 0.412 & 0.127 & 0.204 & 0.218\\
\hspace{1em}Robust pointw. CB & 0.409 & 0.452 & 0.508 & 0.240 & 0.247 & 0.236 & 0.127 & 0.130 & 0.124\\
\addlinespace[0.3em]
\hline
\multicolumn{10}{c}{\textbf{$\phi = 0.5$}}\\
\hline
\hspace{1em}Classical sup-t CB & 0.414 & 0.795 & 0.919 & 0.208 & 0.410 & 0.478 & 0.103 & 0.208 & 0.243\\
\hspace{1em}Classical Bonf. CB & 0.414 & 0.825 & 0.945 & 0.208 & 0.431 & 0.497 & 0.103 & 0.220 & 0.254\\
\hspace{1em}Classical pointw. CB & 0.414 & 0.527 & 0.540 & 0.208 & 0.275 & 0.284 & 0.103 & 0.141 & 0.145\\
\hspace{1em}Robust sup-t CB & 0.368 & 0.706 & 0.894 & 0.216 & 0.401 & 0.443 & 0.112 & 0.217 & 0.240\\
\hspace{1em}Robust Bonf. CB & 0.368 & 0.744 & 0.953 & 0.216 & 0.423 & 0.464 & 0.112 & 0.230 & 0.252\\
\hspace{1em}Robust pointw. CB & 0.368 & 0.475 & 0.545 & 0.216 & 0.270 & 0.265 & 0.112 & 0.147 & 0.144\\
\addlinespace[0.3em]
\hline
\multicolumn{10}{c}{\textbf{$\phi = 0.75$}}\\
\hline
\hspace{1em}Classical sup-t CB & 0.359 & 0.863 & 1.034 & 0.168 & 0.457 & 0.569 & 0.080 & 0.235 & 0.298\\
\hspace{1em}Classical Bonf. CB & 0.359 & 0.935 & 1.102 & 0.168 & 0.515 & 0.630 & 0.080 & 0.271 & 0.337\\
\hspace{1em}Classical pointw. CB & 0.359 & 0.597 & 0.630 & 0.168 & 0.329 & 0.360 & 0.080 & 0.173 & 0.193\\
\hspace{1em}Robust sup-t CB & 0.306 & 0.727 & NA & 0.167 & 0.433 & 0.507 & 0.086 & 0.242 & 0.290\\
\hspace{1em}Robust Bonf. CB & 0.306 & 0.794 & 1.026 & 0.167 & 0.486 & 0.561 & 0.086 & 0.277 & 0.326\\
\hspace{1em}Robust pointw. CB & 0.306 & 0.507 & 0.586 & 0.167 & 0.310 & 0.320 & 0.086 & 0.177 & 0.186\\
\addlinespace[0.3em]
\hline
\multicolumn{10}{c}{\textbf{$\phi = 0.95$}}\\
\hline
\hspace{1em}Classical sup-t CB & 0.290 & 0.925 & 1.169 & 0.109 & 0.458 & 0.690 & 0.046 & 0.212 & 0.362\\
\hspace{1em}Classical Bonf. CB & 0.290 & 1.051 & 1.302 & 0.109 & 0.564 & 0.841 & 0.046 & 0.275 & 0.472\\
\hspace{1em}Classical pointw. CB & 0.290 & 0.671 & 0.744 & 0.109 & 0.360 & 0.481 & 0.046 & 0.176 & 0.270\\
\hspace{1em}Robust sup-t CB & 0.217 & 0.628 & 0.845 & 0.092 & 0.354 & 0.487 & 0.044 & 0.194 & 0.302\\
\hspace{1em}Robust Bonf. CB & 0.217 & 0.721 & 0.961 & 0.092 & 0.431 & 0.591 & 0.044 & 0.246 & 0.384\\
\hspace{1em}Robust pointw. CB & 0.217 & 0.461 & 0.549 & 0.092 & 0.275 & 0.338 & 0.044 & 0.157 & 0.220\\
\bottomrule
\end{tabular}}
\end{table}

\begin{table}[H]
\centering
\caption{Average width of confidence bands (nominal coverage: $1 - \alpha = 0.9 $) -- averaged over the lag $h$ as well as over simulation runs; time series of length $T$ are generated by an AR(1) process with coefficient $\phi$ and $\chi^2(3)$ GARCH errors, length of the bands is denoted by $H$; the bandwidth for variance estimation is $L = T^{1/2}$ for classical bands and $L = (T-H)^{1/3}$ for robust bands. NA: For 0.1\% of Monte Carlo runs, the equicoordinate quantile could not be computed due to non-positive semi-definiteness of the estimated covariance matrix.}
\resizebox{\ifdim\width>\linewidth\linewidth\else\width\fi}{!}{
\begin{tabular}{l|ccc|ccc|cccl|ccc|ccc|cccl|ccc|ccc|cccl|ccc|ccc|cccl|ccc|ccc|cccl|ccc|ccc|cccl|ccc|ccc|cccl|ccc|ccc|cccl|ccc|ccc|cccl|ccc|ccc|ccc}
\toprule
\multicolumn{1}{c}{ } & \multicolumn{3}{c}{$T = 50$} & \multicolumn{3}{c}{$T = 200$} & \multicolumn{3}{c}{$T = 800$} \\
\cmidrule(l{3pt}r{3pt}){2-4} \cmidrule(l{3pt}r{3pt}){5-7} \cmidrule(l{3pt}r{3pt}){8-10}
 & $H = 1$ & $H = 10 $ & $H = 25$ & $H = 1$ & $H = 10 $ & $H = 25$ & $H = 1$ & $H = 10 $ & $H = 25$\\
\midrule
\addlinespace[0.3em]
\hline
\multicolumn{10}{c}{\textbf{$\phi = 0$}}\\
\hline
\hspace{1em}Classical sup-t CB & 0.458 & 0.738 & 0.833 & 0.233 & 0.367 & 0.414 & 0.117 & 0.183 & 0.205\\
\hspace{1em}Classical Bonf. CB & 0.458 & 0.746 & 0.840 & 0.233 & 0.371 & 0.417 & 0.117 & 0.184 & 0.206\\
\hspace{1em}Classical pointw. CB & 0.458 & 0.476 & 0.480 & 0.233 & 0.237 & 0.238 & 0.117 & 0.118 & 0.118\\
\hspace{1em}Robust sup-t CB & 0.401 & 0.654 & 0.814 & 0.246 & 0.368 & 0.390 & 0.135 & 0.200 & 0.213\\
\hspace{1em}Robust Bonf. CB & 0.401 & 0.675 & 0.855 & 0.246 & 0.376 & 0.399 & 0.135 & 0.203 & 0.216\\
\hspace{1em}Robust pointw. CB & 0.401 & 0.431 & 0.488 & 0.246 & 0.240 & 0.228 & 0.135 & 0.130 & 0.123\\
\addlinespace[0.3em]
\hline
\multicolumn{10}{c}{\textbf{$\phi = 0.25$}}\\
\hline
\hspace{1em}Classical sup-t CB & 0.449 & 0.750 & 0.851 & 0.228 & 0.377 & 0.429 & 0.114 & 0.189 & 0.214\\
\hspace{1em}Classical Bonf. CB & 0.449 & 0.763 & 0.862 & 0.228 & 0.384 & 0.434 & 0.114 & 0.192 & 0.217\\
\hspace{1em}Classical pointw. CB & 0.449 & 0.487 & 0.492 & 0.228 & 0.245 & 0.248 & 0.114 & 0.123 & 0.124\\
\hspace{1em}Robust sup-t CB & 0.395 & 0.663 & 0.834 & 0.239 & 0.376 & 0.407 & 0.129 & 0.205 & 0.221\\
\hspace{1em}Robust Bonf. CB & 0.395 & 0.687 & 0.879 & 0.239 & 0.387 & 0.418 & 0.129 & 0.210 & 0.225\\
\hspace{1em}Robust pointw. CB & 0.395 & 0.439 & 0.502 & 0.239 & 0.247 & 0.239 & 0.129 & 0.134 & 0.129\\
\addlinespace[0.3em]
\hline
\multicolumn{10}{c}{\textbf{$\phi = 0.5$}}\\
\hline
\hspace{1em}Classical sup-t CB & 0.415 & 0.796 & 0.921 & 0.206 & 0.409 & 0.477 & 0.103 & 0.208 & 0.243\\
\hspace{1em}Classical Bonf. CB & 0.415 & 0.828 & 0.948 & 0.206 & 0.429 & 0.496 & 0.103 & 0.220 & 0.254\\
\hspace{1em}Classical pointw. CB & 0.415 & 0.529 & 0.542 & 0.206 & 0.274 & 0.283 & 0.103 & 0.140 & 0.145\\
\hspace{1em}Robust sup-t CB & 0.382 & 0.700 & 0.883 & 0.226 & 0.402 & 0.447 & 0.121 & 0.227 & 0.250\\
\hspace{1em}Robust Bonf. CB & 0.382 & 0.740 & 0.941 & 0.226 & 0.425 & 0.469 & 0.121 & 0.241 & 0.262\\
\hspace{1em}Robust pointw. CB & 0.382 & 0.472 & 0.538 & 0.226 & 0.271 & 0.268 & 0.121 & 0.154 & 0.150\\
\addlinespace[0.3em]
\hline
\multicolumn{10}{c}{\textbf{$\phi = 0.75$}}\\
\hline
\hspace{1em}Classical sup-t CB & 0.358 & 0.866 & 1.037 & 0.167 & 0.457 & 0.569 & 0.081 & 0.235 & 0.298\\
\hspace{1em}Classical Bonf. CB & 0.358 & 0.938 & 1.107 & 0.167 & 0.515 & 0.630 & 0.081 & 0.271 & 0.337\\
\hspace{1em}Classical pointw. CB & 0.358 & 0.599 & 0.633 & 0.167 & 0.329 & 0.360 & 0.081 & 0.173 & 0.193\\
\hspace{1em}Robust sup-t CB & 0.335 & 0.736 & 0.940 & 0.186 & 0.442 & 0.521 & 0.097 & 0.252 & 0.301\\
\hspace{1em}Robust Bonf. CB & 0.335 & 0.807 & 1.030 & 0.186 & 0.495 & 0.577 & 0.097 & 0.287 & 0.339\\
\hspace{1em}Robust pointw. CB & 0.335 & 0.516 & 0.589 & 0.186 & 0.316 & 0.330 & 0.097 & 0.184 & 0.194\\
\addlinespace[0.3em]
\hline
\multicolumn{10}{c}{\textbf{$\phi = 0.95$}}\\
\hline
\hspace{1em}Classical sup-t CB & 0.290 & 0.924 & 1.168 & 0.110 & 0.459 & 0.689 & 0.046 & 0.213 & 0.362\\
\hspace{1em}Classical Bonf. CB & 0.290 & 1.050 & 1.300 & 0.110 & 0.565 & 0.840 & 0.046 & 0.276 & 0.471\\
\hspace{1em}Classical pointw. CB & 0.290 & 0.671 & 0.743 & 0.110 & 0.361 & 0.480 & 0.046 & 0.176 & 0.269\\
\hspace{1em}Robust sup-t CB & 0.233 & 0.643 & NA & 0.104 & 0.376 & 0.500 & 0.051 & 0.221 & 0.325\\
\hspace{1em}Robust Bonf. CB & 0.233 & 0.740 & 0.979 & 0.104 & 0.457 & 0.607 & 0.051 & 0.280 & 0.413\\
\hspace{1em}Robust pointw. CB & 0.233 & 0.472 & 0.560 & 0.104 & 0.292 & 0.347 & 0.051 & 0.179 & 0.236\\
\bottomrule
\end{tabular}}
\end{table}

\section{Detailed Simulation Results for Dynamic Regression Residuals}
\label{Sect:sims_residuals}

\subsection{Testing White Noise under IID Errors}

\begin{table}[H]
\centering
\caption{Relative frequency of rejections for the null hypothesis of white noise of the errors in a dynamic regression for inference bands and hypothesis tests (significance level $\alpha = 0.1 $); time series of length $T$ are generated by an AR(2) process with coefficients $\phi_1=0.5$ and $\phi_2$ and iid normal errors, an AR(1) regression is performed; length of the bands is denoted by $H$.}
\resizebox{\ifdim\width>\linewidth\linewidth\else\width\fi}{!}{
\begin{tabular}{l|ccc|ccc|cccl|ccc|ccc|cccl|ccc|ccc|cccl|ccc|ccc|cccl|ccc|ccc|cccl|ccc|ccc|cccl|ccc|ccc|cccl|ccc|ccc|cccl|ccc|ccc|cccl|ccc|ccc|ccc}
\toprule
\multicolumn{1}{c}{ } & \multicolumn{3}{c}{$T = 50$} & \multicolumn{3}{c}{$T = 200$} & \multicolumn{3}{c}{$T = 800$} \\
\cmidrule(l{3pt}r{3pt}){2-4} \cmidrule(l{3pt}r{3pt}){5-7} \cmidrule(l{3pt}r{3pt}){8-10}
 & $H = 1$ & $H = 10 $ & $H = 25$ & $H = 1$ & $H = 10 $ & $H = 25$ & $H = 1$ & $H = 10 $ & $H = 25$\\
\midrule
\addlinespace[0.3em]
\hline
\multicolumn{10}{c}{\textbf{size: $\phi_2 = 0$}}\\
\hline
\hspace{1em}Corrected classical SB & 0.106 & 0.049 & 0.025 & 0.099 & 0.090 & 0.072 & 0.101 & 0.110 & 0.095\\
\hspace{1em}Corrected robust SB & 0.093 & 0.036 & 0.027 & 0.127 & 0.110 & 0.057 & 0.140 & 0.122 & 0.106\\
\hspace{1em}Naive classical sup-t SB & 0.001 & 0.037 & 0.018 & 0.002 & 0.074 & 0.064 & 0.001 & 0.095 & 0.087\\
\hspace{1em}Naive classical pointw. SB & 0.001 & 0.464 & 0.671 & 0.002 & 0.593 & 0.890 & 0.001 & 0.604 & 0.931\\
\hspace{1em}Ljung-Box & NA & 0.108 & 0.132 & NA & 0.109 & 0.120 & NA & 0.118 & 0.109\\
\hspace{1em}Breusch-Godfrey & 0.131 & 0.068 & 0.000 & 0.100 & 0.111 & 0.068 & 0.103 & 0.110 & 0.099\\
\hspace{1em}Robust Box-Pierce & 0.093 & 0.029 & 0.917 & 0.127 & 0.102 & 0.021 & 0.140 & 0.143 & 0.101\\
\addlinespace[0.3em]
\hline
\multicolumn{10}{c}{\textbf{power: $\phi_2 = 0.125$}}\\
\hline
\hspace{1em}Corrected classical SB & 0.136 & 0.049 & 0.027 & 0.459 & 0.240 & 0.151 & 0.962 & 0.838 & 0.741\\
\hspace{1em}Corrected robust SB & 0.129 & 0.050 & 0.025 & 0.449 & 0.271 & 0.174 & 0.960 & 0.834 & 0.733\\
\hspace{1em}Naive classical sup-t SB & 0.011 & 0.037 & 0.020 & 0.084 & 0.140 & 0.075 & 0.688 & 0.522 & 0.406\\
\hspace{1em}Naive classical pointw. SB & 0.011 & 0.476 & 0.673 & 0.084 & 0.694 & 0.919 & 0.688 & 0.951 & 0.988\\
\hspace{1em}Ljung-Box & NA & 0.134 & 0.146 & NA & 0.241 & 0.186 & NA & 0.740 & 0.569\\
\hspace{1em}Breusch-Godfrey & 0.151 & 0.056 & 0.000 & 0.470 & 0.174 & 0.095 & 0.966 & 0.714 & 0.519\\
\hspace{1em}Robust Box-Pierce & 0.129 & 0.029 & 0.920 & 0.449 & 0.199 & 0.029 & 0.960 & 0.688 & 0.459\\
\addlinespace[0.3em]
\hline
\multicolumn{10}{c}{\textbf{power: $\phi_2 = 0.25$}}\\
\hline
\hspace{1em}Corrected classical SB & 0.352 & 0.142 & 0.070 & 0.950 & 0.801 & 0.718 & 1.000 & 1.000 & 1.000\\
\hspace{1em}Corrected robust SB & 0.296 & 0.121 & 0.067 & 0.940 & 0.770 & 0.604 & 1.000 & 1.000 & 1.000\\
\hspace{1em}Naive classical sup-t SB & 0.080 & 0.080 & 0.039 & 0.750 & 0.526 & 0.396 & 1.000 & 0.999 & 0.998\\
\hspace{1em}Naive classical pointw. SB & 0.080 & 0.570 & 0.742 & 0.750 & 0.951 & 0.989 & 1.000 & 1.000 & 1.000\\
\hspace{1em}Ljung-Box & NA & 0.207 & 0.224 & NA & 0.747 & 0.593 & NA & 1.000 & 1.000\\
\hspace{1em}Breusch-Godfrey & 0.376 & 0.103 & 0.000 & 0.950 & 0.682 & 0.361 & 1.000 & 1.000 & 0.999\\
\hspace{1em}Robust Box-Pierce & 0.296 & 0.060 & 0.926 & 0.940 & 0.499 & 0.120 & 1.000 & 0.871 & 0.836\\
\bottomrule
\end{tabular}}
\label{tab:coverage_dynamic_regression_iid}
\end{table}

\begin{table}[H]
\caption{Average width of significance bands (significance level $\alpha = 0.1 $), averaged over the lag $h$ as well as over simulation runs, for the null hypothesis of white noise of the errors in a dynamic regression; time series of length $T$ are generated by an AR(2) process with coefficients $\phi_1=0.5$ and $\phi_2$ and iid normal errors, an AR(1) regression is performed; length of the bands is denoted by $H$.}
\centering
\resizebox{\ifdim\width>\linewidth\linewidth\else\width\fi}{!}{
\begin{tabular}{l|ccc|ccc|cccl|ccc|ccc|cccl|ccc|ccc|cccl|ccc|ccc|cccl|ccc|ccc|cccl|ccc|ccc|cccl|ccc|ccc|cccl|ccc|ccc|cccl|ccc|ccc|cccl|ccc|ccc|ccc}
\toprule
\multicolumn{1}{c}{ } & \multicolumn{3}{c}{$T = 50$} & \multicolumn{3}{c}{$T = 200$} & \multicolumn{3}{c}{$T = 800$} \\
\cmidrule(l{3pt}r{3pt}){2-4} \cmidrule(l{3pt}r{3pt}){5-7} \cmidrule(l{3pt}r{3pt}){8-10}
 & $H = 1$ & $H = 10 $ & $H = 25$ & $H = 1$ & $H = 10 $ & $H = 25$ & $H = 1$ & $H = 10 $ & $H = 25$\\
\midrule
\addlinespace[0.3em]
\hline
\multicolumn{10}{c}{\textbf{$\phi_2 = 0$}}\\
\hline
\hspace{1em}Corrected classical SB & 0.224 & 0.676 & 0.788 & 0.115 & 0.337 & 0.394 & 0.058 & 0.169 & 0.197\\
\hspace{1em}Corrected robust SB & 0.316 & 0.764 & 1.089 & 0.131 & 0.353 & 0.423 & 0.056 & 0.172 & 0.201\\
\hspace{1em}Naive classical sup-t SB & 0.465 & 0.724 & 0.810 & 0.233 & 0.362 & 0.405 & 0.116 & 0.181 & \vphantom{2} 0.202\\
\hspace{1em}Naive classical pointw. SB & 0.465 & 0.465 & 0.465 & 0.233 & 0.233 & 0.233 & 0.116 & 0.116 & \vphantom{2} 0.116\\
\addlinespace[0.3em]
\hline
\multicolumn{10}{c}{\textbf{$\phi_2 = 0.125$}}\\
\hline
\hspace{1em}Corrected classical SB & 0.249 & 0.678 & 0.789 & 0.130 & 0.339 & 0.395 & 0.066 & 0.170 & 0.197\\
\hspace{1em}Corrected robust SB & 0.310 & 0.760 & 1.091 & 0.133 & 0.353 & 0.423 & 0.065 & 0.173 & 0.202\\
\hspace{1em}Naive classical sup-t SB & 0.465 & 0.724 & 0.810 & 0.233 & 0.362 & 0.405 & 0.116 & 0.181 & \vphantom{1} 0.202\\
\hspace{1em}Naive classical pointw. SB & 0.465 & 0.465 & 0.465 & 0.233 & 0.233 & 0.233 & 0.116 & 0.116 & \vphantom{1} 0.116\\
\addlinespace[0.3em]
\hline
\multicolumn{10}{c}{\textbf{$\phi_2 = 0.25$}}\\
\hline
\hspace{1em}Corrected classical SB & 0.276 & 0.681 & 0.790 & 0.149 & 0.340 & 0.395 & 0.077 & 0.170 & 0.198\\
\hspace{1em}Corrected robust SB & 0.317 & 0.759 & 1.089 & 0.151 & 0.354 & 0.423 & 0.081 & 0.175 & 0.203\\
\hspace{1em}Naive classical sup-t SB & 0.465 & 0.724 & 0.810 & 0.233 & 0.362 & 0.405 & 0.116 & 0.181 & 0.202\\
\hspace{1em}Naive classical pointw. SB & 0.465 & 0.465 & 0.465 & 0.233 & 0.233 & 0.233 & 0.116 & 0.116 & 0.116\\
\bottomrule
\end{tabular}}
\label{tab:width_dynamic_regression_iid}
\end{table}

\begin{table}[H]
\centering
\caption{Fraction of adjusted homoskedastic and heteroskedastic covariance estimators via Algorithm \ref{alg:shrinkage} in a dynamic regression; time series of length $T$ are generated by an AR(2) process with coefficients $\phi_1=0.5$ and $\phi_2$ and iid normal errors, an AR(1) regression is performed; length of the bands is denoted by $H$.}
\resizebox{\ifdim\width>\linewidth\linewidth\else\width\fi}{!}{
\begin{tabular}{l|ccc|ccc|cccl|ccc|ccc|cccl|ccc|ccc|cccl|ccc|ccc|cccl|ccc|ccc|cccl|ccc|ccc|cccl|ccc|ccc|cccl|ccc|ccc|cccl|ccc|ccc|cccl|ccc|ccc|ccc}
\toprule
\multicolumn{1}{c}{ } & \multicolumn{3}{c}{$T = 50$} & \multicolumn{3}{c}{$T = 200$} & \multicolumn{3}{c}{$T = 800$} \\
\cmidrule(l{3pt}r{3pt}){2-4} \cmidrule(l{3pt}r{3pt}){5-7} \cmidrule(l{3pt}r{3pt}){8-10}
 & $H = 1$ & $H = 10 $ & $H = 25$ & $H = 1$ & $H = 10 $ & $H = 25$ & $H = 1$ & $H = 10 $ & $H = 25$\\
\midrule
\addlinespace[0.3em]
\hline
\multicolumn{10}{c}{\textbf{$\phi_2 = 0$}}\\
\hline
\hspace{1em}hom & 0.000 & 0.822 & 0.983 & 0.000 & 0.819 & 0.992 & 0.000 & 0.843 & 0.995\\
\hspace{1em}het & 0.401 & 0.865 & 0.994 & 0.161 & 0.702 & 0.855 & 0.010 & 0.613 & 0.702\\
\addlinespace[0.3em]
\hline
\multicolumn{10}{c}{\textbf{$\phi_2 = 0.125$}}\\
\hline
\hspace{1em}hom & 0.002 & 0.756 & 0.955 & 0.000 & 0.834 & 0.985 & 0.000 & 0.955 & 0.999\\
\hspace{1em}het & 0.311 & 0.848 & 0.999 & 0.083 & 0.707 & 0.867 & 0.002 & 0.637 & 0.738\\
\addlinespace[0.3em]
\hline
\multicolumn{10}{c}{\textbf{$\phi_2 = 0.25$}}\\
\hline
\hspace{1em}hom & 0.002 & 0.696 & 0.922 & 0.000 & 0.875 & 0.987 & 0.000 & 0.999 & 1.000\\
\hspace{1em}het & 0.253 & 0.871 & 0.999 & 0.023 & 0.716 & 0.880 & 0.000 & 0.623 & 0.725\\
\bottomrule
\end{tabular}}
\label{tab:dyn_iid_adjust_frac}
\end{table}

\begin{table}[H]
\centering
\caption{Median lag, at which the homoskedastic and heteroskedastic covariance estimators are adjusted via Algorithm \ref{alg:shrinkage} in a dynamic regression; time series of length $T$ are generated by an AR(2) process with coefficients $\phi_1=0.5$ and $\phi_2$ and iid normal errors, an AR(1) regression is performed; length of the bands is denoted by $H$. NA: no adjustment.}
\resizebox{\ifdim\width>\linewidth\linewidth\else\width\fi}{!}{
\begin{tabular}{l|ccc|ccc|cccl|ccc|ccc|cccl|ccc|ccc|cccl|ccc|ccc|cccl|ccc|ccc|cccl|ccc|ccc|cccl|ccc|ccc|cccl|ccc|ccc|cccl|ccc|ccc|cccl|ccc|ccc|ccc}
\toprule
\multicolumn{1}{c}{ } & \multicolumn{3}{c}{$T = 50$} & \multicolumn{3}{c}{$T = 200$} & \multicolumn{3}{c}{$T = 800$} \\
\cmidrule(l{3pt}r{3pt}){2-4} \cmidrule(l{3pt}r{3pt}){5-7} \cmidrule(l{3pt}r{3pt}){8-10}
 & $H = 1$ & $H = 10 $ & $H = 25$ & $H = 1$ & $H = 10 $ & $H = 25$ & $H = 1$ & $H = 10 $ & $H = 25$\\
\midrule
\addlinespace[0.3em]
\hline
\multicolumn{10}{c}{\textbf{$\phi_2 = 0$}}\\
\hline
\hspace{1em}hom & NA & 4 & 5 & NA & 5 & 6 & NA & 6 & 6\\
\hspace{1em}het & 1 & 2 & 2 & 1 & 2 & 2 & 1 & 2 & 2\\
\addlinespace[0.3em]
\hline
\multicolumn{10}{c}{\textbf{$\phi_2 = 0.125$}}\\
\hline
\hspace{1em}hom & 1 & 5 & 6 & NA & 6 & 6 & NA & 6 & 6\\
\hspace{1em}het & 1 & 2 & 2 & 1 & 2 & 3 & 1 & 3 & 3\\
\addlinespace[0.3em]
\hline
\multicolumn{10}{c}{\textbf{$\phi_2 = 0.25$}}\\
\hline
\hspace{1em}hom & 1 & 5 & 6 & NA & 6 & 7 & NA & 6 & 6\\
\hspace{1em}het & 1 & 2 & 2 & 1 & 3 & 3 & NA & 4 & 4\\
\bottomrule
\end{tabular}}
\label{tab:dyn_iid_adjust_lag}
\end{table}

\subsection{Testing White Noise under Normal GARCH Errors}

\begin{table}[H]
\centering
\caption{Relative frequency of rejections for the null hypothesis of white noise of the errors in a dynamic regression for inference bands and hypothesis tests (significance level $\alpha = 0.1 $); time series of length $T$ are generated by an AR(2) process with coefficients $\phi_1=0.5$ and $\phi_2$ and normal GARCH errors, an AR(1) regression is performed; length of the bands is denoted by $H$.}
\resizebox{\ifdim\width>\linewidth\linewidth\else\width\fi}{!}{
\begin{tabular}{l|ccc|ccc|cccl|ccc|ccc|cccl|ccc|ccc|cccl|ccc|ccc|cccl|ccc|ccc|cccl|ccc|ccc|cccl|ccc|ccc|cccl|ccc|ccc|cccl|ccc|ccc|cccl|ccc|ccc|ccc}
\toprule
\multicolumn{1}{c}{ } & \multicolumn{3}{c}{$T = 50$} & \multicolumn{3}{c}{$T = 200$} & \multicolumn{3}{c}{$T = 800$} \\
\cmidrule(l{3pt}r{3pt}){2-4} \cmidrule(l{3pt}r{3pt}){5-7} \cmidrule(l{3pt}r{3pt}){8-10}
 & $H = 1$ & $H = 10 $ & $H = 25$ & $H = 1$ & $H = 10 $ & $H = 25$ & $H = 1$ & $H = 10 $ & $H = 25$\\
\midrule
\addlinespace[0.3em]
\hline
\multicolumn{10}{c}{\textbf{size: $\phi_2 = 0$}}\\
\hline
\hspace{1em}Corrected classical SB & 0.139 & 0.071 & 0.039 & 0.170 & 0.179 & 0.138 & 0.142 & 0.220 & 0.204\\
\hspace{1em}Corrected robust SB & 0.106 & 0.084 & 0.133 & 0.078 & 0.122 & 0.188 & 0.049 & 0.102 & 0.159\\
\hspace{1em}Naive classical sup-t SB & 0.006 & 0.047 & 0.018 & 0.012 & 0.149 & 0.122 & 0.016 & 0.195 & 0.186\\
\hspace{1em}Naive classical pointw. SB & 0.006 & 0.502 & 0.688 & 0.012 & 0.683 & 0.903 & 0.016 & 0.713 & 0.952\\
\hspace{1em}Ljung-Box & NA & 0.128 & 0.122 & NA & 0.253 & 0.237 & NA & 0.249 & 0.255\\
\hspace{1em}Breusch-Godfrey & 0.160 & 0.103 & 0.000 & 0.174 & 0.243 & 0.161 & 0.151 & 0.243 & 0.227\\
\hspace{1em}Robust Box-Pierce & 0.106 & 0.112 & 0.956 & 0.078 & 0.158 & 0.234 & 0.049 & 0.112 & 0.193\\
\addlinespace[0.3em]
\hline
\multicolumn{10}{c}{\textbf{power: $\phi_2 = 0.125$}}\\
\hline
\hspace{1em}Corrected classical SB & 0.167 & 0.071 & 0.032 & 0.446 & 0.326 & 0.242 & 0.934 & 0.837 & 0.765\\
\hspace{1em}Corrected robust SB & 0.118 & 0.099 & 0.117 & 0.268 & 0.198 & 0.265 & 0.764 & 0.609 & 0.581\\
\hspace{1em}Naive classical sup-t SB & 0.020 & 0.050 & 0.023 & 0.112 & 0.227 & 0.157 & 0.657 & 0.584 & 0.471\\
\hspace{1em}Naive classical pointw. SB & 0.020 & 0.517 & 0.672 & 0.112 & 0.778 & 0.939 & 0.657 & 0.957 & 0.992\\
\hspace{1em}Ljung-Box & NA & 0.149 & 0.137 & NA & 0.352 & 0.320 & NA & 0.780 & 0.661\\
\hspace{1em}Breusch-Godfrey & 0.175 & 0.089 & 0.000 & 0.467 & 0.354 & 0.238 & 0.938 & 0.780 & 0.637\\
\hspace{1em}Robust Box-Pierce & 0.118 & 0.120 & 0.953 & 0.268 & 0.209 & 0.270 & 0.764 & 0.608 & 0.494\\
\addlinespace[0.3em]
\hline
\multicolumn{10}{c}{\textbf{power: $\phi_2 = 0.25$}}\\
\hline
\hspace{1em}Corrected classical SB & 0.363 & 0.171 & 0.100 & 0.925 & 0.805 & 0.715 & 1.000 & 1.000 & 1.000\\
\hspace{1em}Corrected robust SB & 0.271 & 0.176 & 0.159 & 0.779 & 0.676 & 0.614 & 0.986 & 0.988 & 0.998\\
\hspace{1em}Naive classical sup-t SB & 0.105 & 0.093 & 0.054 & 0.712 & 0.573 & 0.450 & 1.000 & 0.996 & 0.995\\
\hspace{1em}Naive classical pointw. SB & 0.105 & 0.608 & 0.749 & 0.712 & 0.954 & 0.987 & 1.000 & 1.000 & 1.000\\
\hspace{1em}Ljung-Box & NA & 0.242 & 0.214 & NA & 0.777 & 0.669 & NA & 1.000 & 0.998\\
\hspace{1em}Breusch-Godfrey & 0.406 & 0.140 & 0.000 & 0.931 & 0.750 & 0.492 & 1.000 & 1.000 & 0.998\\
\hspace{1em}Robust Box-Pierce & 0.271 & 0.151 & 0.948 & 0.779 & 0.516 & 0.350 & 0.986 & 0.970 & 0.942\\
\bottomrule
\end{tabular}}
\label{tab:coverage_dynamic_regression_GARCH}
\end{table}

\begin{table}[H]
\centering
\caption{Average width of significance bands (significance level $\alpha = 0.1 $), averaged over the lag $h$ as well as over simulation runs, for the null hypothesis of white noise of the errors in a dynamic regression; time series of length $T$ are generated by an AR(2) process with coefficients $\phi_1=0.5$ and $\phi_2$ and normal GARCH errors, an AR(1) regression is performed; length of the bands is denoted by $H$.}
\resizebox{\ifdim\width>\linewidth\linewidth\else\width\fi}{!}{
\begin{tabular}{l|ccc|ccc|cccl|ccc|ccc|cccl|ccc|ccc|cccl|ccc|ccc|cccl|ccc|ccc|cccl|ccc|ccc|cccl|ccc|ccc|cccl|ccc|ccc|cccl|ccc|ccc|cccl|ccc|ccc|ccc}
\toprule
\multicolumn{1}{c}{ } & \multicolumn{3}{c}{$T = 50$} & \multicolumn{3}{c}{$T = 200$} & \multicolumn{3}{c}{$T = 800$} \\
\cmidrule(l{3pt}r{3pt}){2-4} \cmidrule(l{3pt}r{3pt}){5-7} \cmidrule(l{3pt}r{3pt}){8-10}
 & $H = 1$ & $H = 10 $ & $H = 25$ & $H = 1$ & $H = 10 $ & $H = 25$ & $H = 1$ & $H = 10 $ & $H = 25$\\
\midrule
\addlinespace[0.3em]
\hline
\multicolumn{10}{c}{\textbf{$\phi_2 = 0$}}\\
\hline
\hspace{1em}Corrected classical SB & 0.229 & 0.677 & 0.789 & 0.116 & 0.338 & 0.394 & 0.058 & 0.169 & 0.197\\
\hspace{1em}Corrected robust SB & 0.335 & 0.741 & 0.963 & 0.195 & 0.378 & 0.408 & 0.103 & 0.194 & 0.208\\
\hspace{1em}Naive classical sup-t SB & 0.465 & 0.724 & 0.810 & 0.233 & 0.362 & 0.405 & 0.116 & 0.181 & \vphantom{2} 0.202\\
\hspace{1em}Naive classical pointw. SB & 0.465 & 0.465 & 0.465 & 0.233 & 0.233 & 0.233 & 0.116 & 0.116 & \vphantom{2} 0.116\\
\addlinespace[0.3em]
\hline
\multicolumn{10}{c}{\textbf{$\phi_2 = 0.125$}}\\
\hline
\hspace{1em}Corrected classical SB & 0.254 & 0.679 & 0.790 & 0.132 & 0.339 & 0.395 & 0.066 & 0.170 & 0.197\\
\hspace{1em}Corrected robust SB & 0.340 & 0.740 & 0.972 & 0.203 & 0.381 & 0.407 & 0.106 & 0.194 & 0.209\\
\hspace{1em}Naive classical sup-t SB & 0.465 & 0.724 & 0.810 & 0.233 & 0.362 & 0.405 & 0.116 & 0.181 & \vphantom{1} 0.202\\
\hspace{1em}Naive classical pointw. SB & 0.465 & 0.465 & 0.465 & 0.233 & 0.233 & 0.233 & 0.116 & 0.116 & \vphantom{1} 0.116\\
\addlinespace[0.3em]
\hline
\multicolumn{10}{c}{\textbf{$\phi_2 = 0.25$}}\\
\hline
\hspace{1em}Corrected classical SB & 0.281 & 0.682 & 0.791 & 0.151 & 0.341 & 0.396 & 0.077 & 0.170 & 0.198\\
\hspace{1em}Corrected robust SB & 0.352 & 0.739 & 0.971 & 0.212 & 0.377 & 0.408 & 0.115 & 0.195 & 0.209\\
\hspace{1em}Naive classical sup-t SB & 0.465 & 0.724 & 0.810 & 0.233 & 0.362 & 0.405 & 0.116 & 0.181 & 0.202\\
\hspace{1em}Naive classical pointw. SB & 0.465 & 0.465 & 0.465 & 0.233 & 0.233 & 0.233 & 0.116 & 0.116 & 0.116\\
\bottomrule
\end{tabular}}
\label{tab:width_coverage_dynamic_regression_iid}
\end{table}

\begin{table}[H]
\centering
\caption{Fraction of adjusted homoskedastic and heteroskedastic covariance estimators via Algorithm \ref{alg:shrinkage} in a dynamic regression; time series of length $T$ are generated by an AR(2) process with coefficients $\phi_1=0.5$ and $\phi_2$ and normal GARCH errors, an AR(1) regression is performed; length of the bands is denoted by $H$.}
\resizebox{\ifdim\width>\linewidth\linewidth\else\width\fi}{!}{
\begin{tabular}{l|ccc|ccc|cccl|ccc|ccc|cccl|ccc|ccc|cccl|ccc|ccc|cccl|ccc|ccc|cccl|ccc|ccc|cccl|ccc|ccc|cccl|ccc|ccc|cccl|ccc|ccc|cccl|ccc|ccc|ccc}
\toprule
\multicolumn{1}{c}{ } & \multicolumn{3}{c}{$T = 50$} & \multicolumn{3}{c}{$T = 200$} & \multicolumn{3}{c}{$T = 800$} \\
\cmidrule(l{3pt}r{3pt}){2-4} \cmidrule(l{3pt}r{3pt}){5-7} \cmidrule(l{3pt}r{3pt}){8-10}
 & $H = 1$ & $H = 10 $ & $H = 25$ & $H = 1$ & $H = 10 $ & $H = 25$ & $H = 1$ & $H = 10 $ & $H = 25$\\
\midrule
\addlinespace[0.3em]
\hline
\multicolumn{10}{c}{\textbf{$\phi_2 = 0$}}\\
\hline
\hspace{1em}hom & 0.000 & 0.770 & 0.944 & 0.000 & 0.740 & 0.945 & 0 & 0.742 & 0.943\\
\hspace{1em}het & 0.272 & 0.842 & 0.982 & 0.044 & 0.364 & 0.640 & 0 & 0.096 & 0.162\\
\addlinespace[0.3em]
\hline
\multicolumn{10}{c}{\textbf{$\phi_2 = 0.125$}}\\
\hline
\hspace{1em}hom & 0.002 & 0.716 & 0.917 & 0.000 & 0.743 & 0.916 & 0 & 0.866 & 0.953\\
\hspace{1em}het & 0.228 & 0.836 & 0.986 & 0.021 & 0.397 & 0.666 & 0 & 0.103 & 0.187\\
\addlinespace[0.3em]
\hline
\multicolumn{10}{c}{\textbf{$\phi_2 = 0.25$}}\\
\hline
\hspace{1em}hom & 0.001 & 0.663 & 0.864 & 0.000 & 0.763 & 0.917 & 0 & 0.918 & 0.961\\
\hspace{1em}het & 0.192 & 0.852 & 0.988 & 0.011 & 0.464 & 0.701 & 0 & 0.114 & 0.216\\
\bottomrule
\end{tabular}}
\label{tab:dyn_garch_adjust_frac}
\end{table}

\begin{table}[H]
\centering
\caption{Median lag, at which the homoskedastic and heteroskedastic covariance estimators are adjusted via Algorithm \ref{alg:shrinkage} in a dynamic regression; time series of length $T$ are generated by an AR(2) process with coefficients $\phi_1=0.5$ and $\phi_2$ and iid normal errors, an AR(1) regression is performed; length of the bands is denoted by $H$. NA: no adjustment.}
\resizebox{\ifdim\width>\linewidth\linewidth\else\width\fi}{!}{
\begin{tabular}{l|ccc|ccc|cccl|ccc|ccc|cccl|ccc|ccc|cccl|ccc|ccc|cccl|ccc|ccc|cccl|ccc|ccc|cccl|ccc|ccc|cccl|ccc|ccc|cccl|ccc|ccc|cccl|ccc|ccc|ccc}
\toprule
\multicolumn{1}{c}{ } & \multicolumn{3}{c}{$T = 50$} & \multicolumn{3}{c}{$T = 200$} & \multicolumn{3}{c}{$T = 800$} \\
\cmidrule(l{3pt}r{3pt}){2-4} \cmidrule(l{3pt}r{3pt}){5-7} \cmidrule(l{3pt}r{3pt}){8-10}
 & $H = 1$ & $H = 10 $ & $H = 25$ & $H = 1$ & $H = 10 $ & $H = 25$ & $H = 1$ & $H = 10 $ & $H = 25$\\
\midrule
\addlinespace[0.3em]
\hline
\multicolumn{10}{c}{\textbf{$\phi_2 = 0$}}\\
\hline
\hspace{1em}hom & NA & 4 & 5 & NA & 5 & 6 & NA & 5 & 6\\
\hspace{1em}het & 1 & 2 & 1 & 1 & 2 & 3 & NA & 3 & 3\\
\addlinespace[0.3em]
\hline
\multicolumn{10}{c}{\textbf{$\phi_2 = 0.125$}}\\
\hline
\hspace{1em}hom & 1 & 5 & 6 & NA & 6 & 6 & NA & 6 & 6\\
\hspace{1em}het & 1 & 2 & 1 & 1 & 3 & 3 & NA & 3 & 5\\
\addlinespace[0.3em]
\hline
\multicolumn{10}{c}{\textbf{$\phi_2 = 0.25$}}\\
\hline
\hspace{1em}hom & 1 & 5 & 6 & NA & 6 & 7 & NA & 7 & 7\\
\hspace{1em}het & 1 & 2 & 1 & 1 & 3 & 3 & NA & 4 & 5\\
\bottomrule
\end{tabular}}
\label{tab:dyn_garch_adjust_lag}
\end{table}

			\section{Additional Material for the Case Studies} \label{appsubsec:additional_material_case_studies}
			
									\begin{figure}[H]
								\noindent \centering{}\includegraphics[scale=0.5]{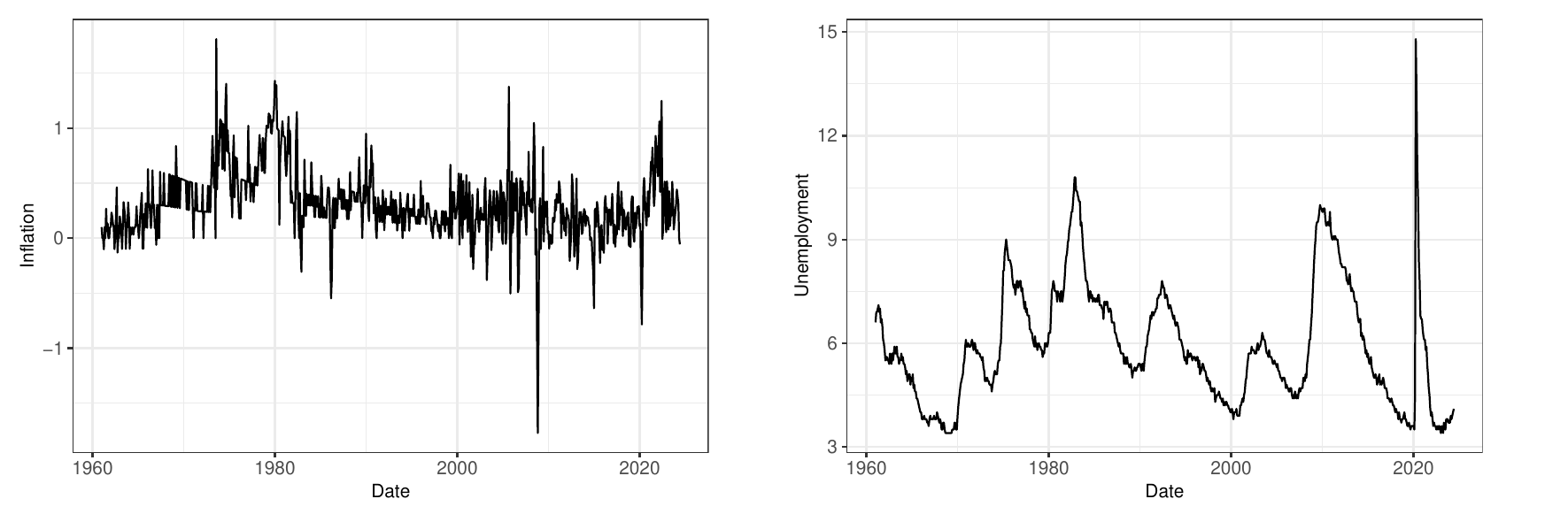}
								\vspace{-0.5cm}
								\caption{Monthly inflation (left) and unemployment rate (right) in percent.}	
								\label{fig:monthyl_inflation_unemployment}
							\end{figure}	

            \begin{figure}[H]
				\noindent \centering{}\includegraphics[scale=0.5]{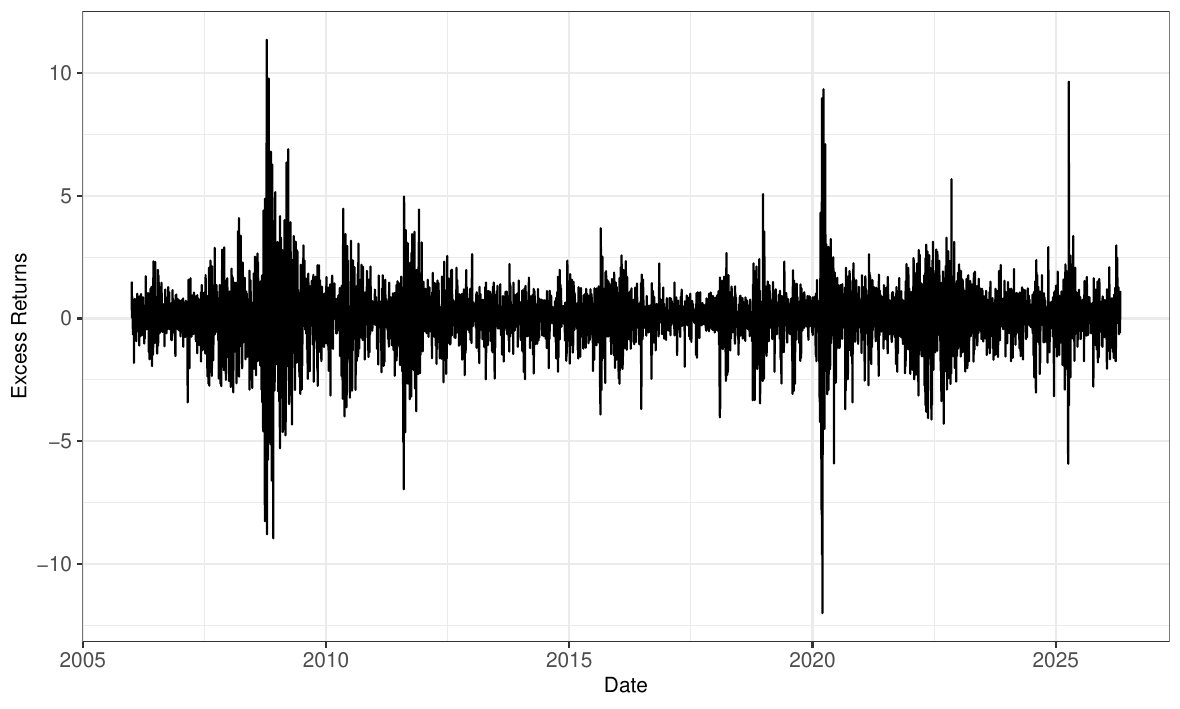}
				\vspace{-0.5cm}
				\caption{Daily Fama-French excess returns in percent.}	
				\label{fig:fama_return_series}
							\end{figure}	
							
																				\begin{figure}[H]
								\noindent \centering{}\includegraphics[scale=0.5]{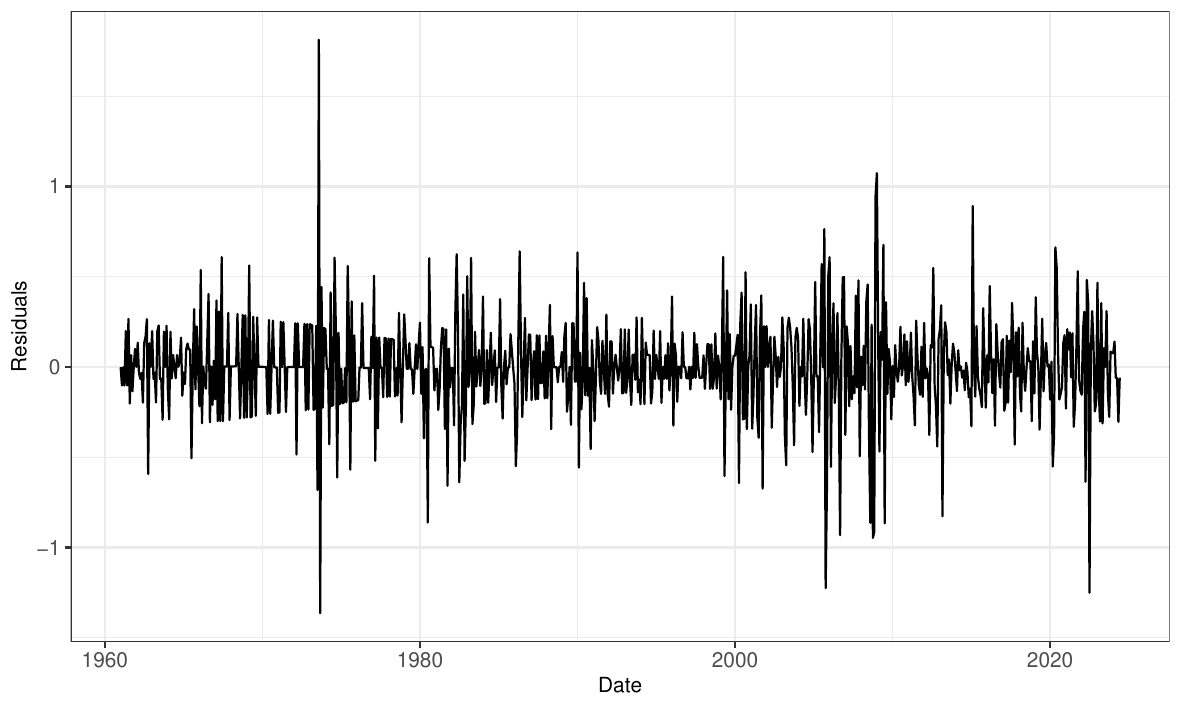}
								\vspace{-0.5cm}
								\caption{Regression residuals from the static Phillips curve estimated by LS.}	
								\label{fig:phillips_res}
							\end{figure}

										\begin{figure}[H]
				\noindent \centering{}\includegraphics[scale=0.5]{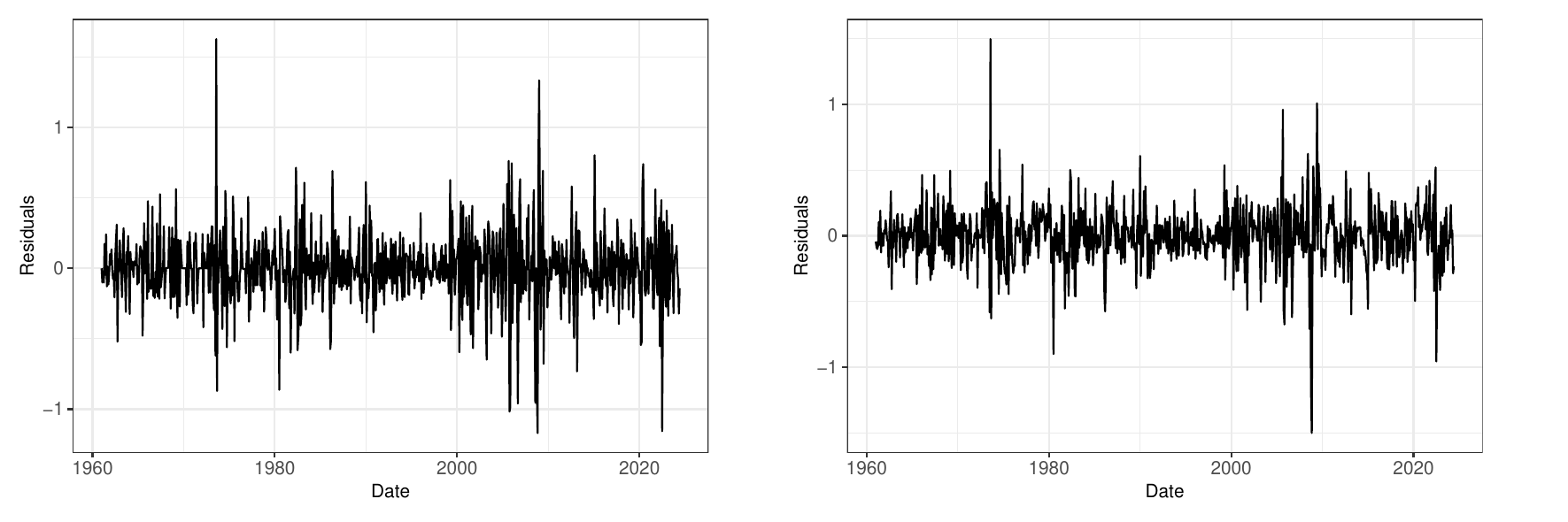}
				\vspace{-0.5cm}
				\caption{Regression residuals from a Phillips curve regression in differences as in Equation \eqref{eq:phillips_dyn_diff}. Left: $p=1$ and $r=0$. Right: $p=r=12$.}	
				\label{fig:phillips_res_dyn_diff}
			\end{figure}	
			
						\begin{figure}[H]
				\noindent \centering{}\includegraphics[scale=0.5]{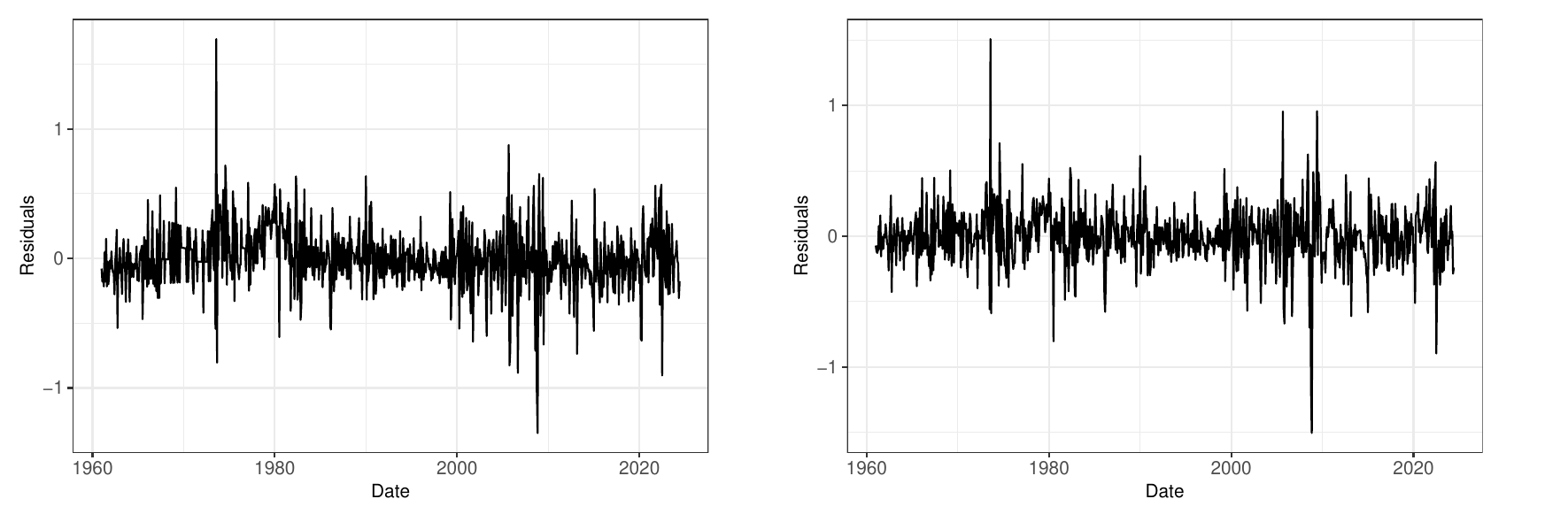}
				\vspace{-0.5cm}
				\caption{Regression residuals from a Phillips curve regression in levels as in Equation \eqref{eq:phillips_dyn_level}. Left: $p=1$ and $r=0$. Right: $p=r=12$.}	
				\label{fig:phillips_res_dyn_level}
			\end{figure}	
			
												\begin{figure}[H]
				\noindent \centering{}\includegraphics[scale=0.5]{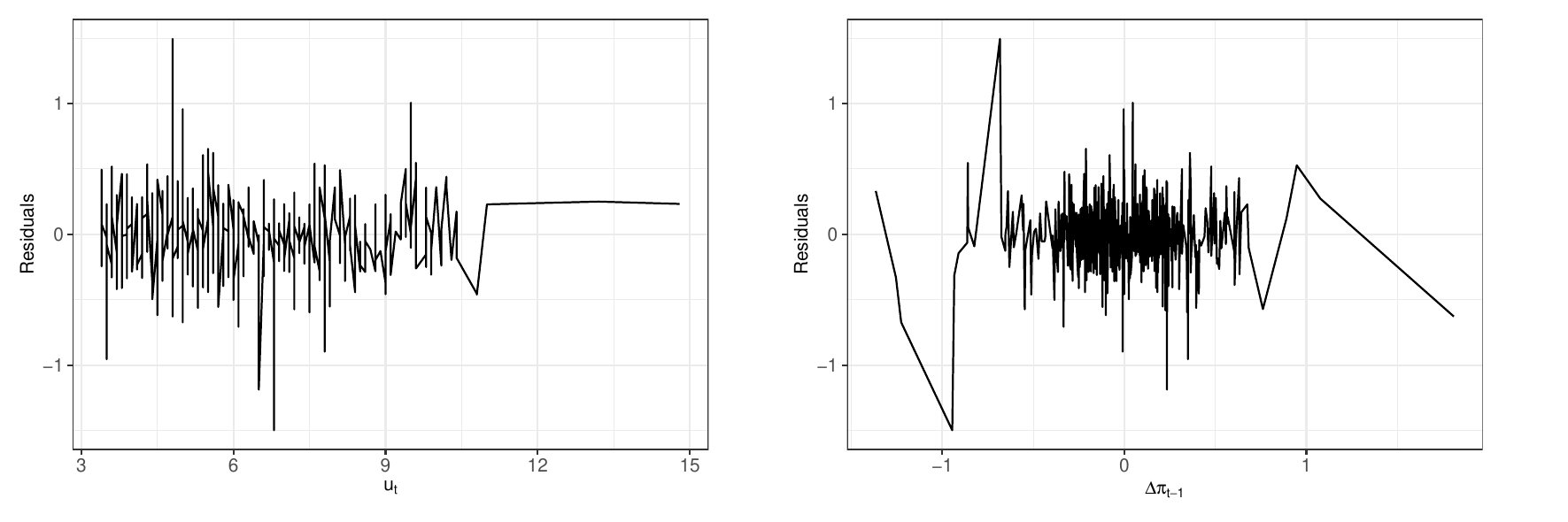}
				\vspace{-0.5cm}
				\caption{Regression residuals from a Phillips curve regression in differences as in Equation \eqref{eq:phillips_dyn_diff} with $p=r=12$ against the most recent regressors, i.\,e., against $u_t$ (left) and $\Delta \pi_{t-1}$ (right).}	
				\label{fig:phillips_res_against_x_diff}
			\end{figure}	
			
\begin{table}[H] \centering 
  \caption{Coefficient estimates from the static Phillips curve regression  and the dynamic regression in differences \eqref{eq:phillips_dyn_diff}. } 
  \label{tab:phillips_estimates_diff} 
\small 
\begin{tabular*}{\textwidth}{c @{\extracolsep{\fill}}lD{.}{.}{-3} D{.}{.}{-3} D{.}{.}{-3} }
\\[-1.8ex]\hline 
\hline \\[-1.8ex] 
 & \multicolumn{3}{c}{Dependent variable: $\Delta \pi_t$} \\ 
\cline{2-4} 
\\[-1.8ex] & \multicolumn{1}{c}{(1)} & \multicolumn{1}{c}{(2)} & \multicolumn{1}{c}{(3)}\\ 
\hline \\[-1.8ex] 
 Intercept & -0.011$ $(0.036) & -0.008$ $(0.034) & 0.011$ $(0.033) \\ 
  $ u_t $ & 0.002$ $(0.006) & 0.001$ $(0.006) & -0.082^{***}$ $(0.020) \\ 
  $u_{t-1}$ &  &  & 0.060^{**}$ $(0.028) \\ 
  $u_{t-2}$ &  &  & 0.018$ $(0.028) \\ 
  $u_{t-3}$ &  &  & -0.010$ $(0.028) \\ 
  $u_{t-4}$ &  &  & 0.003$ $(0.028) \\ 
  $u_{t-5}$ &  &  & -0.015$ $(0.028) \\ 
  $u_{t-6}$ &  &  & 0.010$ $(0.028) \\ 
  $u_{t-7}$ &  &  & 0.008$ $(0.028) \\ 
  $u_{t-8}$ &  &  & 0.026$ $(0.028) \\ 
  $u_{t-9}$ &  &  & -0.038$ $(0.028) \\ 
  $u_{t-10}$ &  &  & 0.034$ $(0.028) \\ 
  $u_{t-11}$ &  &  & -0.015$ $(0.028) \\ 
  $u_{t-12}$ &  &  & -0.0004$ $(0.020) \\ 
  $\Delta \pi_{t-1}$ &  & -0.272^{***}$ $(0.035) & -0.543^{***}$ $(0.037) \\ 
  $\Delta \pi_{t-2}$ &  &  & -0.543^{***}$ $(0.042) \\ 
  $\Delta \pi_{t-3}$ &  &  & -0.507^{***}$ $(0.047) \\ 
  $\Delta \pi_{t-4}$ &  &  & -0.415^{***}$ $(0.050) \\ 
  $\Delta \pi_{t-5}$ &  &  & -0.370^{***}$ $(0.052) \\ 
  $\Delta \pi_{t-6}$ &  &  & -0.334^{***}$ $(0.053) \\ 
  $\Delta \pi_{t-7}$ &  &  & -0.238^{***}$ $(0.053) \\ 
  $\Delta \pi_{t-8}$ &  &  & -0.212^{***}$ $(0.052) \\ 
  $\Delta \pi_{t-9}$ &  &  & -0.124^{**}$ $(0.050) \\ 
  $\Delta \pi_{t-10}$ &  &  & -0.036$ $(0.047) \\ 
  $\Delta \pi_{t-11}$ &  &  & 0.105^{**}$ $(0.042) \\ 
  $\Delta \pi_{t-12}$ &  &  & -0.014$ $(0.037) \\ 
 \hline \\[-1.8ex] 
Observations & \multicolumn{1}{c}{762} & \multicolumn{1}{c}{762} & \multicolumn{1}{c}{762} \\ 
R$^{2}$ & \multicolumn{1}{c}{0.0001} & \multicolumn{1}{c}{0.074} & \multicolumn{1}{c}{0.303} \\ 
Adjusted R$^{2}$ & \multicolumn{1}{c}{-0.001} & \multicolumn{1}{c}{0.072} & \multicolumn{1}{c}{0.280} \\ 
\hline 
\hline \\[-1.8ex] 
\textit{Note:}  & \multicolumn{3}{r}{$^{*}$p$<$0.1; $^{**}$p$<$0.05; $^{***}$p$<$0.01} \\ 
\end{tabular*} 
\end{table} 

\begin{table}[H] \centering 
  \caption{Coefficient estimates from the dynamic Phillips curve regression in levels \eqref{eq:phillips_dyn_level}. } 
  \label{tab:phillips_estimates_level} 
\small 
\begin{tabular}{@{\extracolsep{5pt}}lD{.}{.}{-3} D{.}{.}{-3} } 
\\[-1.8ex]\hline 
\hline \\[-1.8ex] 
 & \multicolumn{2}{c}{Dependent variable: $\pi_t$} \\ 
\cline{2-3} 
\\[-1.8ex] & \multicolumn{1}{c}{(1)} & \multicolumn{1}{c}{(2)}\\ 
\hline \\[-1.8ex] 
 Intercept & 0.116^{***}$ $(0.034) & 0.043$ $(0.035) \\ 
  $ u_t $ & -0.00003$ $(0.005) & -0.077^{***}$ $(0.020) \\ 
  $u_{t-1}$ &  & 0.059^{**}$ $(0.028) \\ 
  $u_{t-2}$ &  & 0.018$ $(0.028) \\ 
  $u_{t-3}$ &  & -0.010$ $(0.028) \\ 
  $u_{t-4}$ &  & 0.003$ $(0.028) \\ 
  $u_{t-5}$ &  & -0.016$ $(0.028) \\ 
  $u_{t-6}$ &  & 0.010$ $(0.028) \\ 
  $u_{t-7}$ &  & 0.007$ $(0.028) \\ 
  $u_{t-8}$ &  & 0.025$ $(0.028) \\ 
  $u_{t-9}$ &  & -0.039$ $(0.028) \\ 
  $u_{t-10}$ &  & 0.033$ $(0.028) \\ 
  $u_{t-11}$ &  & -0.015$ $(0.028) \\ 
  $u_{t-12}$ &  & 0.001$ $(0.020) \\ 
  $ \pi_{t-1}$ & 0.626^{***}$ $(0.028) & 0.444^{***}$ $(0.037) \\ 
  $ \pi_{t-2}$ &  & -0.005$ $(0.040) \\ 
  $ \pi_{t-3}$ &  & 0.029$ $(0.040) \\ 
  $ \pi_{t-4}$ &  & 0.083^{**}$ $(0.040) \\ 
  $ \pi_{t-5}$ &  & 0.036$ $(0.040) \\ 
  $ \pi_{t-6}$ &  & 0.027$ $(0.040) \\ 
  $ \pi_{t-7}$ &  & 0.086^{**}$ $(0.040) \\ 
  $ \pi_{t-8}$ &  & 0.017$ $(0.040) \\ 
  $ \pi_{t-9}$ &  & 0.078^{**}$ $(0.039) \\ 
  $ \pi_{t-10}$ &  & 0.078^{**}$ $(0.040) \\ 
  $ \pi_{t-11}$ &  & 0.132^{***}$ $(0.040) \\ 
  $ \pi_{t-12}$ &  & -0.128^{***}$ $(0.037) \\ 
 \hline \\[-1.8ex] 
Observations & \multicolumn{1}{c}{762} & \multicolumn{1}{c}{762} \\ 
R$^{2}$ & \multicolumn{1}{c}{0.391} & \multicolumn{1}{c}{0.486} \\ 
Adjusted R$^{2}$ & \multicolumn{1}{c}{0.390} & \multicolumn{1}{c}{0.469} \\ 
\hline 
\hline \\[-1.8ex] 
\textit{Note:}  & \multicolumn{2}{r}{$^{*}$p$<$0.1; $^{**}$p$<$0.05; $^{***}$p$<$0.01} \\ 
\end{tabular} 
\end{table}

			
						\begin{figure}
				\noindent \centering{}\includegraphics[scale=0.5]{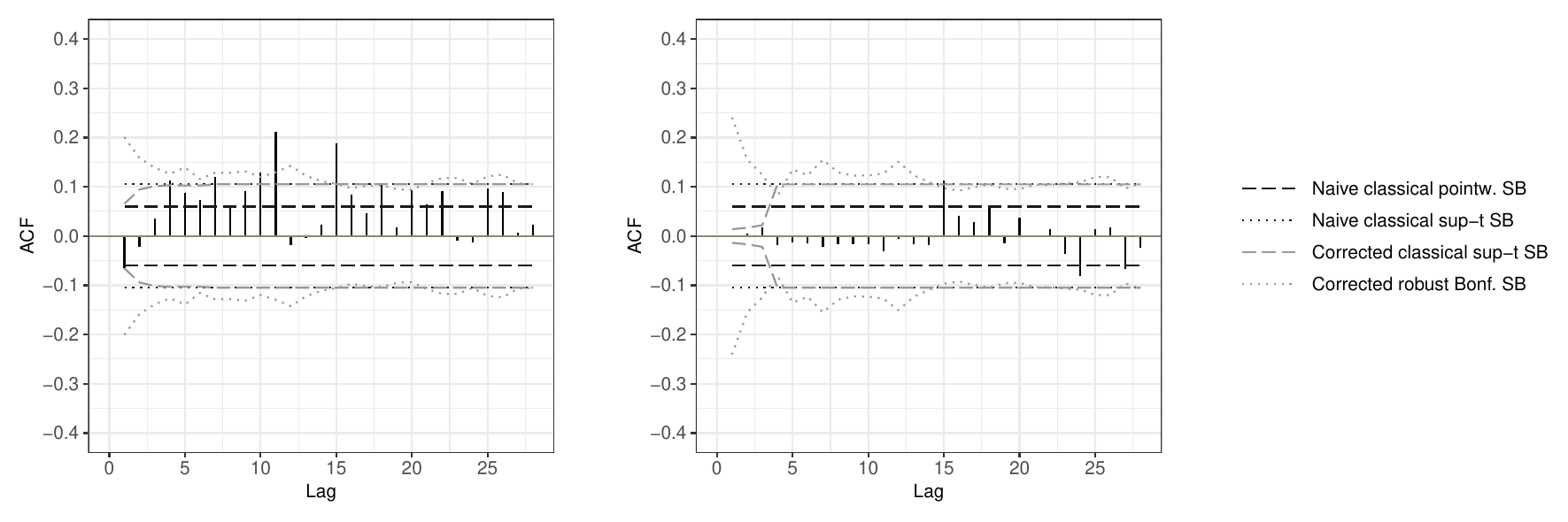}
				\vspace{-0.5cm}
				\caption{Empirical autocorrelations with  $90\%$ significance bands for regression residuals from a Phillips curve regression in levels as in Equation \eqref{eq:phillips_dyn_level}. Left: $p=1$ and $r=0$. Right: $p=r=12$.}	
				\label{fig:phillips_sig_dyn_level}
			\end{figure}	
						
						\begin{figure}[H]
								\noindent \centering{}\includegraphics[scale=0.5]{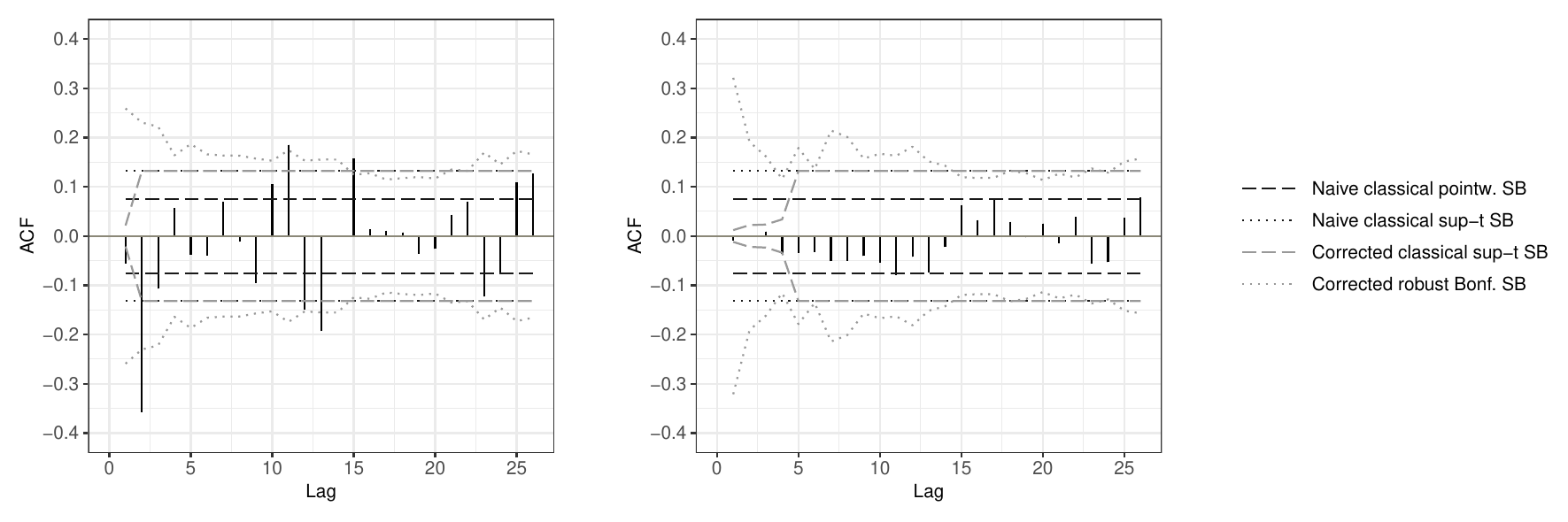}
								\vspace{-0.5cm}
								\caption{Data from 1985 onwards: Empirical autocorrelations with $90\%$ significance bands for regression residuals from a Phillips curve regression in differences as in Equation \eqref{eq:phillips_dyn_diff}. Left: $p=1$ and $r=0$. Right: $p=r=12$.}	
								\label{fig:phillips_sig_dyn_diff_1985}
							\end{figure}	
							
						\begin{figure}[H]
								\noindent \centering{}\includegraphics[scale=0.5]{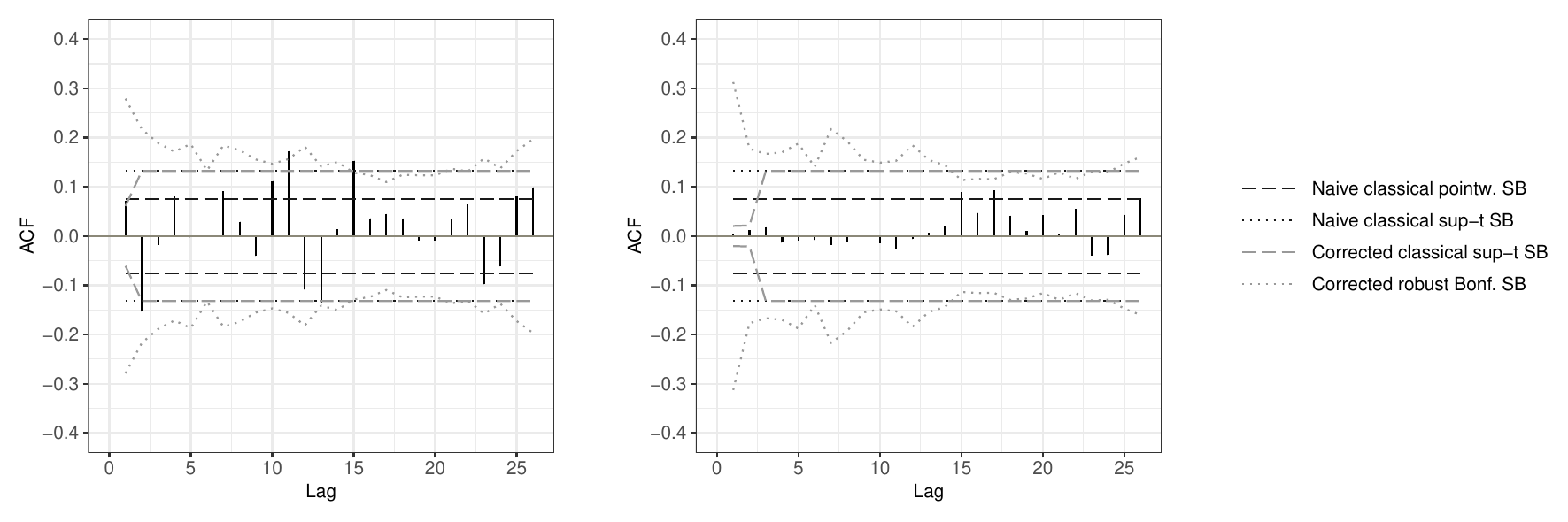}
								\vspace{-0.5cm}
								\caption{Data from 1985 onwards: Empirical autocorrelations with $90\%$ significance bands for regression residuals from a Phillips curve regression in levels as in Equation \eqref{eq:phillips_dyn_level}. Left: $p=1$ and $r=0$. Right: $p=r=12$.}	
								\label{fig:phillips_sig_dyn_level_1985}
							\end{figure}

While the forecasting literature tends to formulate the Phillips curve in terms of differences, the traditional Phillips curve is usually studied with inflation in levels. Thus, Figure~\ref{fig:phillips_sig_dyn_level} shows the autocorrelation function of the residuals with significance bands for the following model, again estimated by LS:
			\begin{equation} \label{eq:phillips_dyn_level}
				\pi_t = a + \sum_{i=1}^{p} \phi_i \pi_{t-i} + \sum_{j=0}^{r} \gamma_j u_{t-j}  + \varepsilon_t \, .
			\end{equation}
			Again, for $p=1$ and $r=0$ on the left-hand side there are significant (on the significance level $10\%$) dynamics in the error, likely invalidating the LS estimates. As before, adding lags ($p=r=12$) reduces the magnitude of the autocorrelation function. However, it slightly exceeds the upper bound of all simultaneous significance bands at lag 15 leading to an overall rejection of the null hypotheses at $10\%$.
			
			Finally, we perform some graphical checks of the assumptions underlying the construction of our inference bands for dynamic regressions (assumptions \ref{Ass:MDS} and \ref{Ass:homoskedastic_errors}) and a robustness check with respect to the sampling window. The right-hand side of Figure \ref{fig:phillips_res_dyn_diff}  shows the time plot of the residuals for the case for which the null of white noise is not rejected, i.\,e., $p=r=12$ in equation \eqref{eq:phillips_dyn_diff}. The plot suggests that there is virtually no autoregressive conditional heteroskedasticity. We also plot the residuals from this regression against the most relevant (recent) regressors, i.\,e., against $\Delta \pi_{t-1}$ and $u_{t}$, in Figure \ref{fig:phillips_res_against_x_diff}. Again, there seems to be virtually no conditional heteroskedasticity. Thus, the  assumption of conditional homoskedasticity seems to be a good approximation of reality.
					
			
			
			As a robustness check with respect to the time period, we run the same dynamic regression models from 1985 onwards instead of 1961, see  Figures~\ref{fig:phillips_sig_dyn_diff_1985} and  \ref{fig:phillips_sig_dyn_level_1985}. The empirical autocorrelations and significance bands look very similar to the ones above. For both models the null of white noise is again clearly rejected in the case $p=1$ and $r=0$. Interestingly, for the model in differences we have a different test decision of the naive and the corrected classical band 
            for $p=r=12$ due to lag 4 on the right-hand side of Figure~\ref{fig:phillips_sig_dyn_diff_1985}, which falls a tiny bit out of the corrected band. Lastly, tables \ref{tab:phillips_estimates_diff} and \ref{tab:phillips_estimates_level} show that most lags of unemployment are insignificant at $10\%$ significance level. Discarding them in equations \eqref{eq:phillips_dyn_diff} and \eqref{eq:phillips_dyn_level} yields very similar plots of the empirical autocorrelation functions and the respective inference bands, which we do not report here.
			
\section{Extensions to other Time Series Correlations}
\label{Sect:extensions_time_series_correlations}

                         Extensions may address further classical time series correlations. First, asymptotic significance bands for cross-correlations are immediately available. Hypotheses of interest could be ``no cross-correlation up to a certain lead and lag'', ``no cross-correlations and one process is WN'', ``no cross-correlation and both processes are WN''. Limiting normality arises, where the shape of the Bartlett covariance matrix depends on the specified null hypothesis; see \citet[Sect. 11.1.3]{Box1994} for details. Second, bands for partial autocorrelations up to some order $H$ are straightforward, see \citet[Sect. 3.2.7]{Box1994} for limiting normality and references. The asymptotic covariance matrix depends on the null hypothesis under test, e.\,g., AR($p$) for some specified $p$. Third, our approach can be carried to a multivariate framework. \citet[Prop. 4.6]{Lutkepohl2005} establishes limiting normality and provides the covariance matrix for a vector of residual autocorrelations from  vector autoregressions under the null hypothesis of WN. Since vector autoregressions contain lagged endogenous regressors, the covariance matrix displays a similar structure like in Proposition \ref{Prop:ARDL}.

\end{document}